\documentclass[
 reprint,
unsortedaddress,
nobibnotes,
 amsmath,amssymb,
 aps,
prl
]{revtex4-2}

\usepackage{graphicx}%
\usepackage{dcolumn}%
\usepackage{bm}%
\usepackage{xcolor}
\usepackage{siunitx}
\usepackage{xr-hyper}
\usepackage{hyperref}
\usepackage{dsfont}
\usepackage{braket}
\usepackage{xspace}

\newcommand{\Ca}[2]{$^{#1}$Ca${}^{#2}$\xspace}
\newcommand{\Ar}[2]{$^{#1}$Ar${}^{#2}$\xspace}
\newcommand{\Be}[2]{$^{#1}$Be${}^{#2}$\xspace}
\newcommand{\Yb}{Yb${}^{+}$\xspace}

\newcommand{\CaStoD}[1]{${{}^{2}S_{1/2} \rightarrow {}^{2}D_{#1}}$\xspace}
\newcommand{\CaDtoD}{${{}^{2}D_{3/2} \rightarrow {}^{2}D_{5/2}}$\xspace}
\newcommand{\CaPtoP}{${{}^{3}P_{0}~\rightarrow~{}^{3}P_{1}}$\xspace}

\begin{document}

\title{Nonlinear calcium King plot constrains new bosons and nuclear properties}%

 \author{Alexander Wilzewski}
 \thanks{These authors contributed equally.}
 \author{Lukas J. Spie{\ss}}
 \author{Malte Wehrheim}
 \author{Shuying Chen}
 \author{Steven A. King}
 \author{Peter Micke}
 \author{Melina Filzinger}
 \author{Martin R. Steinel}
 \author{Nils Huntemann}
 \author{Erik Benkler}
 \author{Piet O. Schmidt$^\star$}
 \affiliation{Physikalisch-Technische Bundesanstalt, Bundesallee 100, 38116 Braunschweig, Germany\\
 $^\star$Also at Institut f\"{u}r Quantenoptik, Leibniz Universit\"{a}t Hannover, Welfengarten 1, 30167 Hannover, Germany}%
 
\author{Luca I. Huber}
\thanks{These authors contributed equally.}
\author{Jeremy Flannery}
\author{Roland Matt}
\thanks{Current address Oxford Ionics, Kidlington, OX5 1GN , UK}
\author{Martin Stadler}
\author{Robin Oswald}
\author{Fabian Schmid}
\author{Daniel Kienzler}
\author{Jonathan Home}
\author{Diana P. L. \surname{Aude Craik}}
\affiliation{Institute for Quantum Electronics, Department of Physics, Eidgenössische Technische Hochschule Z\"{u}rich, Otto-Stern-Weg 1, 8093 Zurich, Switzerland}%

\author{Menno Door}
\thanks{These authors contributed equally.}
\author{Sergey Eliseev}
\author{Pavel Filianin}
\author{Jost Herkenhoff}
\author{Kathrin Kromer}
\author{Klaus Blaum}
\affiliation{Max-Planck-Institut f\"{u}r Kernphysik, Saupfercheckweg 1, 69117 Heidelberg, Germany}

\author{Vladimir A. Yerokhin}
\author{Igor A. Valuev}
\author{Natalia S. Oreshkina}
\affiliation{Max-Planck-Institut f\"{u}r Kernphysik, Saupfercheckweg 1, 69117 Heidelberg, Germany}

\author{Chunhai Lyu}
\author{Sreya Banerjee}
\author{Christoph H. Keitel}
\author{Zolt\'{a}n Harman}
\affiliation{Max-Planck-Institut f\"{u}r Kernphysik, Saupfercheckweg 1, 69117 Heidelberg, Germany}

\author{Julian C. Berengut}%
\affiliation{School of Physics, University of New South Wales, Sydney, NSW 2052, Australia\\
and UNSW Nuclear Innovation Centre, UNSW Sydney, Kensington, NSW 2052, Australia}

\author{Anna Viatkina}%
\author{Jan Gilles}
\author{Andrey Surzhykov}
\affiliation{Physikalisch-Technische Bundesanstalt, Bundesallee 100, 38116 Braunschweig, Germany\\
and Institut f\"ur Mathematische Physik, Technische Universit\"at Braunschweig, D-38116 Braunschweig, Germany}

\author{Michael K. Rosner}
\author{Jos\'{e} R. {Crespo L\'{o}pez-Urrutia}}
\affiliation{Max-Planck-Institut f\"{u}r Kernphysik, Saupfercheckweg 1, 69117 Heidelberg, Germany}

\author{Jan Richter$^{\mathsection}$}
\thanks{These authors contributed equally.}
\author{Agnese Mariotti}
\thanks{These authors contributed equally.}
\author{Elina Fuchs$^\mathsection$}

\affiliation{Institut f\"{u}r Theoretische Physik, Leibniz Universit\"{a}t Hannover, Appelstra{\ss}e 2, 30167 Hannover, Germany\\
$^\mathsection$Also at Physikalisch-Technische Bundesanstalt, Bundesallee 100, 38116 Braunschweig, Germany}

\date{\today}%
\begin{abstract}
    Nonlinearities in King plots (KP) of isotope shifts (IS) can reveal the existence of beyond-Standard-Model (BSM) interactions that couple electrons and neutrons.
    However, it is crucial to distinguish higher-order Standard Model (SM) effects from BSM physics. 
    We measure the IS of the transitions \CaPtoP in \Ca{}{14+} and \CaStoD{5/2} in \Ca{}{+} with sub-Hz precision as well as the nuclear mass ratios with relative uncertainties below \SI{4e-11}{} for the five stable, even isotopes of calcium (\Ca{40,42,44,46,48}{}).
    Combined, these measurements yield a calcium KP nonlinearity with a significance of $\sim 900 \sigma$. 
    Precision calculations show that the nonlinearity cannot be fully accounted for by the expected largest higher-order SM effect, the second-order mass shift, and identify the little-studied nuclear polarization as the only remaining SM contribution that may be large enough to explain it.
    Despite the observed nonlinearity, we improve existing KP-based constraints on a hypothetical Yukawa interaction for most of the new boson masses between $10~\mathrm{eV/c^2}$ and $10^7~\mathrm{eV/c^2}$. 

\end{abstract}

\maketitle

    \textit{Introduction}.---The Standard Model of particle physics (SM) is incomplete: it cannot, for instance, explain the observed matter-antimatter asymmetry of the universe~\cite{Gavela:1993ts,Huet:1994jb}, does not contain a viable dark matter candidate (see e.g.~\cite{Feng:2010gw} for a review) and lacks neutrino masses inferred from the observation of  neutrino oscillations~\cite{Super-Kamiokande:1998kpq}. 
    The quest to discover beyond-Standard-Model (BSM) physics requires searches across different frontiers and technologies, including accelerators at the energy frontier, rare processes at the intensity frontier, observations at the cosmic frontier, and low-energy experiments at the precision frontier.  
    Precision isotope shift (IS) spectroscopy has been used to search for a new fundamental interaction mediated by a hypothetical scalar boson $\phi$ that would couple electrons and neutrons~\cite{delaunay_probing_2017, berengut_probing_2018}.
    To lowest order, the SM predicts a linear relation between the IS of two transitions in a so-called King plot (KP)~\cite{king_comments_1963}, but an additional bosonic field would introduce a nonlinearity (NL) there.
    However, higher-order effects within the SM can also lead to a NL~\cite{berengut_probing_2018,yerokhin_nonlinear_2020,muller_nonlinearities_2021,viatkina_calculation_2023}.
    Recently, distinct sources of NL were uncovered in the KP of ytterbium, which were attributed to effects of its large and deformed nucleus rather than to a new boson~\cite{counts_evidence_2020,figueroa_precision_2022,ono_observation_2022,hur_evidence_2022,door_search_2024}.
    In contrast to Yb, the KP of the calcium isotopic chain thus far appeared to be linear within measurement uncertainties~\cite{solaro_improved_2020,chang_systematic-free_2024} since the lower mass and spherical nucleus of Ca suppress finite nuclear-size effects by three orders of magnitude relative to Yb~\cite{solaro_improved_2020,chang_systematic-free_2024}.\\
    In this Letter, we report on the observation of a nonlinear KP in Ca by combining measurements of (i) the absolute frequencies of the \CaPtoP optical transition in highly charged \Ca{A}{14+} with an IS uncertainty $<\SI{150}{\milli\hertz}$, (ii) the \CaStoD{5/2} transition in \Ca{}{+} with its IS uncertainties reduced to $<\SI{70}{\milli\hertz}$ by correlation spectroscopy, and (iii) the relative nuclear masses in a Penning trap to better than \SI{4e-11}{}.
    These measurements include all five stable and even isotopes $(A=40,42,44,46,48)$ of calcium.
    Our calculations of the second-order mass shift and nuclear polarization show that the observed NL are compatible with higher-order SM contributions.
    For most of the parameter space, our data places the most stringent KP-derived constraints to date on the existence of a new boson.
    Our bounds are limited by the current theoretical uncertainty on nuclear polarization and on the second-order mass shift in \Ca{}{+}.\\
        {\renewcommand{\arraystretch}{1.4}
        \begin{table*}
        \caption{\label{tab:experimental_results}Nuclear mass ratios $R_{AA^\prime}$ and isotope shift (IS) frequencies in \SI{}{\hertz} and their 1-$\sigma$ uncertainties in parentheses. The IS of the \CaPtoP transition in \Ca{}{14+} is denoted as $\delta\nu_{570}$, the IS of the \CaStoD{5/2} and \CaDtoD in \Ca{}{+} as $\delta\nu_{729}$ and $\delta\nu_{\textbf{DD}}$, respectively. The reference isotope is $A^\prime = 40$.}
        \begin{ruledtabular}
        \begin{tabular}{cllll}
         $A$ & $R_{AA^\prime}$ & \multicolumn{1}{l}{$\delta\nu_{570}^{AA^\prime}$}  &  \multicolumn{1}{l}{$\delta\nu_{729}^{AA^\prime}$} &\multicolumn{1}{l}{$\delta\nu_{\textbf{DD}}^{AA^\prime}$}  \\ \hline
                 
         42 & 1.049 961 066 498(15) & \;\, 539 088 421.24(12) & 2 771 872 430.217(27)\footnote{Statistical and systematic uncertainty added in quadrature.}
         & -3 519 944.6(60)\footnote{Calculated using $\delta\nu_{729}$ from this work and $\delta\nu_{732}$ from Ref.~\cite{chang_systematic-free_2024}} \\
         
         44 & 1.099 943 105 797(15) & 1 030 447 731.64(11) & 5 340 887 395.288(38)\footnotemark[1]
         & -6 792 440.1(59)\footnotemark[2]\\
         
         46 & 1.149 958 773 895(30) & 1 481 135 946.74(14) & 7 768 401 432.916(63)\footnotemark[1]
         & -9 901 524(21)\footnote{Taken from Ref.~\cite{solaro_improved_2020}}\\
         
         48 & 1.199 990 087 090(40)& 1 894 297 294.53(14) & 9 990 382 526.834(55)\footnotemark[1]
         & -12 746 588.2(57)\footnotemark[2] \\
         
        \end{tabular}
        \end{ruledtabular}
        \end{table*}
        }
    \indent \textit{Precision measurements}.---In \Ca{}{14+}, we measured the IS of the recently found magnetic-dipole-allowed \CaPtoP fine-structure transition at \SI{570}{\nano\meter}~\cite{rehbehn_sensitivity_2021} with an excited-state lifetime of \SI{11}{\milli\second}, using similar methods to those of our previous work in \Ar{}{13+}~\cite{king_optical_2022}.
    Highly charged ions (HCI) are produced in an electron beam ion trap starting from ablation-loaded isotopically enriched samples~\cite{schweiger_production_2019}.
    The ions are extracted, decelerated and re-trapped in a cryogenic Paul trap~\cite{schmoger_coulomb_2015, schmoger_deceleration_2015, micke_heidelberg_2018, micke_closed-cycle_2019, leopold_cryogenic_2019, king_algorithmic_2021}.
    Since HCI lack fast-cycling laser cooling transitions, we employ sympathetic cooling and quantum-logic spectroscopy using a co-trapped \Be{}{+} logic ion~\cite{schmidt_spectroscopy_2005, micke_coherent_2020}. 
    The \Ca{}{14+} transition was sequentially robed with a laser that was pre-stabilized to the ultra-stable Si2 cavity at PTB~\cite{matei_15text_2017} resulting in a Hz-level laser linewidth.
    The \Ca{}{14+} transition was sequentially probed with a laser that was pre-stabilized to the ultra-stable Si2 cavity at PTB~\cite{matei_15text_2017} resulting in a Hz-level laser linewidth.
   We probed the three transitions to the excited-state Zeeman manifold with this laser.
   The laser was steered to the average frequency of the three transitions, which is then free of the linear Zeeman shift and the quadrupole shift.
     From the measured frequency ratios and the \Yb\ frequency, we extract the absolute frequencies and the IS $\delta\nu_{570}^{A A^\prime} = (\rho^A - \rho^{A^\prime}) \times \nu_\text{Yb}$ between isotope $A$ and a reference isotope $A^\prime = 40$, shown in Tab.~\ref{tab:experimental_results}.
    We attain an overall uncertainty of the IS of about \SI{150}{\milli\hertz}, which is dominated by the statistical uncertainty.
    Details are discussed in the Supplemental Material~\cite{noauthor_see_nodate} and Refs.~\cite{peik_laser_2005,webster_force-insensitive_2011,leopold_tunable_2016,dawel_coherent_2024,monroe_resolved-sideband_1995,wilzewski_isotope_2024,spies_excited-state_2024,gilles_quadratic_2024,keller_precise_2015,yu_investigating_2019,kramida_nist_nodate,denker_ergebnisse_2022,jonsson_energies_2011}.\\
    To perform IS measurements of the \CaStoD{5/2} electric-quadrupole transition in \Ca{}{+}, we employed correlation spectroscopy~\cite{roos_precision_2005,chwalla_precision_2007,manovitz_precision_2019}. 
    We co-trapped the reference \Ca{A^\prime}{+} ion with another \Ca{A}{+} isotope in a linear Paul trap. After an initial $\pi/2$-pulse on both ions, the two-ion state is in the superposition
    \begin{equation}
        \label{eq:mixedState}
        \ket{\psi}  = \frac{1}{2}(\ket{g_A g_{A^\prime}} + \ket{e_A e_{A^\prime}}) + \frac{1}{2} (\ket{g_A e_{A^\prime}} + \ket{e_A g_{A^\prime}}),
    \end{equation}
    where $\ket{g}$ is a Zeeman state in the $S_{1/2}$-submanifold and $\ket{e}$ is a state in the $D_{5/2}$-submanifold. 
    The state $\frac{1}{\sqrt{2}}(\ket{g_A g_{A^\prime}} + \ket{e_A e_{A^\prime}})$ is subject to decoherence and leads to a constant background in the final signal for the measurement times used in our experiment. 
    In contrast, the state $\ket{\Psi^+} = \frac{1}{\sqrt{2}}(\ket{g_A e_{A^\prime}} + \ket{e_A g_{A^\prime}})$ is in a decoherence-free subspace that rejects common-mode noise and precesses at the IS frequency during the Ramsey waiting time.
    A final $\pi/2$-pulse on both ions imprints the accumulated phase onto the parity of the two-ion state, which was then extracted by optical state detection.
    We conducted measurements at five different Ramsey waiting times: \SI{30}{\milli\second}, \SI{50}{\milli\second}, \SI{100}{\milli\second}, \SI{300}{\milli\second}, and \SI{325}{\milli\second}.
    For each of them, we scanned the relative phase in the second $\pi/2$-pulse between the two ions to unambiguously determine the phase accumulated during the Ramsey waiting time.
    From a linear fit to the measured phases we extract the IS frequency.
    The two isotopes were addressed simultaneously by driving a broadband electro-optical modulator at the two frequencies required by the pair of Ca ions. 
    The frequency difference was generated by a microwave signal generator referenced to a GPS-disciplined rubidium clock.
    The IS was measured for two transitions, $|S_{1/2}, m_J=-1/2\rangle \rightarrow |D_{5/2}, m_J=-1/2\rangle$~ and $|S_{1/2}, m_J=+1/2\rangle \rightarrow |D_{5/2}, m_J=+1/2\rangle$, as well as with the ions in two different spatial configurations in which the ions were aligned in the order $A-A^\prime$ or $A^\prime-A$ along the trap axis. 
    Averaging over these four measurements suppresses the differential first-order Zeeman shift caused by differential Land\'{e} $g$-factors, a magnetic field gradient and other potentially position-dependent systematic shifts such as electric-quadrupole shifts or AC Stark shifts due to laser light leakage~\cite{manovitz_precision_2019}. 
    Details of the setup, methods, and uncertainty budget are discussed in the Supplemental Material~\cite{noauthor_see_nodate} and Refs.~\cite{lucas_isotope-selective_2004,roos_designer_2006,matt_roland_parallel_2023,chwalla_absolute_2009,berkeland_minimization_1998,noauthor_58503b_2003,knollmann_part-per-billion_2019}.\\
    The nuclear mass ratios were determined with the Penning-trap mass spectrometer PENTATRAP by measuring the cyclotron frequency ratios between highly charged Ca isotopes in four stacked identical Penning traps, combined with high-precision calculations of their electronic binding energies, as in our recent work on Yb isotope mass ratios~\cite{door_search_2024}.
    A single ion is loaded in each of the three trapping regions in alternating isotope sequence $A~-~A^\prime~-~A$, while leaving the top trap empty.
    In the two central traps, the free cyclotron frequency $\omega_c^{A} = q_AB/M_A$ of the trapped ions with charge-to-mass ratio $q_A/M_A$ were determined in the magnetic field $B$. 
    Note that $M_A$ includes the electron masses and their binding energies. 
    We obtain $\omega_c^{A}$ by measuring the axial frequency $\omega_z^A$, the modified cyclotron frequency $\omega_+^A$ and the magnetron frequency $\omega_-^A$ of the ions and applying the invariance theorem $(\omega_c^{A})^2=(\omega_+^{A})^2+(\omega_z^{A})^2+(\omega_-^{A})^2$~\cite{brown_precision_1982}. 
    The three eigenfrequencies were directly measured using an axial detection system ($\omega_z^A$), or indirectly using mode coupling~\cite{cornell_mode_1990} to the axial motion ($\omega_-^A$) and additional phase-sensitive methods ($\omega_+^A$) for highest precision~\cite{cornell_single-ion_1989}. 
    By shuttling the ions after each measurement sequence between neighboring traps, the ion species (and isotopes) were permuted in the central traps to determine the cyclotron-frequency ratio of the two ions in the same magnetic field.
    The axial motion with its eigenfrequency $\omega_z^A$ is proportional to $\sqrt{q_AU_0/M_A}$, with the voltage $U_0$ being the axial trap depth.
    We tune $U_0$ to bring the axial motion in resonance with the detection system.
    We chose charge states between $10+$ and $15+$ for the different isotopes to reduce the difference in $q_A/M_A$, and tuned the axial detection system to keep the same potential in the central traps between measurements of ion pairs~\cite{door_search_2024}.
    Through the measured ratios $\textrm{R}_{\omega_c}=\omega_c^{A^\prime}/\omega_c^{A}=q_{A^\prime}/q_A \times M_A/M_{A^\prime}$, we determine the ratio of the nuclear masses $m_A$ and $m_{A^\prime}$ as
     \begin{equation}
     \begin{split}
         R_{AA^\prime} &= \frac{m_A}{m_{A^\prime}} \\
         &= \textrm{R}_{\omega_c} + \frac{\textrm{R}_{\omega_c} (\bar{q}_{A^\prime} m_e - E^{\bar{q}_{A^\prime}}_{A^\prime}) - (\bar{q}_A m_e - E^{\bar{q}_A}_A)}{m(^{A^\prime}\textrm{Ca}) - 20m_e + E^{20}_{A^\prime}} \;.
    \end{split}
    \end{equation}
    Here, $A^\prime = 40$ is the reference isotope and $\bar{q}_A=20-q_A$ the number of bound electrons in the ionic charge state for isotope $A$.
    Then, $E^{20}_{A^\prime}$ and $E^{\bar{q}_A}_A$ are the total electronic binding energies of the corresponding neutral and highly charged isotopes, respectively. 
    We calculated those with uncertainties below \SI{0.2}{\electronvolt} using an \textit{ab initio} method implemented in the GRASP2018 code~\cite{froese_fischer_grasp2018fortran_2019} (see Supplemental Material ~\cite{noauthor_see_nodate} and Refs.~\cite{Grant1970,Desclaux1971,grant2007relativistic,GRASP2018,Jonsson2023-2,PhysRevA.90.062517,PhysRevLett.126.183001,Jonsson2023,lyu2023extreme}). 
    The neutral mass $m(^{40}\text{Ca})$~\cite{wang_ame_2021} and the electron mass $m_e$~\cite{tiesinga_codata_2021} are taken from literature.
    The final nuclear mass ratios are given in Tab.~\ref{tab:experimental_results}.\\
    \indent \textit{Nonlinear King plot.}---To leading order, the IS of a transition $i$ is the sum of mass shift (MS) and field shift, each factorizable into electronic and nuclear parts~\cite{king_comments_1963}:
        \begin{equation}\label{eq:is}
            \delta\nu_i^{A A^\prime} = K_i \, \mu^{A A^\prime} + F_i \, \delta \langle r^2 \rangle^{A A^\prime}\,.
        \end{equation}
    The first term reflects the mass-dependent nonrelativistic recoil of the nucleus, proportional to $\mu^{A A^\prime}~=~\frac{m_A - m_{A^\prime}}{m_A m_{A^\prime}} =  \frac{1}{m_{A^\prime}}\frac{R_{AA^\prime}-1}{R_{AA^\prime}}$
    and to the electronic coefficient $K_i$.
    The second one originates from the finite size of the nucleus and is proportional to the difference in mean-squared nuclear charge radii, $\delta \langle r^2 \rangle^{A A^\prime} = \langle r^2 \rangle^A - \langle r^2 \rangle^{A^\prime}$ and to the electronic coefficient $F_i$.
    The relatively poorly known radii can be eliminated with an IS measurement of a second transition $j$, yielding the linear King relation~\cite{king_comments_1963}:
        \begin{equation}\label{eq:king_relation}
            \delta \bar{\nu}_j^{A A^\prime} = K_{ji} + F_{ji} \delta \bar{\nu}_i^{A A^\prime},
        \end{equation}
    where $\delta \bar{\nu}_{i,j}^{A A^\prime} = \delta \nu_{i,j}^{A A^\prime} / \mu^{A A^\prime} $ are the modified isotope shifts, $F_{ji} \equiv {F_j}/{F_i}$ and $K_{ji} \equiv K_j - F_{ji}K_i$.
    In combination with our mass measurements, our IS measurements of \CaPtoP in \Ca{}{14+} and \CaStoD{5/2} in \Ca{}{+},which we label "570" and "729" in Tab.~\ref{tab:experimental_results}, respectively, reveal a significant NL of $\sim900\sigma$ of the KP, shown in Fig.~\ref{fig:king_plot} (definition of our NL measure in the Supplemental Material~\cite{noauthor_see_nodate}). 
    \\
    \begin{figure}
        \includegraphics[width=0.4\textwidth]{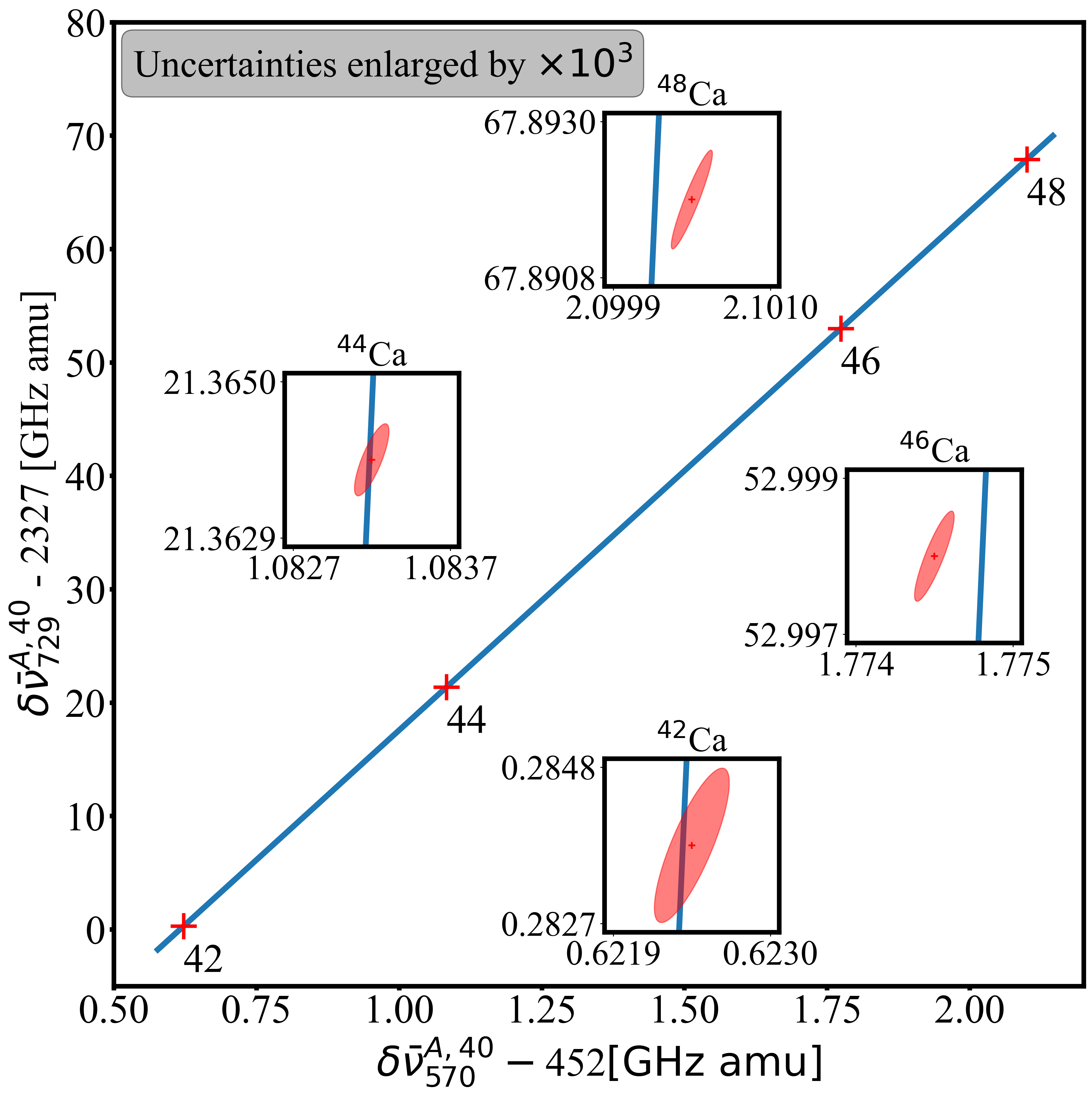}
        \caption{A nonlinearity with a significance of $\sim900\sigma$ appears in the King plot displaying the modified isotope shifts of the transitions \CaStoD{5/2} in \Ca{}{+} ($\delta \bar{\nu}_{729}$) and those of \CaPtoP in \Ca{}{14+} ($\delta \bar{\nu}_{570}$).
        The blue line displays a linear orthogonal-distance regression to the data (red crosses) of Tab.~\ref{tab:experimental_results}. 
        Insets:  Zoomed-in region around the data points, in which the uncertainties displayed by the red ellipses were magnified by a factor 1000 to aid visibility.}
        \label{fig:king_plot}
    \end{figure}
    \begin{figure}[b]
        \centering
        \includegraphics[width=\linewidth]{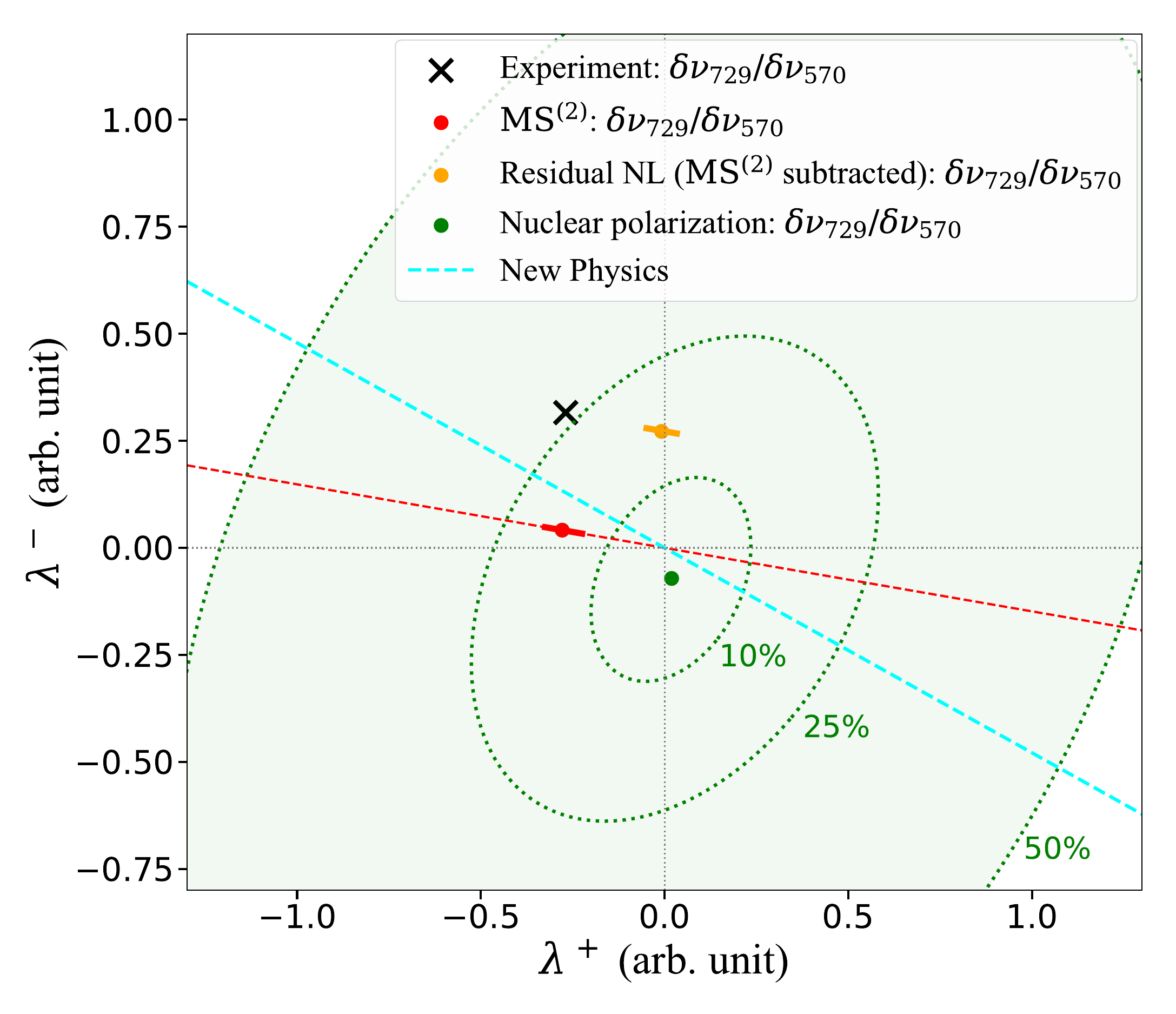}
        \caption{Decomposition plot comparing the measured NL (black cross, uncertainty not visible at this scale) with predictions of different NL sources. Red dot (solid bars denote uncertainties): Calculated second-order MS. Orange dot: Residual NL after subtracting the second-order MS from the data. Green dot: Nuclear polarization contribution, with ellipses bounding possible values assuming theoretical uncertainties of $50\%$, $25\%$ and $10\%$. %
        Cyan line: Direction of new boson contribution.}
        \label{fig:NLdecomposition}
    \end{figure}
    \indent \textit{Nonlinearity decomposition.}---A NL can result from either higher-order SM contributions, or BSM physics.
    To account for both of these, we extend Eq.~\eqref{eq:king_relation} following Ref.~\cite{berengut_generalized_2020}:
        \begin{align}\label{eq:king_plus_NL}
            \vec{\delta \bar{\nu}}_j = K_{ji}\vec{1} + F_{ji} \vec{\delta \bar{\nu}}_i + \sum_{\ell=1}^\infty G_{ji}^{(\ell)}\vec{\bar{\eta}}_{\,(\ell)}\,,
        \end{align}
    where $\ell$ runs over the possible higher-order terms.
    With the possible exception of the nuclear polarizability contribution (discussed later), these terms factorize into an electronic part, $G_{ji}^{(\ell)} = G_{j}^{(\ell)}-F_{ji}G_{i}^{(\ell)}$ with isotope-independent coefficients $G_{i}^{(\ell)}$, and nuclear parameters $\vec{\bar{\eta}}_{\,(\ell)}$~\cite{viatkina_calculation_2023}, which we express as mass-normalized vectors in isotope-pair space (four-dimensional, in our case)~\cite{berengut_generalized_2020}.
    The vectors $\vec{1}$ and $\vec{\delta \bar{\nu}}_i$ in Eq.~\eqref{eq:king_plus_NL} span the plane of King linearity in isotope-pair space. 
    Each source of NL produces a signature pattern of residuals in the KP and has a finite projection on two out-of plane vectors, $\vec\Lambda_+$ and $\vec\Lambda_-$.
    Together, the four vectors form a basis such that 
        \begin{equation}\label{eq:NL_dec}
            \vec{\delta \bar{\nu}}_j=K_{ji} \vec{1}+F_{ji}\vec{\delta \bar{\nu}}_i+ \left(\lambda_+ \vec\Lambda_+ + \lambda_- \vec\Lambda_-\right)\,.
        \end{equation}
    The vectors $\vec\Lambda_+ \propto (\delta\bar{\nu}^3_i-\delta\bar{\nu}^2_i,\delta\bar{\nu}^1_i-\delta\bar{\nu}^4_i, \delta\bar{\nu}^4_i-\delta\bar{\nu}^1_i,\delta\bar{\nu}^2_i-\delta\bar{\nu}^3_i)$ and $\vec\Lambda_- \propto (\delta\bar{\nu}^4_i-\delta\bar{\nu}^2_i,\delta\bar{\nu}^1_i-\delta\bar{\nu}^3_i, \delta\bar{\nu}^2_i-\delta\bar{\nu}^4_i,\delta\bar{\nu}^3_i-\delta\bar{\nu}^1_i)$ span the patterns of residuals for the four points around the straight-line fit, where the superscripts (1, 2, 3, 4) label the isotope pairs (40-42, 40-44, 40-46, 40-48).
    We can hence decompose any pattern of residuals in the KP into a vector sum of the contributing effects, using the decomposition analysis pioneered in Refs.~\cite{counts_evidence_2020, hur_evidence_2022}.
    This analysis translates the KP residuals into a point in ($\lambda_+$, $\lambda_-$)-space; our data yields the black cross in Fig.~\ref{fig:NLdecomposition}.
    Each potential source of NL can only contribute a vector along a specific direction in this plot, because the nuclear dependence of the effect fixes the ratio $\lambda_+/\lambda_-$.
    The electronic coefficient, $G_{ji}^{(\ell)}$ in Eq.~\eqref{eq:king_plus_NL}, determines the magnitude of the effect's vector contribution. 
    Based on previous~\cite{viatkina_calculation_2023} and our current calculations, two SM effects contribute to the NL we observe: second-order MS and nuclear polarization, discussed later. 
    Other higher-order effects, like nuclear deformation~\cite{sun_nuclear_2024} are expected to be significantly smaller (see Supplemental Material~\cite{noauthor_see_nodate}). 
    A scalar boson, $\phi$, coupling neutrons and electrons could also contribute to the NL, by generating a Yukawa potential: 
    \begin{equation}
        V_{\phi}(r,m_\phi) = -\alpha_\mathrm{BSM} (A-Z) \frac{e^{-m_\phi r}}{r}\,,
    \end{equation}
    where $m_\phi$ is the boson mass, and $\alpha_\mathrm{BSM} =\frac{y_e y_n}{4\pi}$ represents the coupling strength of the boson to electrons ($y_e$) and neutrons ($y_n$). 
     The contribution to the IS is parametrized by an electronic coefficient $G^{(1)}_{i} = \alpha_\mathrm{BSM} X_{i}$ and nuclear parameter $ \bar\eta_{\,(1)}^{A A^\prime} = \gamma^{A A^\prime}/\mu^{A A^\prime}=(A-A^\prime)/\mu^{A A^\prime}$ (which determines the direction of the new boson contribution to be along the cyan line in Fig.~\ref{fig:NLdecomposition}).
    The $X_{i}$ are differences of the overlap of the Yukawa potential with the electronic wave-functions of the involved electrons and were calculated using the AMBiT code (see Supplemental Material~\cite{noauthor_see_nodate} and Ref.~\cite{kahl_ambit_2019}). %
    The second-order MS has electronic coefficient $G^{(2)}_{i}~=~K^{(2)}_{i}$ and nuclear parameter $\bar{\eta}_{\,(2)}^{A A^\prime}~=~\mu_{(2)}^{AA^\prime}/\mu^{A A^\prime}~=~\frac{1}{m_{A^\prime}^2} \frac{R_{AA^\prime}^2-1}{R_{AA^\prime}^2}/\mu^{A A^\prime}$ (red line in Fig.~\ref{fig:NLdecomposition}). 
    We calculated $K^{(2)}_{570} = \SI{-1.0(1)}{\giga\hertz}\times\mathrm{amu}^2$ (see Supplemental Material~\cite{noauthor_see_nodate} and Refs.~\cite{yerokhin_hyperfine_2008,naze_isotope_2014,yerokhin_nonlinear_2020}) and combined it with the previously calculated value $K^{(2)}_{729} = \SI{0.59}{\giga\hertz}\times\mathrm{amu}^2$~\cite{viatkina_calculation_2023} to predict the magnitude of this effect, shown as the red dot in Fig.~\ref{fig:NLdecomposition}.
    Evidently, this effect alone cannot account for our observed NL, since the direction of the red line does not intersect the black cross.
    The calculated values of $K^{(2)}_{i}$ enable us to subtract the second-order MS from the data, obtaining the residual NL (orange dot in Fig.~\ref{fig:NLdecomposition}).
    This cannot be explained by a new-boson contribution alone, since it's not intersected by the cyan line.
    To estimate whether it can be explained instead by nuclear polarization, we use existing calculations for \Ca{}{+}~\cite{viatkina_calculation_2023} and perform new ones for \Ca{}{14+} using the framework and nuclear model of Ref.~\cite{viatkina_calculation_2023} (see Supplemental Material~\cite{noauthor_see_nodate} and Ref.~\cite{plunien_nuclear_1991,nefiodov_nuclear_1996}).
    We treat the nuclear polarization in the Coulomb approximation with contributions from the dominant nuclear rotational transition and estimates for the giant resonances up to octupole order~\cite{valuev_full_2024}.
    Assigning an uncertainty to the calculated value is difficult, especially for differential quantities, since they are inevitably phenomenological and depend on various model assumptions and parameters.
    Since this effect essentially mixes nuclear and electronic subsystems and is not \textit{a priori} factorizable, we cannot assign to it a definitive direction on the plot.
    Instead, we show in Fig.~\ref{fig:NLdecomposition} the center value of the NL contributed by the calculated nuclear polarization (green dot), alongside uncertainty ellipses bounding its possible values. 
    We estimate $50\%$ uncertainty on the IS of the calculated ratio functions $g_{ab}$ introduced in the Supplemental Material~\cite{noauthor_see_nodate}. 
    Since this uncertainty is dominated by the nuclear model, we treat it here as correlated across the two transitions, but uncorrelated across isotope pairs. 
    The resulting uncertainty ellipse overlaps with our measured NL, making it compatible with the SM.
    A reduction to 10\% uncertainty would allow us to elucidate whether the residual NL is entirely explained by nuclear polarization.\\
    \begin{figure}
        \centering
        \includegraphics[width=\linewidth]{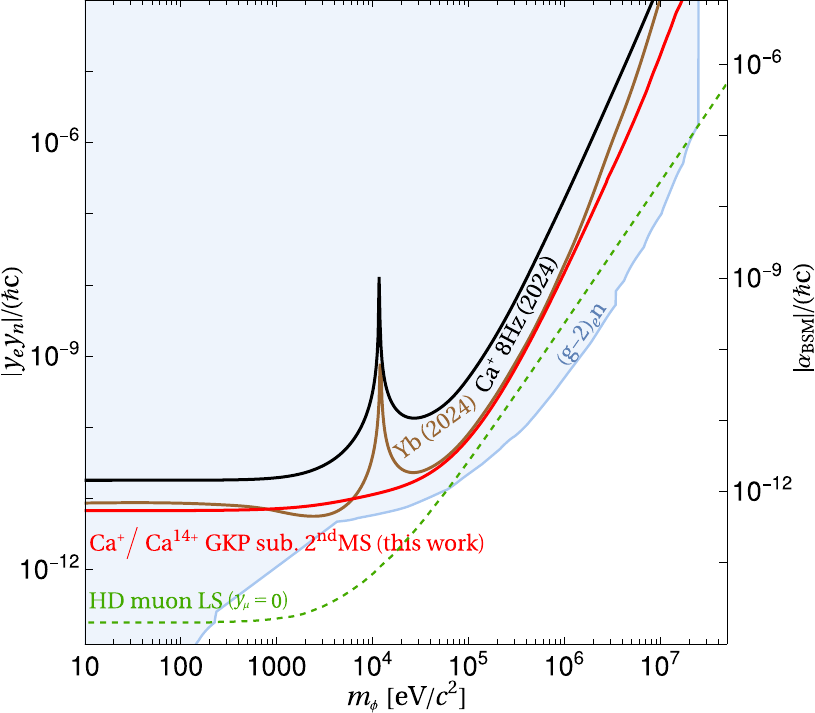}
        \caption{Laboratory bounds on the BSM coupling $y_ey_n = 4 \pi \, \alpha_\mathrm{BSM}$: recent bounds from King plots (KP) in \Ca{}{+}~\cite{chang_systematic-free_2024} (black) and Yb~\cite{door_search_2024} (brown). 
        The red curve results from our generalized KP analysis~\cite{berengut_generalized_2020} with the \CaStoD{5/2}, \CaDtoD~\cite{solaro_improved_2020, chang_systematic-free_2024} and \CaPtoP transitions, after subtracting the theoretical prediction for the second-order MS from the data, limiting the effect of known NL from the SM.
        The light-blue region is excluded by bounds on $y_e$ from $(g-2)_e$~\cite{hanneke_new_2008, fan_measurement_2023,Pospelov:2008zw} combined with constraints
        ~\cite{Frugiuele:2016rii} on $y_n$ from neutron scattering experiments~\cite{leeb_constraint_1992, nesvizhevsky_neutron_2008, pokotilovski_constraints_2006, barbieri_evidence_1975}.
        Isotope-shift bounds between H and D using the nuclear charge radius extracted from Lamb shift (LS) measurements in muonic atoms~\cite{Delaunay:2017dku} are shown in green.
        For a detailed  comparison of these bounds, see the Supplemental Material~\cite{noauthor_see_nodate} and Refs.~\cite{gebert_precision_2015,raffelt_limits_2012,grifols_constraints_1986,grifols_constraint_1989,michael_bordag_advances_2009}
        }
        \label{fig:exclusion_plot}
    \end{figure} 
    \indent \textit{Constraints on new bosons.}---Despite the observed NL, we improve the constraints on BSM physics with our combined study.
    We conduct a generalized KP (GKP) analysis~\cite{berengut_generalized_2020}, in which higher-order NL sources are eliminated by including additional IS data, akin to the elimination of the charge radii in the original KP. 
    Combining our data with the \CaDtoD fine-structure transition in \Ca{}{+} from  Refs.~\cite{solaro_improved_2020, chang_systematic-free_2024} ($\delta \nu_{\mathrm{DD}}$ in Tab.~\ref{tab:experimental_results}), we can produce a three-transition GKP and account for one NL source in Eq.~\eqref{eq:king_plus_NL}.
    We hence assign the NL sources as follows: the first source is the second-order MS and the second is nuclear polarization. %
    We remove the first by subtracting it from the IS data.
    For this, we calculate $K^{(2)}_{570}$ to 10\,\% uncertainty~\cite{noauthor_see_nodate} and set the uncertainty for the existing values for $K^{(2)}_{729}$ and $K^{(2)}_\mathrm{DD}$~\cite{viatkina_calculation_2023} to 100\%.
    We eliminate a second NL source by performing a GKP analysis. 
    The resulting GKP is linear (NL significance $<1\sigma$), which implies that no further sources of NL contribute significantly.
    The linear GKP thus allows us to exclude, with 95\% confidence, the parameter space above the red line in Fig.~\ref{fig:exclusion_plot} (method in Supplemental Material~\cite{noauthor_see_nodate}).
    This results in the most stringent King plot-based bound to date on a new boson $\phi$ for most of the parameter space.
    Our bound is limited only by the uncertainty on the second-order MS coefficients of \Ca{}{+} and by the measurement precision of $\delta \nu_\mathrm{DD}$ ($\sim20$\,Hz, from Ref.~\cite{solaro_improved_2020,chang_systematic-free_2024}).\\
    \indent \textit{Conclusion.}---The fact that the GKP becomes linear after subtraction of the second-order MS
    (the NL volume decreases by a factor of five while its uncertainty only increases by about a factor of 1.5)
    indicates that the remaining source of NL, probably nuclear polarization, is dominated by one factorizable term. 
    Improved calculations of the \Ca{}{+} second-order MS and a sub-Hz measurement of the \CaDtoD transition in the \Ca{}{+} (or any other narrow-linewidth transition in Ca) would confirm the factorizablity of nuclear polarization if the GKP linearity holds within the reduced uncertainties. 
    Moreover, adding another accurately measured transition for all isotopes would permit a second NL decomposition.
    If the direction of the residual NL in both plots coincide, it would strengthen the evidence that the residual NL is caused by nuclear polarization and is dominated by one factorizable term.
    After subtraction of the second-order MS, the GKP analysis would then fully eliminate the NL introduced by nuclear polarization,
    and, as we show in the Supplemental Material~\cite{noauthor_see_nodate}, BSM-sensitivity would breach the $(g-2)_e \cdot n$ bound into parameter space unconstrained by other laboratory probes.
    
    \indent \textit{Acknowledgments.}---We thank Sonia Bacca, Xavier Rocca Maza, Krzysztof Pachucki and Aldo Antognini for useful discussions on nuclear polarization.
    The isotopes \Ca{40,42,44,48}{} used in this research were supplied by the U.S. Department of Energy Isotope Program, managed by the Office of Isotope R\&D and Production.
    The project was supported by the Physikalisch-Technische Bundesanstalt, the Max-Planck Society, the Max-Planck-Riken-PTB-Center for Time, Constants and Fundamental Symmetries, and the Deutsche Forschungsgemeinschaft (DFG, German Research Foundation) through SCHM2678/5-2, SU 658/4-2, 544815538, the collaborative research centers SFB 1225 ISOQUANT and SFB 1227 DQ-\textit{mat}, and under Germany's Excellence Strategy -- EXC-2123 QuantumFrontiers -- 390837967. This research was supported by the Munich Institute for Astro-, Particle and BioPhysics (MIAPbP), which is funded by the Deutsche Forschungsgemeinschaft (DFG, German Research Foundation) under Germany's Excellence Strategy -- EXC-2094 -- 390783311. The project 20FUN01 TSCAC has received funding from the EMPIR programme co-financed by the Participating States and from the European Union's Horizon 2020 research and innovation program. This project has received funding from the European Research Council (ERC) under the European Union's Horizon 2020 research and innovation program (grant agreement numbers 101019987-FunClocks and 832848-FunI). S.A.K. acknowledges financial support from the Alexander von Humboldt Foundation.
    This project has received funding from the European Union's Horizon 2020 research and innovation programme under the Marie Skłodowska-Curie Grant Agreement No. 795121.
    This project reveived funding from the Intelligence Advanced Research Projects Activity (IARPA), via the US Army Research Office grant W911NF-16-1-0070.

    \indent \textit{Author contribution}.---For IS measurements of \Ca{}{14+}, A.W., M.W., S.C., L.J.S., E.B., M.F., M.R.S. and N.H. carried out the experiment; A.W., M.W. and L.J.S. analyzed the data; A.W., M.K.R., P.O.S. and J.R.C.L.-U. developed and modified the experimental setup; S.A.K., P.M., J.R.C.L.-U. and P.O.S. conceived and supervised the study; A.W. wrote the respective part of the manuscript with input from all co-authors.
    For IS measurements of \Ca{}{+}, L.I.H., J.F., R.M. and D.P.L.A.C. carried out the experiment. M.S. built the control system. R.O. and R.M. built the foundations of the experimental setup. L.I.H. and R.M. calculated the systematic effects, F.S. contributed to calculating the clock systematic, and L.I.H., J.F. and D.P.L.A.C. analysed the data. D.P.L.A.C. performed the first nonlinearity decomposition analysis of our King plot, identified that the 2nd order MS could not account for the full observed nonlinearity and, with theory input from V.A.Y., I.A.V. and N.S.O., established that the nuclear polarization could plausibly  account for the residual nonlinearity. L.I.H, J.F., J.H. and D.P.L.A.C. wrote parts of the manuscript with input from all authors. R.M. proposed the measurement of IS in \Ca{}{+} at high precision. J.H. conceived the dual-isotope apparatus and D.K., J.H. and D.P.L.A.C. supervised the study.
    For the mass-ratio measurements, S.E. and P.F. carried out the experiment; P.F. and M.D. developed the data analysis software; M.D. and S.E. analyzed the data; J.H. and K.K. gave professional advices; K.B. conceived and supervised the study; M.D. wrote the respective part of the manuscript with input from all co-authors.
    A.V., J.G. and A.S. have contributed to the theoretical analysis. I.A.V., V.A.Y. and N.S.O. performed calculations of isotope shift constants. C.L., S.B., C.H.K., and Z.H. performed the electron binding-energy calculations to derive nuclear mass ratios.
    A.M., J.R., E.F. and J.C.B. performed calculations of the BSM physics sensitivities and contributed to the decomposition analysis and the data analysis to generate limits on BSM physics.


\begin{thebibliography}{93}%
	\makeatletter
	\providecommand \@ifxundefined [1]{%
		\@ifx{#1\undefined}
	}%
	\providecommand \@ifnum [1]{%
		\ifnum #1\expandafter \@firstoftwo
		\else \expandafter \@secondoftwo
		\fi
	}%
	\providecommand \@ifx [1]{%
		\ifx #1\expandafter \@firstoftwo
		\else \expandafter \@secondoftwo
		\fi
	}%
	\providecommand \natexlab [1]{#1}%
	\providecommand \enquote  [1]{``#1''}%
	\providecommand \bibnamefont  [1]{#1}%
	\providecommand \bibfnamefont [1]{#1}%
	\providecommand \citenamefont [1]{#1}%
	\providecommand \href@noop [0]{\@secondoftwo}%
	\providecommand \href [0]{\begingroup \@sanitize@url \@href}%
	\providecommand \@href[1]{\@@startlink{#1}\@@href}%
	\providecommand \@@href[1]{\endgroup#1\@@endlink}%
	\providecommand \@sanitize@url [0]{\catcode `\\12\catcode `\$12\catcode
		`\&12\catcode `\#12\catcode `\^12\catcode `\_12\catcode `\%12\relax}%
	\providecommand \@@startlink[1]{}%
	\providecommand \@@endlink[0]{}%
	\providecommand \url  [0]{\begingroup\@sanitize@url \@url }%
	\providecommand \@url [1]{\endgroup\@href {#1}{\urlprefix }}%
	\providecommand \urlprefix  [0]{URL }%
	\providecommand \Eprint [0]{\href }%
	\providecommand \doibase [0]{https://doi.org/}%
	\providecommand \selectlanguage [0]{\@gobble}%
	\providecommand \bibinfo  [0]{\@secondoftwo}%
	\providecommand \bibfield  [0]{\@secondoftwo}%
	\providecommand \translation [1]{[#1]}%
	\providecommand \BibitemOpen [0]{}%
	\providecommand \bibitemStop [0]{}%
	\providecommand \bibitemNoStop [0]{.\EOS\space}%
	\providecommand \EOS [0]{\spacefactor3000\relax}%
	\providecommand \BibitemShut  [1]{\csname bibitem#1\endcsname}%
	\let\auto@bib@innerbib\@empty
	\bibitem [{\citenamefont {Gavela}\ \emph {et~al.}(1994)\citenamefont {Gavela},
		\citenamefont {Hernandez}, \citenamefont {Orloff},\ and\ \citenamefont
		{Pene}}]{Gavela:1993ts}%
	\BibitemOpen
	\bibfield  {author} {\bibinfo {author} {\bibfnamefont {M.~B.}\ \bibnamefont
			{Gavela}}, \bibinfo {author} {\bibfnamefont {P.}~\bibnamefont {Hernandez}},
		\bibinfo {author} {\bibfnamefont {J.}~\bibnamefont {Orloff}},\ and\ \bibinfo
		{author} {\bibfnamefont {O.}~\bibnamefont {Pene}},\ }\bibfield  {title}
	{\bibinfo {title} {{Standard model CP violation and baryon asymmetry}},\
	}\href {https://doi.org/10.1142/S0217732394000629} {\bibfield  {journal}
		{\bibinfo  {journal} {Mod. Phys. Lett. A}\ }\textbf {\bibinfo {volume} {9}},\
		\bibinfo {pages} {795} (\bibinfo {year} {1994})},\ \Eprint
	{https://arxiv.org/abs/hep-ph/9312215} {arXiv:hep-ph/9312215} \BibitemShut
	{NoStop}%
	\bibitem [{\citenamefont {Huet}\ and\ \citenamefont
		{Sather}(1995)}]{Huet:1994jb}%
	\BibitemOpen
	\bibfield  {author} {\bibinfo {author} {\bibfnamefont {P.}~\bibnamefont
			{Huet}}\ and\ \bibinfo {author} {\bibfnamefont {E.}~\bibnamefont {Sather}},\
	}\bibfield  {title} {\bibinfo {title} {{Electroweak baryogenesis and standard
				model CP violation}},\ }\href {https://doi.org/10.1103/PhysRevD.51.379}
	{\bibfield  {journal} {\bibinfo  {journal} {Phys. Rev. D}\ }\textbf {\bibinfo
			{volume} {51}},\ \bibinfo {pages} {379} (\bibinfo {year} {1995})},\ \Eprint
	{https://arxiv.org/abs/hep-ph/9404302} {arXiv:hep-ph/9404302} \BibitemShut
	{NoStop}%
	\bibitem [{\citenamefont {Feng}(2010)}]{Feng:2010gw}%
	\BibitemOpen
	\bibfield  {author} {\bibinfo {author} {\bibfnamefont {J.~L.}\ \bibnamefont
			{Feng}},\ }\bibfield  {title} {\bibinfo {title} {{Dark Matter Candidates from
				Particle Physics and Methods of Detection}},\ }\href
	{https://doi.org/10.1146/annurev-astro-082708-101659} {\bibfield  {journal}
		{\bibinfo  {journal} {Ann. Rev. Astron. Astrophys.}\ }\textbf {\bibinfo
			{volume} {48}},\ \bibinfo {pages} {495} (\bibinfo {year} {2010})},\ \Eprint
	{https://arxiv.org/abs/1003.0904} {arXiv:1003.0904 [astro-ph.CO]}
	\BibitemShut {NoStop}%
	\bibitem [{\citenamefont {Fukuda}\ \emph {et~al.}(1998)\citenamefont {Fukuda}
		\emph {et~al.}}]{Super-Kamiokande:1998kpq}%
	\BibitemOpen
	\bibfield  {author} {\bibinfo {author} {\bibfnamefont {Y.}~\bibnamefont
			{Fukuda}} \emph {et~al.} (\bibinfo {collaboration} {Super-Kamiokande}),\
	}\bibfield  {title} {\bibinfo {title} {{Evidence for oscillation of
				atmospheric neutrinos}},\ }\href
	{https://doi.org/10.1103/PhysRevLett.81.1562} {\bibfield  {journal} {\bibinfo
			{journal} {Phys. Rev. Lett.}\ }\textbf {\bibinfo {volume} {81}},\ \bibinfo
		{pages} {1562} (\bibinfo {year} {1998})},\ \Eprint
	{https://arxiv.org/abs/hep-ex/9807003} {arXiv:hep-ex/9807003} \BibitemShut
	{NoStop}%
	\bibitem [{\citenamefont {Delaunay}\ \emph
		{et~al.}(2017{\natexlab{a}})\citenamefont {Delaunay}, \citenamefont {Ozeri},
		\citenamefont {Perez},\ and\ \citenamefont {Soreq}}]{delaunay_probing_2017}%
	\BibitemOpen
	\bibfield  {author} {\bibinfo {author} {\bibfnamefont {C.}~\bibnamefont
			{Delaunay}}, \bibinfo {author} {\bibfnamefont {R.}~\bibnamefont {Ozeri}},
		\bibinfo {author} {\bibfnamefont {G.}~\bibnamefont {Perez}},\ and\ \bibinfo
		{author} {\bibfnamefont {Y.}~\bibnamefont {Soreq}},\ }\bibfield  {title}
	{\bibinfo {title} {Probing atomic {Higgs}-like forces at the precision
			frontier},\ }\href {https://doi.org/10.1103/PhysRevD.96.093001} {\bibfield
		{journal} {\bibinfo  {journal} {Physical Review D}\ }\textbf {\bibinfo
			{volume} {96}},\ \bibinfo {pages} {093001} (\bibinfo {year}
		{2017}{\natexlab{a}})}\BibitemShut {NoStop}%
	\bibitem [{\citenamefont {Berengut}\ \emph {et~al.}(2018)\citenamefont
		{Berengut}, \citenamefont {Budker}, \citenamefont {Delaunay}, \citenamefont
		{Flambaum}, \citenamefont {Frugiuele}, \citenamefont {Fuchs}, \citenamefont
		{Grojean}, \citenamefont {Harnik}, \citenamefont {Ozeri}, \citenamefont
		{Perez},\ and\ \citenamefont {Soreq}}]{berengut_probing_2018}%
	\BibitemOpen
	\bibfield  {author} {\bibinfo {author} {\bibfnamefont {J.~C.}\ \bibnamefont
			{Berengut}}, \bibinfo {author} {\bibfnamefont {D.}~\bibnamefont {Budker}},
		\bibinfo {author} {\bibfnamefont {C.}~\bibnamefont {Delaunay}}, \bibinfo
		{author} {\bibfnamefont {V.~V.}\ \bibnamefont {Flambaum}}, \bibinfo {author}
		{\bibfnamefont {C.}~\bibnamefont {Frugiuele}}, \bibinfo {author}
		{\bibfnamefont {E.}~\bibnamefont {Fuchs}}, \bibinfo {author} {\bibfnamefont
			{C.}~\bibnamefont {Grojean}}, \bibinfo {author} {\bibfnamefont
			{R.}~\bibnamefont {Harnik}}, \bibinfo {author} {\bibfnamefont
			{R.}~\bibnamefont {Ozeri}}, \bibinfo {author} {\bibfnamefont
			{G.}~\bibnamefont {Perez}},\ and\ \bibinfo {author} {\bibfnamefont
			{Y.}~\bibnamefont {Soreq}},\ }\bibfield  {title} {\bibinfo {title} {Probing
			{New} {Long}-{Range} {Interactions} by {Isotope} {Shift} {Spectroscopy}},\
	}\href {https://doi.org/10.1103/PhysRevLett.120.091801} {\bibfield  {journal}
		{\bibinfo  {journal} {Physical Review Letters}\ }\textbf {\bibinfo {volume}
			{120}},\ \bibinfo {pages} {091801} (\bibinfo {year} {2018})}\BibitemShut
	{NoStop}%
	\bibitem [{\citenamefont {King}(1963)}]{king_comments_1963}%
	\BibitemOpen
	\bibfield  {author} {\bibinfo {author} {\bibfnamefont {W.~H.}\ \bibnamefont
			{King}},\ }\bibfield  {title} {\bibinfo {title} {Comments on the {Article}
			``{Peculiarities} of the {Isotope} {Shift} in the {Samarium} {Spectrum}''},\
	}\href {https://doi.org/10.1364/JOSA.53.000638} {\bibfield  {journal}
		{\bibinfo  {journal} {JOSA}\ }\textbf {\bibinfo {volume} {53}},\ \bibinfo
		{pages} {638} (\bibinfo {year} {1963})}\BibitemShut {NoStop}%
	\bibitem [{\citenamefont {Yerokhin}\ \emph {et~al.}(2020)\citenamefont
		{Yerokhin}, \citenamefont {M\"{u}ller}, \citenamefont {Surzhykov},
		\citenamefont {Micke},\ and\ \citenamefont
		{Schmidt}}]{yerokhin_nonlinear_2020}%
	\BibitemOpen
	\bibfield  {author} {\bibinfo {author} {\bibfnamefont {V.~A.}\ \bibnamefont
			{Yerokhin}}, \bibinfo {author} {\bibfnamefont {R.~A.}\ \bibnamefont
			{M\"{u}ller}}, \bibinfo {author} {\bibfnamefont {A.}~\bibnamefont
			{Surzhykov}}, \bibinfo {author} {\bibfnamefont {P.}~\bibnamefont {Micke}},\
		and\ \bibinfo {author} {\bibfnamefont {P.~O.}\ \bibnamefont {Schmidt}},\
	}\bibfield  {title} {\bibinfo {title} {Nonlinear isotope-shift effects in
			be-like, {B}-like, and {C}-like argon},\ }\href
	{https://doi.org/10.1103/PhysRevA.101.012502} {\bibfield  {journal} {\bibinfo
			{journal} {Physical Review A}\ }\textbf {\bibinfo {volume} {101}},\ \bibinfo
		{pages} {012502} (\bibinfo {year} {2020})}\BibitemShut {NoStop}%
	\bibitem [{\citenamefont {M\"{u}ller}\ \emph {et~al.}(2021)\citenamefont
		{M\"{u}ller}, \citenamefont {Yerokhin}, \citenamefont {Artemyev},\ and\
		\citenamefont {Surzhykov}}]{muller_nonlinearities_2021}%
	\BibitemOpen
	\bibfield  {author} {\bibinfo {author} {\bibfnamefont {R.~A.}\ \bibnamefont
			{M\"{u}ller}}, \bibinfo {author} {\bibfnamefont {V.~A.}\ \bibnamefont
			{Yerokhin}}, \bibinfo {author} {\bibfnamefont {A.~N.}\ \bibnamefont
			{Artemyev}},\ and\ \bibinfo {author} {\bibfnamefont {A.}~\bibnamefont
			{Surzhykov}},\ }\bibfield  {title} {\bibinfo {title} {Nonlinearities of
			{King}'s plot and their dependence on nuclear radii},\ }\href
	{https://doi.org/10.1103/PhysRevA.104.L020802} {\bibfield  {journal}
		{\bibinfo  {journal} {Physical Review A}\ }\textbf {\bibinfo {volume}
			{104}},\ \bibinfo {pages} {L020802} (\bibinfo {year} {2021})}\BibitemShut
	{NoStop}%
	\bibitem [{\citenamefont {Viatkina}\ \emph {et~al.}(2023)\citenamefont
		{Viatkina}, \citenamefont {Yerokhin},\ and\ \citenamefont
		{Surzhykov}}]{viatkina_calculation_2023}%
	\BibitemOpen
	\bibfield  {author} {\bibinfo {author} {\bibfnamefont {A.~V.}\ \bibnamefont
			{Viatkina}}, \bibinfo {author} {\bibfnamefont {V.~A.}\ \bibnamefont
			{Yerokhin}},\ and\ \bibinfo {author} {\bibfnamefont {A.}~\bibnamefont
			{Surzhykov}},\ }\bibfield  {title} {\bibinfo {title} {Calculation of isotope
			shifts and {King}-plot nonlinearities in $\mathrm{Ca}^+$},\ }\href
	{https://doi.org/10.1103/PhysRevA.108.022802} {\bibfield  {journal} {\bibinfo
			{journal} {Physical Review A}\ }\textbf {\bibinfo {volume} {108}},\ \bibinfo
		{pages} {022802} (\bibinfo {year} {2023})}\BibitemShut {NoStop}%
	\bibitem [{\citenamefont {Counts}\ \emph {et~al.}(2020)\citenamefont {Counts},
		\citenamefont {Hur}, \citenamefont {Aude~Craik}, \citenamefont {Jeon},
		\citenamefont {Leung}, \citenamefont {Berengut}, \citenamefont {Geddes},
		\citenamefont {Kawasaki}, \citenamefont {Jhe},\ and\ \citenamefont
		{Vuletić}}]{counts_evidence_2020}%
	\BibitemOpen
	\bibfield  {author} {\bibinfo {author} {\bibfnamefont {I.}~\bibnamefont
			{Counts}}, \bibinfo {author} {\bibfnamefont {J.}~\bibnamefont {Hur}},
		\bibinfo {author} {\bibfnamefont {D.~P.~L.}\ \bibnamefont {Aude~Craik}},
		\bibinfo {author} {\bibfnamefont {H.}~\bibnamefont {Jeon}}, \bibinfo {author}
		{\bibfnamefont {C.}~\bibnamefont {Leung}}, \bibinfo {author} {\bibfnamefont
			{J.~C.}\ \bibnamefont {Berengut}}, \bibinfo {author} {\bibfnamefont
			{A.}~\bibnamefont {Geddes}}, \bibinfo {author} {\bibfnamefont
			{A.}~\bibnamefont {Kawasaki}}, \bibinfo {author} {\bibfnamefont
			{W.}~\bibnamefont {Jhe}},\ and\ \bibinfo {author} {\bibfnamefont
			{V.}~\bibnamefont {Vuletić}},\ }\bibfield  {title} {\bibinfo {title}
		{Evidence for {Nonlinear} {Isotope} {Shift} in $\mathrm{Yb}^+$ {Search} for
			{New} {Boson}},\ }\href {https://doi.org/10.1103/PhysRevLett.125.123002}
	{\bibfield  {journal} {\bibinfo  {journal} {Physical Review Letters}\
		}\textbf {\bibinfo {volume} {125}},\ \bibinfo {pages} {123002} (\bibinfo
		{year} {2020})}\BibitemShut {NoStop}%
	\bibitem [{\citenamefont {Figueroa}\ \emph {et~al.}(2022)\citenamefont
		{Figueroa}, \citenamefont {Berengut}, \citenamefont {Dzuba}, \citenamefont
		{Flambaum}, \citenamefont {Budker},\ and\ \citenamefont
		{Antypas}}]{figueroa_precision_2022}%
	\BibitemOpen
	\bibfield  {author} {\bibinfo {author} {\bibfnamefont {N.~L.}\ \bibnamefont
			{Figueroa}}, \bibinfo {author} {\bibfnamefont {J.~C.}\ \bibnamefont
			{Berengut}}, \bibinfo {author} {\bibfnamefont {V.~A.}\ \bibnamefont {Dzuba}},
		\bibinfo {author} {\bibfnamefont {V.~V.}\ \bibnamefont {Flambaum}}, \bibinfo
		{author} {\bibfnamefont {D.}~\bibnamefont {Budker}},\ and\ \bibinfo {author}
		{\bibfnamefont {D.}~\bibnamefont {Antypas}},\ }\bibfield  {title} {\bibinfo
		{title} {Precision {Determination} of {Isotope} {Shifts} in {Ytterbium} and
			{Implications} for {New} {Physics}},\ }\href
	{https://doi.org/10.1103/PhysRevLett.128.073001} {\bibfield  {journal}
		{\bibinfo  {journal} {Physical Review Letters}\ }\textbf {\bibinfo {volume}
			{128}},\ \bibinfo {pages} {073001} (\bibinfo {year} {2022})}\BibitemShut
	{NoStop}%
	\bibitem [{\citenamefont {Ono}\ \emph {et~al.}(2022)\citenamefont {Ono},
		\citenamefont {Saito}, \citenamefont {Ishiyama}, \citenamefont {Higomoto},
		\citenamefont {Takano}, \citenamefont {Takasu}, \citenamefont {Yamamoto},
		\citenamefont {Tanaka},\ and\ \citenamefont
		{Takahashi}}]{ono_observation_2022}%
	\BibitemOpen
	\bibfield  {author} {\bibinfo {author} {\bibfnamefont {K.}~\bibnamefont
			{Ono}}, \bibinfo {author} {\bibfnamefont {Y.}~\bibnamefont {Saito}}, \bibinfo
		{author} {\bibfnamefont {T.}~\bibnamefont {Ishiyama}}, \bibinfo {author}
		{\bibfnamefont {T.}~\bibnamefont {Higomoto}}, \bibinfo {author}
		{\bibfnamefont {T.}~\bibnamefont {Takano}}, \bibinfo {author} {\bibfnamefont
			{Y.}~\bibnamefont {Takasu}}, \bibinfo {author} {\bibfnamefont
			{Y.}~\bibnamefont {Yamamoto}}, \bibinfo {author} {\bibfnamefont
			{M.}~\bibnamefont {Tanaka}},\ and\ \bibinfo {author} {\bibfnamefont
			{Y.}~\bibnamefont {Takahashi}},\ }\bibfield  {title} {\bibinfo {title}
		{Observation of {Nonlinearity} of {Generalized} {King} {Plot} in the {Search}
			for {New} {Boson}},\ }\href {https://doi.org/10.1103/PhysRevX.12.021033}
	{\bibfield  {journal} {\bibinfo  {journal} {Physical Review X}\ }\textbf
		{\bibinfo {volume} {12}},\ \bibinfo {pages} {021033} (\bibinfo {year}
		{2022})}\BibitemShut {NoStop}%
	\bibitem [{\citenamefont {Hur}\ \emph {et~al.}(2022)\citenamefont {Hur},
		\citenamefont {Aude~Craik}, \citenamefont {Counts}, \citenamefont {Knyazev},
		\citenamefont {Caldwell}, \citenamefont {Leung}, \citenamefont {Pandey},
		\citenamefont {Berengut}, \citenamefont {Geddes}, \citenamefont {Nazarewicz},
		\citenamefont {Reinhard}, \citenamefont {Kawasaki}, \citenamefont {Jeon},
		\citenamefont {Jhe},\ and\ \citenamefont {Vuletić}}]{hur_evidence_2022}%
	\BibitemOpen
	\bibfield  {author} {\bibinfo {author} {\bibfnamefont {J.}~\bibnamefont
			{Hur}}, \bibinfo {author} {\bibfnamefont {D.~P.~L.}\ \bibnamefont
			{Aude~Craik}}, \bibinfo {author} {\bibfnamefont {I.}~\bibnamefont {Counts}},
		\bibinfo {author} {\bibfnamefont {E.}~\bibnamefont {Knyazev}}, \bibinfo
		{author} {\bibfnamefont {L.}~\bibnamefont {Caldwell}}, \bibinfo {author}
		{\bibfnamefont {C.}~\bibnamefont {Leung}}, \bibinfo {author} {\bibfnamefont
			{S.}~\bibnamefont {Pandey}}, \bibinfo {author} {\bibfnamefont {J.~C.}\
			\bibnamefont {Berengut}}, \bibinfo {author} {\bibfnamefont {A.}~\bibnamefont
			{Geddes}}, \bibinfo {author} {\bibfnamefont {W.}~\bibnamefont {Nazarewicz}},
		\bibinfo {author} {\bibfnamefont {P.-G.}\ \bibnamefont {Reinhard}}, \bibinfo
		{author} {\bibfnamefont {A.}~\bibnamefont {Kawasaki}}, \bibinfo {author}
		{\bibfnamefont {H.}~\bibnamefont {Jeon}}, \bibinfo {author} {\bibfnamefont
			{W.}~\bibnamefont {Jhe}},\ and\ \bibinfo {author} {\bibfnamefont
			{V.}~\bibnamefont {Vuletić}},\ }\bibfield  {title} {\bibinfo {title}
		{Evidence of {Two}-{Source} {King} {Plot} {Nonlinearity} in {Spectroscopic}
			{Search} for {New} {Boson}},\ }\href
	{https://doi.org/10.1103/PhysRevLett.128.163201} {\bibfield  {journal}
		{\bibinfo  {journal} {Physical Review Letters}\ }\textbf {\bibinfo {volume}
			{128}},\ \bibinfo {pages} {163201} (\bibinfo {year} {2022})}\BibitemShut
	{NoStop}%
	\bibitem [{\citenamefont {Door}\ \emph {et~al.}(2024)\citenamefont {Door},
		\citenamefont {Yeh}, \citenamefont {Heinz}, \citenamefont {Kirk},
		\citenamefont {Lyu}, \citenamefont {Miyagi}, \citenamefont {Berengut},
		\citenamefont {Bieroń}, \citenamefont {Blaum}, \citenamefont {Dreissen},
		\citenamefont {Eliseev}, \citenamefont {Filianin}, \citenamefont {Filzinger},
		\citenamefont {Fuchs}, \citenamefont {F\"{u}rst}, \citenamefont {Gaigalas},
		\citenamefont {Harman}, \citenamefont {Herkenhoff}, \citenamefont
		{Huntemann}, \citenamefont {Keitel}, \citenamefont {Kromer}, \citenamefont
		{Lange}, \citenamefont {Rischka}, \citenamefont {Schweiger}, \citenamefont
		{Schwenk}, \citenamefont {Shimizu},\ and\ \citenamefont
		{Mehlst\"{a}ubler}}]{door_search_2024}%
	\BibitemOpen
	\bibfield  {author} {\bibinfo {author} {\bibfnamefont {M.}~\bibnamefont
			{Door}}, \bibinfo {author} {\bibfnamefont {C.-H.}\ \bibnamefont {Yeh}},
		\bibinfo {author} {\bibfnamefont {M.}~\bibnamefont {Heinz}}, \bibinfo
		{author} {\bibfnamefont {F.}~\bibnamefont {Kirk}}, \bibinfo {author}
		{\bibfnamefont {C.}~\bibnamefont {Lyu}}, \bibinfo {author} {\bibfnamefont
			{T.}~\bibnamefont {Miyagi}}, \bibinfo {author} {\bibfnamefont {J.~C.}\
			\bibnamefont {Berengut}}, \bibinfo {author} {\bibfnamefont {J.}~\bibnamefont
			{Bieroń}}, \bibinfo {author} {\bibfnamefont {K.}~\bibnamefont {Blaum}},
		\bibinfo {author} {\bibfnamefont {L.~S.}\ \bibnamefont {Dreissen}}, \bibinfo
		{author} {\bibfnamefont {S.}~\bibnamefont {Eliseev}}, \bibinfo {author}
		{\bibfnamefont {P.}~\bibnamefont {Filianin}}, \bibinfo {author}
		{\bibfnamefont {M.}~\bibnamefont {Filzinger}}, \bibinfo {author}
		{\bibfnamefont {E.}~\bibnamefont {Fuchs}}, \bibinfo {author} {\bibfnamefont
			{H.~A.}\ \bibnamefont {F\"{u}rst}}, \bibinfo {author} {\bibfnamefont
			{G.}~\bibnamefont {Gaigalas}}, \bibinfo {author} {\bibfnamefont
			{Z.}~\bibnamefont {Harman}}, \bibinfo {author} {\bibfnamefont
			{J.}~\bibnamefont {Herkenhoff}}, \bibinfo {author} {\bibfnamefont
			{N.}~\bibnamefont {Huntemann}}, \bibinfo {author} {\bibfnamefont {C.~H.}\
			\bibnamefont {Keitel}}, \bibinfo {author} {\bibfnamefont {K.}~\bibnamefont
			{Kromer}}, \bibinfo {author} {\bibfnamefont {D.}~\bibnamefont {Lange}},
		\bibinfo {author} {\bibfnamefont {A.}~\bibnamefont {Rischka}}, \bibinfo
		{author} {\bibfnamefont {C.}~\bibnamefont {Schweiger}}, \bibinfo {author}
		{\bibfnamefont {A.}~\bibnamefont {Schwenk}}, \bibinfo {author} {\bibfnamefont
			{N.}~\bibnamefont {Shimizu}},\ and\ \bibinfo {author} {\bibfnamefont {T.~E.}\
			\bibnamefont {Mehlst\"{a}ubler}},\ }\href {http://arxiv.org/abs/2403.07792}
	{\bibinfo {title} {Search for new bosons with ytterbium isotope shifts}}
	(\bibinfo {year} {2024})\BibitemShut {NoStop}%
	\bibitem [{\citenamefont {Solaro}\ \emph {et~al.}(2020)\citenamefont {Solaro},
		\citenamefont {Meyer}, \citenamefont {Fisher}, \citenamefont {Berengut},
		\citenamefont {Fuchs},\ and\ \citenamefont {Drewsen}}]{solaro_improved_2020}%
	\BibitemOpen
	\bibfield  {author} {\bibinfo {author} {\bibfnamefont {C.}~\bibnamefont
			{Solaro}}, \bibinfo {author} {\bibfnamefont {S.}~\bibnamefont {Meyer}},
		\bibinfo {author} {\bibfnamefont {K.}~\bibnamefont {Fisher}}, \bibinfo
		{author} {\bibfnamefont {J.~C.}\ \bibnamefont {Berengut}}, \bibinfo {author}
		{\bibfnamefont {E.}~\bibnamefont {Fuchs}},\ and\ \bibinfo {author}
		{\bibfnamefont {M.}~\bibnamefont {Drewsen}},\ }\bibfield  {title} {\bibinfo
		{title} {Improved {Isotope}-{Shift}-{Based} {Bounds} on {Bosons} beyond the
			{Standard} {Model} through {Measurements} of the
			${}^{2}\mathrm{D}_{3/2}-{}^{2}\mathrm{D}_{5/2}$ {Interval} in
			$\mathrm{Ca}^+$},\ }\href {https://doi.org/10.1103/PhysRevLett.125.123003}
	{\bibfield  {journal} {\bibinfo  {journal} {Physical Review Letters}\
		}\textbf {\bibinfo {volume} {125}},\ \bibinfo {pages} {123003} (\bibinfo
		{year} {2020})}\BibitemShut {NoStop}%
	\bibitem [{\citenamefont {Chang}\ \emph {et~al.}(2024)\citenamefont {Chang},
		\citenamefont {Awazi}, \citenamefont {Berengut}, \citenamefont {Fuchs},\ and\
		\citenamefont {Doret}}]{chang_systematic-free_2024}%
	\BibitemOpen
	\bibfield  {author} {\bibinfo {author} {\bibfnamefont {T.~T.}\ \bibnamefont
			{Chang}}, \bibinfo {author} {\bibfnamefont {B.~B.}\ \bibnamefont {Awazi}},
		\bibinfo {author} {\bibfnamefont {J.~C.}\ \bibnamefont {Berengut}}, \bibinfo
		{author} {\bibfnamefont {E.}~\bibnamefont {Fuchs}},\ and\ \bibinfo {author}
		{\bibfnamefont {S.~C.}\ \bibnamefont {Doret}},\ }\href
	{https://doi.org/10.48550/arXiv.2311.17337} {\bibinfo {title}
		{Systematic-free limit on new light scalar bosons via isotope shift
			spectroscopy in $\mathrm{Ca}^+$}} (\bibinfo {year} {2024}),\ \bibinfo {note}
	{10.48550/arXiv.2311.17337}\BibitemShut {NoStop}%
	\bibitem [{\citenamefont {Rehbehn}\ \emph {et~al.}(2021)\citenamefont
		{Rehbehn}, \citenamefont {Rosner}, \citenamefont {Bekker}, \citenamefont
		{Berengut}, \citenamefont {Schmidt}, \citenamefont {King}, \citenamefont
		{Micke}, \citenamefont {Gu}, \citenamefont {M\"{u}ller}, \citenamefont
		{Surzhykov},\ and\ \citenamefont
		{L\'{o}pez-Urrutia}}]{rehbehn_sensitivity_2021}%
	\BibitemOpen
	\bibfield  {author} {\bibinfo {author} {\bibfnamefont {N.-H.}\ \bibnamefont
			{Rehbehn}}, \bibinfo {author} {\bibfnamefont {M.~K.}\ \bibnamefont {Rosner}},
		\bibinfo {author} {\bibfnamefont {H.}~\bibnamefont {Bekker}}, \bibinfo
		{author} {\bibfnamefont {J.~C.}\ \bibnamefont {Berengut}}, \bibinfo {author}
		{\bibfnamefont {P.~O.}\ \bibnamefont {Schmidt}}, \bibinfo {author}
		{\bibfnamefont {S.~A.}\ \bibnamefont {King}}, \bibinfo {author}
		{\bibfnamefont {P.}~\bibnamefont {Micke}}, \bibinfo {author} {\bibfnamefont
			{M.~F.}\ \bibnamefont {Gu}}, \bibinfo {author} {\bibfnamefont
			{R.}~\bibnamefont {M\"{u}ller}}, \bibinfo {author} {\bibfnamefont
			{A.}~\bibnamefont {Surzhykov}},\ and\ \bibinfo {author} {\bibfnamefont
			{J.~R.~C.}\ \bibnamefont {L\'{o}pez-Urrutia}},\ }\bibfield  {title} {\bibinfo
		{title} {Sensitivity to new physics of isotope-shift studies using the
			coronal lines of highly charged calcium ions},\ }\href
	{https://doi.org/10.1103/PhysRevA.103.L040801} {\bibfield  {journal}
		{\bibinfo  {journal} {Physical Review A}\ }\textbf {\bibinfo {volume}
			{103}},\ \bibinfo {pages} {L040801} (\bibinfo {year} {2021})}\BibitemShut
	{NoStop}%
	\bibitem [{\citenamefont {King}\ \emph {et~al.}(2022)\citenamefont {King},
		\citenamefont {Spie{\ss}}, \citenamefont {Micke}, \citenamefont {Wilzewski},
		\citenamefont {Leopold}, \citenamefont {Benkler}, \citenamefont {Lange},
		\citenamefont {Huntemann}, \citenamefont {Surzhykov}, \citenamefont
		{Yerokhin}, \citenamefont {Crespo L\'{o}pez-Urrutia},\ and\ \citenamefont
		{Schmidt}}]{king_optical_2022}%
	\BibitemOpen
	\bibfield  {author} {\bibinfo {author} {\bibfnamefont {S.~A.}\ \bibnamefont
			{King}}, \bibinfo {author} {\bibfnamefont {L.~J.}\ \bibnamefont {Spie{\ss}}},
		\bibinfo {author} {\bibfnamefont {P.}~\bibnamefont {Micke}}, \bibinfo
		{author} {\bibfnamefont {A.}~\bibnamefont {Wilzewski}}, \bibinfo {author}
		{\bibfnamefont {T.}~\bibnamefont {Leopold}}, \bibinfo {author} {\bibfnamefont
			{E.}~\bibnamefont {Benkler}}, \bibinfo {author} {\bibfnamefont
			{R.}~\bibnamefont {Lange}}, \bibinfo {author} {\bibfnamefont
			{N.}~\bibnamefont {Huntemann}}, \bibinfo {author} {\bibfnamefont
			{A.}~\bibnamefont {Surzhykov}}, \bibinfo {author} {\bibfnamefont {V.~A.}\
			\bibnamefont {Yerokhin}}, \bibinfo {author} {\bibfnamefont {J.~R.}\
			\bibnamefont {Crespo L\'{o}pez-Urrutia}},\ and\ \bibinfo {author}
		{\bibfnamefont {P.~O.}\ \bibnamefont {Schmidt}},\ }\bibfield  {title}
	{\bibinfo {title} {An optical atomic clock based on a highly charged ion},\
	}\href {https://doi.org/10.1038/s41586-022-05245-4} {\bibfield  {journal}
		{\bibinfo  {journal} {Nature}\ }\textbf {\bibinfo {volume} {611}},\ \bibinfo
		{pages} {43} (\bibinfo {year} {2022})}\BibitemShut {NoStop}%
	\bibitem [{\citenamefont {Schweiger}\ \emph {et~al.}(2019)\citenamefont
		{Schweiger}, \citenamefont {König}, \citenamefont {Crespo
			L\'{o}pez-Urrutia}, \citenamefont {Door}, \citenamefont {Dorrer},
		\citenamefont {D\"{u}llmann}, \citenamefont {Eliseev}, \citenamefont
		{Filianin}, \citenamefont {Huang}, \citenamefont {Kromer}, \citenamefont
		{Micke}, \citenamefont {M\"{u}ller}, \citenamefont {Renisch}, \citenamefont
		{Rischka}, \citenamefont {Sch\"{u}ssler},\ and\ \citenamefont
		{Blaum}}]{schweiger_production_2019}%
	\BibitemOpen
	\bibfield  {author} {\bibinfo {author} {\bibfnamefont {C.}~\bibnamefont
			{Schweiger}}, \bibinfo {author} {\bibfnamefont {C.~M.}\ \bibnamefont
			{König}}, \bibinfo {author} {\bibfnamefont {J.~R.}\ \bibnamefont {Crespo
				L\'{o}pez-Urrutia}}, \bibinfo {author} {\bibfnamefont {M.}~\bibnamefont
			{Door}}, \bibinfo {author} {\bibfnamefont {H.}~\bibnamefont {Dorrer}},
		\bibinfo {author} {\bibfnamefont {C.~E.}\ \bibnamefont {D\"{u}llmann}},
		\bibinfo {author} {\bibfnamefont {S.}~\bibnamefont {Eliseev}}, \bibinfo
		{author} {\bibfnamefont {P.}~\bibnamefont {Filianin}}, \bibinfo {author}
		{\bibfnamefont {W.}~\bibnamefont {Huang}}, \bibinfo {author} {\bibfnamefont
			{K.}~\bibnamefont {Kromer}}, \bibinfo {author} {\bibfnamefont
			{P.}~\bibnamefont {Micke}}, \bibinfo {author} {\bibfnamefont
			{M.}~\bibnamefont {M\"{u}ller}}, \bibinfo {author} {\bibfnamefont
			{D.}~\bibnamefont {Renisch}}, \bibinfo {author} {\bibfnamefont
			{A.}~\bibnamefont {Rischka}}, \bibinfo {author} {\bibfnamefont {R.~X.}\
			\bibnamefont {Sch\"{u}ssler}},\ and\ \bibinfo {author} {\bibfnamefont
			{K.}~\bibnamefont {Blaum}},\ }\bibfield  {title} {\bibinfo {title}
		{Production of highly charged ions of rare species by laser-induced
			desorption inside an electron beam ion trap},\ }\href
	{https://doi.org/10.1063/1.5128331} {\bibfield  {journal} {\bibinfo
			{journal} {Review of Scientific Instruments}\ }\textbf {\bibinfo {volume}
			{90}},\ \bibinfo {pages} {123201} (\bibinfo {year} {2019})}\BibitemShut
	{NoStop}%
	\bibitem [{\citenamefont {Schmöger}\ \emph
		{et~al.}(2015{\natexlab{a}})\citenamefont {Schmöger}, \citenamefont
		{Versolato}, \citenamefont {Schwarz}, \citenamefont {Kohnen}, \citenamefont
		{Windberger}, \citenamefont {Piest}, \citenamefont {Feuchtenbeiner},
		\citenamefont {Pedregosa-Gutierrez}, \citenamefont {Leopold}, \citenamefont
		{Micke}, \citenamefont {Hansen}, \citenamefont {Baumann}, \citenamefont
		{Drewsen}, \citenamefont {Ullrich}, \citenamefont {Schmidt},\ and\
		\citenamefont {L\'{o}pez-Urrutia}}]{schmoger_coulomb_2015}%
	\BibitemOpen
	\bibfield  {author} {\bibinfo {author} {\bibfnamefont {L.}~\bibnamefont
			{Schmöger}}, \bibinfo {author} {\bibfnamefont {O.~O.}\ \bibnamefont
			{Versolato}}, \bibinfo {author} {\bibfnamefont {M.}~\bibnamefont {Schwarz}},
		\bibinfo {author} {\bibfnamefont {M.}~\bibnamefont {Kohnen}}, \bibinfo
		{author} {\bibfnamefont {A.}~\bibnamefont {Windberger}}, \bibinfo {author}
		{\bibfnamefont {B.}~\bibnamefont {Piest}}, \bibinfo {author} {\bibfnamefont
			{S.}~\bibnamefont {Feuchtenbeiner}}, \bibinfo {author} {\bibfnamefont
			{J.}~\bibnamefont {Pedregosa-Gutierrez}}, \bibinfo {author} {\bibfnamefont
			{T.}~\bibnamefont {Leopold}}, \bibinfo {author} {\bibfnamefont
			{P.}~\bibnamefont {Micke}}, \bibinfo {author} {\bibfnamefont {A.~K.}\
			\bibnamefont {Hansen}}, \bibinfo {author} {\bibfnamefont {T.~M.}\
			\bibnamefont {Baumann}}, \bibinfo {author} {\bibfnamefont {M.}~\bibnamefont
			{Drewsen}}, \bibinfo {author} {\bibfnamefont {J.}~\bibnamefont {Ullrich}},
		\bibinfo {author} {\bibfnamefont {P.~O.}\ \bibnamefont {Schmidt}},\ and\
		\bibinfo {author} {\bibfnamefont {J.~R.~C.}\ \bibnamefont
			{L\'{o}pez-Urrutia}},\ }\bibfield  {title} {\bibinfo {title} {Coulomb
			crystallization of highly charged ions},\ }\href
	{https://doi.org/10.1126/science.aaa2960} {\bibfield  {journal} {\bibinfo
			{journal} {Science}\ }\textbf {\bibinfo {volume} {347}},\ \bibinfo {pages}
		{1233} (\bibinfo {year} {2015}{\natexlab{a}})}\BibitemShut {NoStop}%
	\bibitem [{\citenamefont {Schmöger}\ \emph
		{et~al.}(2015{\natexlab{b}})\citenamefont {Schmöger}, \citenamefont
		{Schwarz}, \citenamefont {Baumann}, \citenamefont {Versolato}, \citenamefont
		{Piest}, \citenamefont {Pfeifer}, \citenamefont {Ullrich}, \citenamefont
		{Schmidt},\ and\ \citenamefont {Crespo
			L\'{o}pez-Urrutia}}]{schmoger_deceleration_2015}%
	\BibitemOpen
	\bibfield  {author} {\bibinfo {author} {\bibfnamefont {L.}~\bibnamefont
			{Schmöger}}, \bibinfo {author} {\bibfnamefont {M.}~\bibnamefont {Schwarz}},
		\bibinfo {author} {\bibfnamefont {T.~M.}\ \bibnamefont {Baumann}}, \bibinfo
		{author} {\bibfnamefont {O.~O.}\ \bibnamefont {Versolato}}, \bibinfo {author}
		{\bibfnamefont {B.}~\bibnamefont {Piest}}, \bibinfo {author} {\bibfnamefont
			{T.}~\bibnamefont {Pfeifer}}, \bibinfo {author} {\bibfnamefont
			{J.}~\bibnamefont {Ullrich}}, \bibinfo {author} {\bibfnamefont {P.~O.}\
			\bibnamefont {Schmidt}},\ and\ \bibinfo {author} {\bibfnamefont {J.~R.}\
			\bibnamefont {Crespo L\'{o}pez-Urrutia}},\ }\bibfield  {title} {\bibinfo
		{title} {Deceleration, precooling, and multi-pass stopping of highly charged
			ions in $\mathrm{Be}^+$ {Coulomb} crystals},\ }\href
	{https://doi.org/10.1063/1.4934245} {\bibfield  {journal} {\bibinfo
			{journal} {Review of Scientific Instruments}\ }\textbf {\bibinfo {volume}
			{86}},\ \bibinfo {pages} {103111} (\bibinfo {year}
		{2015}{\natexlab{b}})}\BibitemShut {NoStop}%
	\bibitem [{\citenamefont {Micke}\ \emph {et~al.}(2018)\citenamefont {Micke},
		\citenamefont {K\"{u}hn}, \citenamefont {Buchauer}, \citenamefont {Harries},
		\citenamefont {B\"{u}cking}, \citenamefont {Blaum}, \citenamefont {Cieluch},
		\citenamefont {Egl}, \citenamefont {Hollain}, \citenamefont {Kraemer},
		\citenamefont {Pfeifer}, \citenamefont {Schmidt}, \citenamefont
		{Sch\"{u}ssler}, \citenamefont {Schweiger}, \citenamefont {Stöhlker},
		\citenamefont {Sturm}, \citenamefont {Wolf}, \citenamefont {Bernitt},\ and\
		\citenamefont {Crespo L\'{o}pez-Urrutia}}]{micke_heidelberg_2018}%
	\BibitemOpen
	\bibfield  {author} {\bibinfo {author} {\bibfnamefont {P.}~\bibnamefont
			{Micke}}, \bibinfo {author} {\bibfnamefont {S.}~\bibnamefont {K\"{u}hn}},
		\bibinfo {author} {\bibfnamefont {L.}~\bibnamefont {Buchauer}}, \bibinfo
		{author} {\bibfnamefont {J.~R.}\ \bibnamefont {Harries}}, \bibinfo {author}
		{\bibfnamefont {T.~M.}\ \bibnamefont {B\"{u}cking}}, \bibinfo {author}
		{\bibfnamefont {K.}~\bibnamefont {Blaum}}, \bibinfo {author} {\bibfnamefont
			{A.}~\bibnamefont {Cieluch}}, \bibinfo {author} {\bibfnamefont
			{A.}~\bibnamefont {Egl}}, \bibinfo {author} {\bibfnamefont {D.}~\bibnamefont
			{Hollain}}, \bibinfo {author} {\bibfnamefont {S.}~\bibnamefont {Kraemer}},
		\bibinfo {author} {\bibfnamefont {T.}~\bibnamefont {Pfeifer}}, \bibinfo
		{author} {\bibfnamefont {P.~O.}\ \bibnamefont {Schmidt}}, \bibinfo {author}
		{\bibfnamefont {R.~X.}\ \bibnamefont {Sch\"{u}ssler}}, \bibinfo {author}
		{\bibfnamefont {C.}~\bibnamefont {Schweiger}}, \bibinfo {author}
		{\bibfnamefont {T.}~\bibnamefont {Stöhlker}}, \bibinfo {author}
		{\bibfnamefont {S.}~\bibnamefont {Sturm}}, \bibinfo {author} {\bibfnamefont
			{R.~N.}\ \bibnamefont {Wolf}}, \bibinfo {author} {\bibfnamefont
			{S.}~\bibnamefont {Bernitt}},\ and\ \bibinfo {author} {\bibfnamefont {J.~R.}\
			\bibnamefont {Crespo L\'{o}pez-Urrutia}},\ }\bibfield  {title} {\bibinfo
		{title} {The {Heidelberg} compact electron beam ion traps},\ }\href
	{https://doi.org/10.1063/1.5026961} {\bibfield  {journal} {\bibinfo
			{journal} {Review of Scientific Instruments}\ }\textbf {\bibinfo {volume}
			{89}},\ \bibinfo {pages} {063109} (\bibinfo {year} {2018})}\BibitemShut
	{NoStop}%
	\bibitem [{\citenamefont {Micke}\ \emph {et~al.}(2019)\citenamefont {Micke},
		\citenamefont {Stark}, \citenamefont {King}, \citenamefont {Leopold},
		\citenamefont {Pfeifer}, \citenamefont {Schmöger}, \citenamefont {Schwarz},
		\citenamefont {Spie{\ss}}, \citenamefont {Schmidt},\ and\ \citenamefont
		{Crespo L\'{o}pez-Urrutia}}]{micke_closed-cycle_2019}%
	\BibitemOpen
	\bibfield  {author} {\bibinfo {author} {\bibfnamefont {P.}~\bibnamefont
			{Micke}}, \bibinfo {author} {\bibfnamefont {J.}~\bibnamefont {Stark}},
		\bibinfo {author} {\bibfnamefont {S.~A.}\ \bibnamefont {King}}, \bibinfo
		{author} {\bibfnamefont {T.}~\bibnamefont {Leopold}}, \bibinfo {author}
		{\bibfnamefont {T.}~\bibnamefont {Pfeifer}}, \bibinfo {author} {\bibfnamefont
			{L.}~\bibnamefont {Schmöger}}, \bibinfo {author} {\bibfnamefont
			{M.}~\bibnamefont {Schwarz}}, \bibinfo {author} {\bibfnamefont {L.~J.}\
			\bibnamefont {Spie{\ss}}}, \bibinfo {author} {\bibfnamefont {P.~O.}\
			\bibnamefont {Schmidt}},\ and\ \bibinfo {author} {\bibfnamefont {J.~R.}\
			\bibnamefont {Crespo L\'{o}pez-Urrutia}},\ }\bibfield  {title} {\bibinfo
		{title} {Closed-cycle, low-vibration 4 {K} cryostat for ion traps and other
			applications},\ }\href {https://doi.org/10.1063/1.5088593} {\bibfield
		{journal} {\bibinfo  {journal} {Review of Scientific Instruments}\ }\textbf
		{\bibinfo {volume} {90}},\ \bibinfo {pages} {065104} (\bibinfo {year}
		{2019})}\BibitemShut {NoStop}%
	\bibitem [{\citenamefont {Leopold}\ \emph {et~al.}(2019)\citenamefont
		{Leopold}, \citenamefont {King}, \citenamefont {Micke}, \citenamefont
		{Bautista-Salvador}, \citenamefont {Heip}, \citenamefont {Ospelkaus},
		\citenamefont {Crespo L\'{o}pez-Urrutia},\ and\ \citenamefont
		{Schmidt}}]{leopold_cryogenic_2019}%
	\BibitemOpen
	\bibfield  {author} {\bibinfo {author} {\bibfnamefont {T.}~\bibnamefont
			{Leopold}}, \bibinfo {author} {\bibfnamefont {S.~A.}\ \bibnamefont {King}},
		\bibinfo {author} {\bibfnamefont {P.}~\bibnamefont {Micke}}, \bibinfo
		{author} {\bibfnamefont {A.}~\bibnamefont {Bautista-Salvador}}, \bibinfo
		{author} {\bibfnamefont {J.~C.}\ \bibnamefont {Heip}}, \bibinfo {author}
		{\bibfnamefont {C.}~\bibnamefont {Ospelkaus}}, \bibinfo {author}
		{\bibfnamefont {J.~R.}\ \bibnamefont {Crespo L\'{o}pez-Urrutia}},\ and\
		\bibinfo {author} {\bibfnamefont {P.~O.}\ \bibnamefont {Schmidt}},\
	}\bibfield  {title} {\bibinfo {title} {A cryogenic radio-frequency ion trap
			for quantum logic spectroscopy of highly charged ions},\ }\href
	{https://doi.org/10.1063/1.5100594} {\bibfield  {journal} {\bibinfo
			{journal} {Review of Scientific Instruments}\ }\textbf {\bibinfo {volume}
			{90}},\ \bibinfo {pages} {073201} (\bibinfo {year} {2019})}\BibitemShut
	{NoStop}%
	\bibitem [{\citenamefont {King}\ \emph {et~al.}(2021)\citenamefont {King},
		\citenamefont {Spie{\ss}}, \citenamefont {Micke}, \citenamefont {Wilzewski},
		\citenamefont {Leopold}, \citenamefont {Crespo L\'{o}pez-Urrutia},\ and\
		\citenamefont {Schmidt}}]{king_algorithmic_2021}%
	\BibitemOpen
	\bibfield  {author} {\bibinfo {author} {\bibfnamefont {S.~A.}\ \bibnamefont
			{King}}, \bibinfo {author} {\bibfnamefont {L.~J.}\ \bibnamefont {Spie{\ss}}},
		\bibinfo {author} {\bibfnamefont {P.}~\bibnamefont {Micke}}, \bibinfo
		{author} {\bibfnamefont {A.}~\bibnamefont {Wilzewski}}, \bibinfo {author}
		{\bibfnamefont {T.}~\bibnamefont {Leopold}}, \bibinfo {author} {\bibfnamefont
			{J.~R.}\ \bibnamefont {Crespo L\'{o}pez-Urrutia}},\ and\ \bibinfo {author}
		{\bibfnamefont {P.~O.}\ \bibnamefont {Schmidt}},\ }\bibfield  {title}
	{\bibinfo {title} {Algorithmic {Ground}-{State} {Cooling} of {Weakly}
			{Coupled} {Oscillators} {Using} {Quantum} {Logic}},\ }\href
	{https://doi.org/10.1103/PhysRevX.11.041049} {\bibfield  {journal} {\bibinfo
			{journal} {Physical Review X}\ }\textbf {\bibinfo {volume} {11}},\ \bibinfo
		{pages} {041049} (\bibinfo {year} {2021})}\BibitemShut {NoStop}%
	\bibitem [{\citenamefont {Schmidt}\ \emph {et~al.}(2005)\citenamefont
		{Schmidt}, \citenamefont {Rosenband}, \citenamefont {Langer}, \citenamefont
		{Itano}, \citenamefont {Bergquist},\ and\ \citenamefont
		{Wineland}}]{schmidt_spectroscopy_2005}%
	\BibitemOpen
	\bibfield  {author} {\bibinfo {author} {\bibfnamefont {P.~O.}\ \bibnamefont
			{Schmidt}}, \bibinfo {author} {\bibfnamefont {T.}~\bibnamefont {Rosenband}},
		\bibinfo {author} {\bibfnamefont {C.}~\bibnamefont {Langer}}, \bibinfo
		{author} {\bibfnamefont {W.~M.}\ \bibnamefont {Itano}}, \bibinfo {author}
		{\bibfnamefont {J.~C.}\ \bibnamefont {Bergquist}},\ and\ \bibinfo {author}
		{\bibfnamefont {D.~J.}\ \bibnamefont {Wineland}},\ }\bibfield  {title}
	{\bibinfo {title} {Spectroscopy {Using} {Quantum} {Logic}},\ }\href
	{https://doi.org/10.1126/science.1114375} {\bibfield  {journal} {\bibinfo
			{journal} {Science}\ }\textbf {\bibinfo {volume} {309}},\ \bibinfo {pages}
		{749} (\bibinfo {year} {2005})}\BibitemShut {NoStop}%
	\bibitem [{\citenamefont {Micke}\ \emph {et~al.}(2020)\citenamefont {Micke},
		\citenamefont {Leopold}, \citenamefont {King}, \citenamefont {Benkler},
		\citenamefont {Spie{\ss}}, \citenamefont {Schmöger}, \citenamefont
		{Schwarz}, \citenamefont {Crespo L\'{o}pez-Urrutia},\ and\ \citenamefont
		{Schmidt}}]{micke_coherent_2020}%
	\BibitemOpen
	\bibfield  {author} {\bibinfo {author} {\bibfnamefont {P.}~\bibnamefont
			{Micke}}, \bibinfo {author} {\bibfnamefont {T.}~\bibnamefont {Leopold}},
		\bibinfo {author} {\bibfnamefont {S.~A.}\ \bibnamefont {King}}, \bibinfo
		{author} {\bibfnamefont {E.}~\bibnamefont {Benkler}}, \bibinfo {author}
		{\bibfnamefont {L.~J.}\ \bibnamefont {Spie{\ss}}}, \bibinfo {author}
		{\bibfnamefont {L.}~\bibnamefont {Schmöger}}, \bibinfo {author}
		{\bibfnamefont {M.}~\bibnamefont {Schwarz}}, \bibinfo {author} {\bibfnamefont
			{J.~R.}\ \bibnamefont {Crespo L\'{o}pez-Urrutia}},\ and\ \bibinfo {author}
		{\bibfnamefont {P.~O.}\ \bibnamefont {Schmidt}},\ }\bibfield  {title}
	{\bibinfo {title} {Coherent laser spectroscopy of highly charged ions using
			quantum logic},\ }\href {https://doi.org/10.1038/s41586-020-1959-8}
	{\bibfield  {journal} {\bibinfo  {journal} {Nature}\ }\textbf {\bibinfo
			{volume} {578}},\ \bibinfo {pages} {60} (\bibinfo {year} {2020})}\BibitemShut
	{NoStop}%
	\bibitem [{\citenamefont {Matei}\ \emph {et~al.}(2017)\citenamefont {Matei},
		\citenamefont {Legero}, \citenamefont {H\"{a}fner}, \citenamefont {Grebing},
		\citenamefont {Weyrich}, \citenamefont {Zhang}, \citenamefont {Sonderhouse},
		\citenamefont {Robinson}, \citenamefont {Ye}, \citenamefont {Riehle},\ and\
		\citenamefont {Sterr}}]{matei_15text_2017}%
	\BibitemOpen
	\bibfield  {author} {\bibinfo {author} {\bibfnamefont {D.~G.}\ \bibnamefont
			{Matei}}, \bibinfo {author} {\bibfnamefont {T.}~\bibnamefont {Legero}},
		\bibinfo {author} {\bibfnamefont {S.}~\bibnamefont {H\"{a}fner}}, \bibinfo
		{author} {\bibfnamefont {C.}~\bibnamefont {Grebing}}, \bibinfo {author}
		{\bibfnamefont {R.}~\bibnamefont {Weyrich}}, \bibinfo {author} {\bibfnamefont
			{W.}~\bibnamefont {Zhang}}, \bibinfo {author} {\bibfnamefont
			{L.}~\bibnamefont {Sonderhouse}}, \bibinfo {author} {\bibfnamefont {J.~M.}\
			\bibnamefont {Robinson}}, \bibinfo {author} {\bibfnamefont {J.}~\bibnamefont
			{Ye}}, \bibinfo {author} {\bibfnamefont {F.}~\bibnamefont {Riehle}},\ and\
		\bibinfo {author} {\bibfnamefont {U.}~\bibnamefont {Sterr}},\ }\bibfield
	{title} {\bibinfo {title} {$1.5 \mathrm{\mu m}$ {Lasers} with {Sub}-10 {mHz}
			{Linewidth}},\ }\href {https://doi.org/10.1103/PhysRevLett.118.263202}
	{\bibfield  {journal} {\bibinfo  {journal} {Physical Review Letters}\
		}\textbf {\bibinfo {volume} {118}},\ \bibinfo {pages} {263202} (\bibinfo
		{year} {2017})}\BibitemShut {NoStop}%
	\bibitem [{\citenamefont {Huntemann}\ \emph {et~al.}(2012)\citenamefont
		{Huntemann}, \citenamefont {Okhapkin}, \citenamefont {Lipphardt},
		\citenamefont {Weyers}, \citenamefont {Tamm},\ and\ \citenamefont
		{Peik}}]{huntemann_high-accuracy_2012}%
	\BibitemOpen
	\bibfield  {author} {\bibinfo {author} {\bibfnamefont {N.}~\bibnamefont
			{Huntemann}}, \bibinfo {author} {\bibfnamefont {M.}~\bibnamefont {Okhapkin}},
		\bibinfo {author} {\bibfnamefont {B.}~\bibnamefont {Lipphardt}}, \bibinfo
		{author} {\bibfnamefont {S.}~\bibnamefont {Weyers}}, \bibinfo {author}
		{\bibfnamefont {C.}~\bibnamefont {Tamm}},\ and\ \bibinfo {author}
		{\bibfnamefont {E.}~\bibnamefont {Peik}},\ }\bibfield  {title} {\bibinfo
		{title} {High-{Accuracy} {Optical} {Clock} {Based} on the {Octupole}
			{Transition} in ${}^{171}\mathrm{Yb}^+$},\ }\href
	{https://doi.org/10.1103/PhysRevLett.108.090801} {\bibfield  {journal}
		{\bibinfo  {journal} {Physical Review Letters}\ }\textbf {\bibinfo {volume}
			{108}},\ \bibinfo {pages} {090801} (\bibinfo {year} {2012})}\BibitemShut
	{NoStop}%
	\bibitem [{\citenamefont {Sanner}\ \emph {et~al.}(2019)\citenamefont {Sanner},
		\citenamefont {Huntemann}, \citenamefont {Lange}, \citenamefont {Tamm},
		\citenamefont {Peik}, \citenamefont {Safronova},\ and\ \citenamefont
		{Porsev}}]{sanner_optical_2019}%
	\BibitemOpen
	\bibfield  {author} {\bibinfo {author} {\bibfnamefont {C.}~\bibnamefont
			{Sanner}}, \bibinfo {author} {\bibfnamefont {N.}~\bibnamefont {Huntemann}},
		\bibinfo {author} {\bibfnamefont {R.}~\bibnamefont {Lange}}, \bibinfo
		{author} {\bibfnamefont {C.}~\bibnamefont {Tamm}}, \bibinfo {author}
		{\bibfnamefont {E.}~\bibnamefont {Peik}}, \bibinfo {author} {\bibfnamefont
			{M.~S.}\ \bibnamefont {Safronova}},\ and\ \bibinfo {author} {\bibfnamefont
			{S.~G.}\ \bibnamefont {Porsev}},\ }\bibfield  {title} {\bibinfo {title}
		{Optical clock comparison for {Lorentz} symmetry testing},\ }\href
	{https://doi.org/10.1038/s41586-019-0972-2} {\bibfield  {journal} {\bibinfo
			{journal} {Nature}\ }\textbf {\bibinfo {volume} {567}},\ \bibinfo {pages}
		{204} (\bibinfo {year} {2019})}\BibitemShut {NoStop}%
	\bibitem [{\citenamefont {Lange}\ \emph {et~al.}(2021)\citenamefont {Lange},
		\citenamefont {Huntemann}, \citenamefont {Rahm}, \citenamefont {Sanner},
		\citenamefont {Shao}, \citenamefont {Lipphardt}, \citenamefont {Tamm},
		\citenamefont {Weyers},\ and\ \citenamefont {Peik}}]{lange_improved_2021}%
	\BibitemOpen
	\bibfield  {author} {\bibinfo {author} {\bibfnamefont {R.}~\bibnamefont
			{Lange}}, \bibinfo {author} {\bibfnamefont {N.}~\bibnamefont {Huntemann}},
		\bibinfo {author} {\bibfnamefont {J.~M.}\ \bibnamefont {Rahm}}, \bibinfo
		{author} {\bibfnamefont {C.}~\bibnamefont {Sanner}}, \bibinfo {author}
		{\bibfnamefont {H.}~\bibnamefont {Shao}}, \bibinfo {author} {\bibfnamefont
			{B.}~\bibnamefont {Lipphardt}}, \bibinfo {author} {\bibfnamefont
			{C.}~\bibnamefont {Tamm}}, \bibinfo {author} {\bibfnamefont {S.}~\bibnamefont
			{Weyers}},\ and\ \bibinfo {author} {\bibfnamefont {E.}~\bibnamefont {Peik}},\
	}\bibfield  {title} {\bibinfo {title} {Improved {Limits} for {Violations} of
			{Local} {Position} {Invariance} from {Atomic} {Clock} {Comparisons}},\ }\href
	{https://doi.org/10.1103/PhysRevLett.126.011102} {\bibfield  {journal}
		{\bibinfo  {journal} {Physical Review Letters}\ }\textbf {\bibinfo {volume}
			{126}},\ \bibinfo {pages} {011102} (\bibinfo {year} {2021})}\BibitemShut
	{NoStop}%
	\bibitem [{noa()}]{noauthor_see_nodate}%
	\BibitemOpen
	\bibfield  {title} {\bibinfo {title} {See {Supplemental} {Material} at ({URL}
			will be inserted by publisher) for more details on the experimental
			procedures and the analysis of systematic uncertainties. {Additionally},
			calculations of electronic coefficients and details on how new physics
			constraints are derived and the nonlinearity of the {King} plot is decomposed
			are found therein},\ }\href@noop {} {\bibinfo  {journal} {Physical Review
			Letters}\ }\BibitemShut {NoStop}%
	\bibitem [{\citenamefont {Peik}\ \emph {et~al.}(2005)\citenamefont {Peik},
		\citenamefont {Schneider},\ and\ \citenamefont {Tamm}}]{peik_laser_2005}%
	\BibitemOpen
	\bibfield  {journal} {  }\bibfield  {author} {\bibinfo {author} {\bibfnamefont
			{E.}~\bibnamefont {Peik}}, \bibinfo {author} {\bibfnamefont {T.}~\bibnamefont
			{Schneider}},\ and\ \bibinfo {author} {\bibfnamefont {C.}~\bibnamefont
			{Tamm}},\ }\bibfield  {title} {\bibinfo {title} {Laser frequency
			stabilization to a single ion},\ }\href
	{https://doi.org/10.1088/0953-4075/39/1/012} {\bibfield  {journal} {\bibinfo
			{journal} {Journal of Physics B: Atomic, Molecular and Optical Physics}\
		}\textbf {\bibinfo {volume} {39}},\ \bibinfo {pages} {145} (\bibinfo {year}
		{2005})}\BibitemShut {NoStop}%
	\bibitem [{\citenamefont {Webster}\ and\ \citenamefont
		{Gill}(2011)}]{webster_force-insensitive_2011}%
	\BibitemOpen
	\bibfield  {author} {\bibinfo {author} {\bibfnamefont {S.}~\bibnamefont
			{Webster}}\ and\ \bibinfo {author} {\bibfnamefont {P.}~\bibnamefont {Gill}},\
	}\bibfield  {title} {\bibinfo {title} {Force-insensitive optical cavity},\
	}\href {https://doi.org/10.1364/OL.36.003572} {\bibfield  {journal} {\bibinfo
			{journal} {Optics Letters}\ }\textbf {\bibinfo {volume} {36}},\ \bibinfo
		{pages} {3572} (\bibinfo {year} {2011})}\BibitemShut {NoStop}%
	\bibitem [{\citenamefont {Leopold}\ \emph {et~al.}(2016)\citenamefont
		{Leopold}, \citenamefont {Schmöger}, \citenamefont {Feuchtenbeiner},
		\citenamefont {Grebing}, \citenamefont {Micke}, \citenamefont {Scharnhorst},
		\citenamefont {Leroux}, \citenamefont {López-Urrutia},\ and\ \citenamefont
		{Schmidt}}]{leopold_tunable_2016}%
	\BibitemOpen
	\bibfield  {author} {\bibinfo {author} {\bibfnamefont {T.}~\bibnamefont
			{Leopold}}, \bibinfo {author} {\bibfnamefont {L.}~\bibnamefont {Schmöger}},
		\bibinfo {author} {\bibfnamefont {S.}~\bibnamefont {Feuchtenbeiner}},
		\bibinfo {author} {\bibfnamefont {C.}~\bibnamefont {Grebing}}, \bibinfo
		{author} {\bibfnamefont {P.}~\bibnamefont {Micke}}, \bibinfo {author}
		{\bibfnamefont {N.}~\bibnamefont {Scharnhorst}}, \bibinfo {author}
		{\bibfnamefont {I.~D.}\ \bibnamefont {Leroux}}, \bibinfo {author}
		{\bibfnamefont {J.~R.~C.}\ \bibnamefont {López-Urrutia}},\ and\ \bibinfo
		{author} {\bibfnamefont {P.~O.}\ \bibnamefont {Schmidt}},\ }\bibfield
	{title} {\bibinfo {title} {A tunable low-drift laser stabilized to an atomic
			reference},\ }\href {https://doi.org/10.1007/s00340-016-6511-z} {\bibfield
		{journal} {\bibinfo  {journal} {Applied Physics B}\ }\textbf {\bibinfo
			{volume} {122}},\ \bibinfo {pages} {236} (\bibinfo {year}
		{2016})}\BibitemShut {NoStop}%
	\bibitem [{\citenamefont {Dawel}\ \emph {et~al.}(2024)\citenamefont {Dawel},
		\citenamefont {Wilzewski}, \citenamefont {Herbers}, \citenamefont {Pelzer},
		\citenamefont {Kramer}, \citenamefont {Hild}, \citenamefont {Dietze},
		\citenamefont {Krinner}, \citenamefont {Spethmann},\ and\ \citenamefont
		{Schmidt}}]{dawel_coherent_2024}%
	\BibitemOpen
	\bibfield  {author} {\bibinfo {author} {\bibfnamefont {F.}~\bibnamefont
			{Dawel}}, \bibinfo {author} {\bibfnamefont {A.}~\bibnamefont {Wilzewski}},
		\bibinfo {author} {\bibfnamefont {S.}~\bibnamefont {Herbers}}, \bibinfo
		{author} {\bibfnamefont {L.}~\bibnamefont {Pelzer}}, \bibinfo {author}
		{\bibfnamefont {J.}~\bibnamefont {Kramer}}, \bibinfo {author} {\bibfnamefont
			{M.~B.}\ \bibnamefont {Hild}}, \bibinfo {author} {\bibfnamefont
			{K.}~\bibnamefont {Dietze}}, \bibinfo {author} {\bibfnamefont
			{L.}~\bibnamefont {Krinner}}, \bibinfo {author} {\bibfnamefont {N.~C.~H.}\
			\bibnamefont {Spethmann}},\ and\ \bibinfo {author} {\bibfnamefont {P.~O.}\
			\bibnamefont {Schmidt}},\ }\bibfield  {title} {\bibinfo {title} {Coherent
			photo-thermal noise cancellation in a dual-wavelength optical cavity for
			narrow-linewidth laser frequency stabilisation},\ }\href
	{https://doi.org/10.1364/OE.516433} {\bibfield  {journal} {\bibinfo
			{journal} {Optics Express}\ }\textbf {\bibinfo {volume} {32}},\ \bibinfo
		{pages} {7276} (\bibinfo {year} {2024})}\BibitemShut {NoStop}%
	\bibitem [{\citenamefont {Monroe}\ \emph {et~al.}(1995)\citenamefont {Monroe},
		\citenamefont {Meekhof}, \citenamefont {King}, \citenamefont {Jefferts},
		\citenamefont {Itano}, \citenamefont {Wineland},\ and\ \citenamefont
		{Gould}}]{monroe_resolved-sideband_1995}%
	\BibitemOpen
	\bibfield  {author} {\bibinfo {author} {\bibfnamefont {C.}~\bibnamefont
			{Monroe}}, \bibinfo {author} {\bibfnamefont {D.~M.}\ \bibnamefont {Meekhof}},
		\bibinfo {author} {\bibfnamefont {B.~E.}\ \bibnamefont {King}}, \bibinfo
		{author} {\bibfnamefont {S.~R.}\ \bibnamefont {Jefferts}}, \bibinfo {author}
		{\bibfnamefont {W.~M.}\ \bibnamefont {Itano}}, \bibinfo {author}
		{\bibfnamefont {D.~J.}\ \bibnamefont {Wineland}},\ and\ \bibinfo {author}
		{\bibfnamefont {P.}~\bibnamefont {Gould}},\ }\bibfield  {title} {\bibinfo
		{title} {Resolved-{Sideband} {Raman} {Cooling} of a {Bound} {Atom} to the
			{3D} {Zero}-{Point} {Energy}},\ }\href
	{https://doi.org/10.1103/PhysRevLett.75.4011} {\bibfield  {journal} {\bibinfo
			{journal} {Physical Review Letters}\ }\textbf {\bibinfo {volume} {75}},\
		\bibinfo {pages} {4011} (\bibinfo {year} {1995})}\BibitemShut {NoStop}%
	\bibitem [{\citenamefont {Wilzewski}(2024)}]{wilzewski_isotope_2024}%
	\BibitemOpen
	\bibfield  {author} {\bibinfo {author} {\bibfnamefont {A.}~\bibnamefont
			{Wilzewski}},\ }\emph {\bibinfo {title} {Isotope shift measurements in highly
			charged calcium}},\ \href {https://doi.org/10.15488/17882} {Ph.D. thesis},\
	\bibinfo  {school} {Hannover : Institutionelles Repositorium der Leibniz
		Universit\"{a}t Hannover} (\bibinfo {year} {2024})\BibitemShut {NoStop}%
	\bibitem [{\citenamefont {Spieß}\ \emph {et~al.}(2024)\citenamefont {Spieß},
		\citenamefont {Chen}, \citenamefont {Wilzewski}, \citenamefont {Wehrheim},
		\citenamefont {Gilles}, \citenamefont {Surzhykov}, \citenamefont {Benkler},
		\citenamefont {Filzinger}, \citenamefont {Steinel}, \citenamefont
		{Huntemann}, \citenamefont {Cheung}, \citenamefont {Porsev}, \citenamefont
		{Bondarev}, \citenamefont {Safronova}, \citenamefont {Crespo
			López-Urrutia},\ and\ \citenamefont {Schmidt}}]{spies_excited-state_2024}%
	\BibitemOpen
	\bibfield  {author} {\bibinfo {author} {\bibfnamefont {L.~J.}\ \bibnamefont
			{Spieß}}, \bibinfo {author} {\bibfnamefont {S.}~\bibnamefont {Chen}},
		\bibinfo {author} {\bibfnamefont {A.}~\bibnamefont {Wilzewski}}, \bibinfo
		{author} {\bibfnamefont {M.}~\bibnamefont {Wehrheim}}, \bibinfo {author}
		{\bibfnamefont {J.}~\bibnamefont {Gilles}}, \bibinfo {author} {\bibfnamefont
			{A.}~\bibnamefont {Surzhykov}}, \bibinfo {author} {\bibfnamefont
			{E.}~\bibnamefont {Benkler}}, \bibinfo {author} {\bibfnamefont
			{M.}~\bibnamefont {Filzinger}}, \bibinfo {author} {\bibfnamefont
			{M.}~\bibnamefont {Steinel}}, \bibinfo {author} {\bibfnamefont
			{N.}~\bibnamefont {Huntemann}}, \bibinfo {author} {\bibfnamefont
			{C.}~\bibnamefont {Cheung}}, \bibinfo {author} {\bibfnamefont {S.~G.}\
			\bibnamefont {Porsev}}, \bibinfo {author} {\bibfnamefont {A.~I.}\
			\bibnamefont {Bondarev}}, \bibinfo {author} {\bibfnamefont {M.~S.}\
			\bibnamefont {Safronova}}, \bibinfo {author} {\bibfnamefont {J.~R.}\
			\bibnamefont {Crespo López-Urrutia}},\ and\ \bibinfo {author} {\bibfnamefont
			{P.~O.}\ \bibnamefont {Schmidt}},\ }\bibfield  {title} {\bibinfo {title}
		{Excited-state magnetic properties of carbon-like calcium
			$\mathrm{Ca}^{14+}$},\ }\href@noop {} {\bibfield  {journal} {\bibinfo
			{journal} {Publication in preparation}\ } (\bibinfo {year}
		{2024})}\BibitemShut {NoStop}%
	\bibitem [{\citenamefont {Gilles}\ \emph {et~al.}(2024)\citenamefont {Gilles},
		\citenamefont {Fritzsche}, \citenamefont {Spieß}, \citenamefont {Schmidt},\
		and\ \citenamefont {Surzhykov}}]{gilles_quadratic_2024}%
	\BibitemOpen
	\bibfield  {author} {\bibinfo {author} {\bibfnamefont {J.}~\bibnamefont
			{Gilles}}, \bibinfo {author} {\bibfnamefont {S.}~\bibnamefont {Fritzsche}},
		\bibinfo {author} {\bibfnamefont {L.~J.}\ \bibnamefont {Spieß}}, \bibinfo
		{author} {\bibfnamefont {P.~O.}\ \bibnamefont {Schmidt}},\ and\ \bibinfo
		{author} {\bibfnamefont {A.}~\bibnamefont {Surzhykov}},\ }\bibfield  {title}
	{\bibinfo {title} {Quadratic {Zeeman} and electric quadrupole shifts in
			highly charged ions},\ }\href {https://doi.org/10.1103/PhysRevA.110.052812}
	{\bibfield  {journal} {\bibinfo  {journal} {Physical Review A}\ }\textbf
		{\bibinfo {volume} {110}},\ \bibinfo {pages} {052812} (\bibinfo {year}
		{2024})}\BibitemShut {NoStop}%
	\bibitem [{\citenamefont {Keller}\ \emph {et~al.}(2015)\citenamefont {Keller},
		\citenamefont {Partner}, \citenamefont {Burgermeister},\ and\ \citenamefont
		{Mehlst\"{a}ubler}}]{keller_precise_2015}%
	\BibitemOpen
	\bibfield  {author} {\bibinfo {author} {\bibfnamefont {J.}~\bibnamefont
			{Keller}}, \bibinfo {author} {\bibfnamefont {H.~L.}\ \bibnamefont {Partner}},
		\bibinfo {author} {\bibfnamefont {T.}~\bibnamefont {Burgermeister}},\ and\
		\bibinfo {author} {\bibfnamefont {T.~E.}\ \bibnamefont {Mehlst\"{a}ubler}},\
	}\bibfield  {title} {\bibinfo {title} {Precise determination of micromotion
			for trapped-ion optical clocks},\ }\href {https://doi.org/10.1063/1.4930037}
	{\bibfield  {journal} {\bibinfo  {journal} {Journal of Applied Physics}\
		}\textbf {\bibinfo {volume} {118}},\ \bibinfo {pages} {104501} (\bibinfo
		{year} {2015})}\BibitemShut {NoStop}%
	\bibitem [{\citenamefont {Yu}\ and\ \citenamefont
		{Sahoo}(2019)}]{yu_investigating_2019}%
	\BibitemOpen
	\bibfield  {author} {\bibinfo {author} {\bibfnamefont {Y.-m.}\ \bibnamefont
			{Yu}}\ and\ \bibinfo {author} {\bibfnamefont {B.~K.}\ \bibnamefont {Sahoo}},\
	}\bibfield  {title} {\bibinfo {title} {Investigating ground-state
			fine-structure properties to explore suitability of boronlike
			$\mathrm{{S}}^{11+}-\mathrm{{K}}^{14+}$ and galliumlike
			$\mathrm{{Nb}}^{14+}-\mathrm{{Ru}}^{13+}$ ions as possible atomic clocks},\
	}\href {https://doi.org/10.1103/PhysRevA.99.022513} {\bibfield  {journal}
		{\bibinfo  {journal} {Physical Review A}\ }\textbf {\bibinfo {volume} {99}},\
		\bibinfo {pages} {022513} (\bibinfo {year} {2019})}\BibitemShut {NoStop}%
	\bibitem [{\citenamefont {Kramida}\ \emph {et~al.}()\citenamefont {Kramida},
		\citenamefont {Ralchenko}, \citenamefont {Reader},\ and\ \citenamefont
		{Team}}]{kramida_nist_nodate}%
	\BibitemOpen
	\bibfield  {author} {\bibinfo {author} {\bibfnamefont {A.}~\bibnamefont
			{Kramida}}, \bibinfo {author} {\bibfnamefont {Y.}~\bibnamefont {Ralchenko}},
		\bibinfo {author} {\bibfnamefont {J.}~\bibnamefont {Reader}},\ and\ \bibinfo
		{author} {\bibfnamefont {N.~A.}\ \bibnamefont {Team}},\ }\href@noop {}
	{\bibinfo {title} {{NIST} {Atomic} {Spectra} {Database} (ver.
			5.11)}}\BibitemShut {NoStop}%
	\bibitem [{\citenamefont {Denker}\ and\ \citenamefont
		{Timmen}(2022)}]{denker_ergebnisse_2022}%
	\BibitemOpen
	\bibfield  {author} {\bibinfo {author} {\bibfnamefont {H.}~\bibnamefont
			{Denker}}\ and\ \bibinfo {author} {\bibfnamefont {L.}~\bibnamefont
			{Timmen}},\ }\bibfield  {title} {\bibinfo {title} {Ergebnisse des
			{Nivellements} an der {PTB} am 12.07.2022.},\ }\href@noop {} {\bibfield
		{journal} {\bibinfo  {journal} {Internal Report}\ } (\bibinfo {year}
		{2022})}\BibitemShut {NoStop}%
	\bibitem [{\citenamefont {Jönsson}\ \emph {et~al.}(2011)\citenamefont
		{Jönsson}, \citenamefont {Rynkun},\ and\ \citenamefont
		{Gaigalas}}]{jonsson_energies_2011}%
	\BibitemOpen
	\bibfield  {author} {\bibinfo {author} {\bibfnamefont {P.}~\bibnamefont
			{Jönsson}}, \bibinfo {author} {\bibfnamefont {P.}~\bibnamefont {Rynkun}},\
		and\ \bibinfo {author} {\bibfnamefont {G.}~\bibnamefont {Gaigalas}},\
	}\bibfield  {title} {\bibinfo {title} {Energies, {E1}, {M1}, and {E2}
			transition rates, hyperfine structures, and {Land\'{e}} $g_j$ factors for
			states of the 2s22p2, 2s2p3, and 2p4 configurations in carbon-like ions
			between {F} {IV} and {Ni} {XXIII}},\ }\href
	{https://doi.org/10.1016/j.adt.2011.05.001} {\bibfield  {journal} {\bibinfo
			{journal} {Atomic Data and Nuclear Data Tables}\ }\textbf {\bibinfo {volume}
			{97}},\ \bibinfo {pages} {648} (\bibinfo {year} {2011})}\BibitemShut
	{NoStop}%
	\bibitem [{\citenamefont {Roos}(2005)}]{roos_precision_2005}%
	\BibitemOpen
	\bibfield  {author} {\bibinfo {author} {\bibfnamefont {C.~F.}\ \bibnamefont
			{Roos}},\ }\href {https://doi.org/10.48550/arXiv.quant-ph/0508148} {\bibinfo
		{title} {Precision frequency measurements with entangled states}} (\bibinfo
	{year} {2005}),\ \bibinfo {note}
	{10.48550/arXiv.quant-ph/0508148}\BibitemShut {NoStop}%
	\bibitem [{\citenamefont {Chwalla}\ \emph {et~al.}(2007)\citenamefont
		{Chwalla}, \citenamefont {Kim}, \citenamefont {Monz}, \citenamefont
		{Schindler}, \citenamefont {Riebe}, \citenamefont {Roos},\ and\ \citenamefont
		{Blatt}}]{chwalla_precision_2007}%
	\BibitemOpen
	\bibfield  {author} {\bibinfo {author} {\bibfnamefont {M.}~\bibnamefont
			{Chwalla}}, \bibinfo {author} {\bibfnamefont {K.}~\bibnamefont {Kim}},
		\bibinfo {author} {\bibfnamefont {T.}~\bibnamefont {Monz}}, \bibinfo {author}
		{\bibfnamefont {P.}~\bibnamefont {Schindler}}, \bibinfo {author}
		{\bibfnamefont {M.}~\bibnamefont {Riebe}}, \bibinfo {author} {\bibfnamefont
			{C.}~\bibnamefont {Roos}},\ and\ \bibinfo {author} {\bibfnamefont
			{R.}~\bibnamefont {Blatt}},\ }\bibfield  {title} {\bibinfo {title} {Precision
			spectroscopy with two correlated atoms},\ }\href
	{https://doi.org/10.1007/s00340-007-2867-4} {\bibfield  {journal} {\bibinfo
			{journal} {Applied Physics B}\ }\textbf {\bibinfo {volume} {89}},\ \bibinfo
		{pages} {483} (\bibinfo {year} {2007})}\BibitemShut {NoStop}%
	\bibitem [{\citenamefont {Manovitz}\ \emph {et~al.}(2019)\citenamefont
		{Manovitz}, \citenamefont {Shaniv}, \citenamefont {Shapira}, \citenamefont
		{Ozeri},\ and\ \citenamefont {Akerman}}]{manovitz_precision_2019}%
	\BibitemOpen
	\bibfield  {author} {\bibinfo {author} {\bibfnamefont {T.}~\bibnamefont
			{Manovitz}}, \bibinfo {author} {\bibfnamefont {R.}~\bibnamefont {Shaniv}},
		\bibinfo {author} {\bibfnamefont {Y.}~\bibnamefont {Shapira}}, \bibinfo
		{author} {\bibfnamefont {R.}~\bibnamefont {Ozeri}},\ and\ \bibinfo {author}
		{\bibfnamefont {N.}~\bibnamefont {Akerman}},\ }\bibfield  {title} {\bibinfo
		{title} {Precision {Measurement} of {Atomic} {Isotope} {Shifts} {Using} a
			{Two}-{Isotope} {Entangled} {State}},\ }\href
	{https://doi.org/10.1103/PhysRevLett.123.203001} {\bibfield  {journal}
		{\bibinfo  {journal} {Physical Review Letters}\ }\textbf {\bibinfo {volume}
			{123}},\ \bibinfo {pages} {203001} (\bibinfo {year} {2019})}\BibitemShut
	{NoStop}%
	\bibitem [{\citenamefont {Lucas}\ \emph {et~al.}(2004)\citenamefont {Lucas},
		\citenamefont {Ramos}, \citenamefont {Home}, \citenamefont {McDonnell},
		\citenamefont {Nakayama}, \citenamefont {Stacey}, \citenamefont {Webster},
		\citenamefont {Stacey},\ and\ \citenamefont
		{Steane}}]{lucas_isotope-selective_2004}%
	\BibitemOpen
	\bibfield  {author} {\bibinfo {author} {\bibfnamefont {D.~M.}\ \bibnamefont
			{Lucas}}, \bibinfo {author} {\bibfnamefont {A.}~\bibnamefont {Ramos}},
		\bibinfo {author} {\bibfnamefont {J.~P.}\ \bibnamefont {Home}}, \bibinfo
		{author} {\bibfnamefont {M.~J.}\ \bibnamefont {McDonnell}}, \bibinfo {author}
		{\bibfnamefont {S.}~\bibnamefont {Nakayama}}, \bibinfo {author}
		{\bibfnamefont {J.~P.}\ \bibnamefont {Stacey}}, \bibinfo {author}
		{\bibfnamefont {S.~C.}\ \bibnamefont {Webster}}, \bibinfo {author}
		{\bibfnamefont {D.~N.}\ \bibnamefont {Stacey}},\ and\ \bibinfo {author}
		{\bibfnamefont {A.~M.}\ \bibnamefont {Steane}},\ }\bibfield  {title}
	{\bibinfo {title} {Isotope-selective photoionization for calcium ion
			trapping},\ }\href {https://doi.org/10.1103/PhysRevA.69.012711} {\bibfield
		{journal} {\bibinfo  {journal} {Physical Review A - Atomic, Molecular, and
				Optical Physics}\ }\textbf {\bibinfo {volume} {69}},\ \bibinfo {pages} {13}
		(\bibinfo {year} {2004})}\BibitemShut {NoStop}%
	\bibitem [{\citenamefont {Roos}\ \emph {et~al.}(2006)\citenamefont {Roos},
		\citenamefont {Chwalla}, \citenamefont {Kim}, \citenamefont {Riebe},\ and\
		\citenamefont {Blatt}}]{roos_designer_2006}%
	\BibitemOpen
	\bibfield  {author} {\bibinfo {author} {\bibfnamefont {C.~F.}\ \bibnamefont
			{Roos}}, \bibinfo {author} {\bibfnamefont {M.}~\bibnamefont {Chwalla}},
		\bibinfo {author} {\bibfnamefont {K.}~\bibnamefont {Kim}}, \bibinfo {author}
		{\bibfnamefont {M.}~\bibnamefont {Riebe}},\ and\ \bibinfo {author}
		{\bibfnamefont {R.}~\bibnamefont {Blatt}},\ }\bibfield  {title} {\bibinfo
		{title} {`{Designer} atoms' for quantum metrology},\ }\href
	{https://doi.org/10.1038/nature05101} {\bibfield  {journal} {\bibinfo
			{journal} {Nature}\ }\textbf {\bibinfo {volume} {443}},\ \bibinfo {pages}
		{316} (\bibinfo {year} {2006})}\BibitemShut {NoStop}%
	\bibitem [{\citenamefont {Matt}(2023)}]{matt_roland_parallel_2023}%
	\BibitemOpen
	\bibfield  {author} {\bibinfo {author} {\bibfnamefont {R.}~\bibnamefont
			{Matt}},\ }\emph {\bibinfo {title} {Parallel control of a dual-isotope
			trapped ion register}},\ \href@noop {} {Ph.D. thesis},\ \bibinfo  {school}
	{ETH Zurich} (\bibinfo {year} {2023})\BibitemShut {NoStop}%
	\bibitem [{\citenamefont {Chwalla}\ \emph {et~al.}(2009)\citenamefont
		{Chwalla}, \citenamefont {Benhelm}, \citenamefont {Kim}, \citenamefont
		{Kirchmair}, \citenamefont {Monz}, \citenamefont {Riebe}, \citenamefont
		{Schindler}, \citenamefont {Villar}, \citenamefont {H\"{a}nsel},
		\citenamefont {Roos}, \citenamefont {Blatt}, \citenamefont {Abgrall},
		\citenamefont {Santarelli}, \citenamefont {Rovera},\ and\ \citenamefont
		{Laurent}}]{chwalla_absolute_2009}%
	\BibitemOpen
	\bibfield  {author} {\bibinfo {author} {\bibfnamefont {M.}~\bibnamefont
			{Chwalla}}, \bibinfo {author} {\bibfnamefont {J.}~\bibnamefont {Benhelm}},
		\bibinfo {author} {\bibfnamefont {K.}~\bibnamefont {Kim}}, \bibinfo {author}
		{\bibfnamefont {G.}~\bibnamefont {Kirchmair}}, \bibinfo {author}
		{\bibfnamefont {T.}~\bibnamefont {Monz}}, \bibinfo {author} {\bibfnamefont
			{M.}~\bibnamefont {Riebe}}, \bibinfo {author} {\bibfnamefont
			{P.}~\bibnamefont {Schindler}}, \bibinfo {author} {\bibfnamefont {A.~S.}\
			\bibnamefont {Villar}}, \bibinfo {author} {\bibfnamefont {W.}~\bibnamefont
			{H\"{a}nsel}}, \bibinfo {author} {\bibfnamefont {C.~F.}\ \bibnamefont
			{Roos}}, \bibinfo {author} {\bibfnamefont {R.}~\bibnamefont {Blatt}},
		\bibinfo {author} {\bibfnamefont {M.}~\bibnamefont {Abgrall}}, \bibinfo
		{author} {\bibfnamefont {G.}~\bibnamefont {Santarelli}}, \bibinfo {author}
		{\bibfnamefont {G.~D.}\ \bibnamefont {Rovera}},\ and\ \bibinfo {author}
		{\bibfnamefont {P.}~\bibnamefont {Laurent}},\ }\bibfield  {title} {\bibinfo
		{title} {Absolute {Frequency} {Measurement} of the ${}^{40}\mathrm{Ca}^{+}$
			$4s^2{S}_{1/2}-3d^2{D}_{5/2}$ {Clock} {Transition}},\ }\href
	{https://doi.org/10.1103/PhysRevLett.102.023002} {\bibfield  {journal}
		{\bibinfo  {journal} {Physical Review Letters}\ }\textbf {\bibinfo {volume}
			{102}},\ \bibinfo {pages} {023002} (\bibinfo {year} {2009})}\BibitemShut
	{NoStop}%
	\bibitem [{\citenamefont {Berkeland}\ \emph {et~al.}(1998)\citenamefont
		{Berkeland}, \citenamefont {Miller}, \citenamefont {Bergquist}, \citenamefont
		{Itano},\ and\ \citenamefont {Wineland}}]{berkeland_minimization_1998}%
	\BibitemOpen
	\bibfield  {author} {\bibinfo {author} {\bibfnamefont {D.~J.}\ \bibnamefont
			{Berkeland}}, \bibinfo {author} {\bibfnamefont {J.~D.}\ \bibnamefont
			{Miller}}, \bibinfo {author} {\bibfnamefont {J.~C.}\ \bibnamefont
			{Bergquist}}, \bibinfo {author} {\bibfnamefont {W.~M.}\ \bibnamefont
			{Itano}},\ and\ \bibinfo {author} {\bibfnamefont {D.~J.}\ \bibnamefont
			{Wineland}},\ }\bibfield  {title} {\bibinfo {title} {Minimization of ion
			micromotion in a {Paul} trap},\ }\href {https://doi.org/10.1063/1.367318}
	{\bibfield  {journal} {\bibinfo  {journal} {Journal of Applied Physics}\
		}\textbf {\bibinfo {volume} {83}},\ \bibinfo {pages} {5025} (\bibinfo {year}
		{1998})}\BibitemShut {NoStop}%
	\bibitem [{noa(2003)}]{noauthor_58503b_2003}%
	\BibitemOpen
	\href {https://accusrc.com/uploads/datasheets/4975_58503b.pdf} {\emph
		{\bibinfo {title} {{58503B} {GPS} {Time} and {Frequency} {Reference}
				{Receiver}}}},\ \bibinfo {type} {Tech. Rep.}\ (\bibinfo  {institution}
	{Symmetricom, Inc.},\ \bibinfo {year} {2003})\BibitemShut {NoStop}%
	\bibitem [{\citenamefont {Knollmann}\ \emph {et~al.}(2019)\citenamefont
		{Knollmann}, \citenamefont {Patel},\ and\ \citenamefont
		{Doret}}]{knollmann_part-per-billion_2019}%
	\BibitemOpen
	\bibfield  {author} {\bibinfo {author} {\bibfnamefont {F.~W.}\ \bibnamefont
			{Knollmann}}, \bibinfo {author} {\bibfnamefont {A.~N.}\ \bibnamefont
			{Patel}},\ and\ \bibinfo {author} {\bibfnamefont {S.~C.}\ \bibnamefont
			{Doret}},\ }\bibfield  {title} {\bibinfo {title} {Part-per-billion
			measurement of the $4^2\mathrm{S}_{1/2} \rightarrow 3^2\mathrm{D}_{5/2}$
			electric-quadrupole-transition isotope shifts between
			${}^{42,44,48}\mathrm{Ca}^+$ and ${}^{40}\mathrm{Ca}^+$},\ }\href
	{https://doi.org/10.1103/PhysRevA.100.022514} {\bibfield  {journal} {\bibinfo
			{journal} {Physical Review A}\ }\textbf {\bibinfo {volume} {100}},\ \bibinfo
		{pages} {022514} (\bibinfo {year} {2019})}\BibitemShut {NoStop}%
	\bibitem [{\citenamefont {Brown}\ and\ \citenamefont
		{Gabrielse}(1982)}]{brown_precision_1982}%
	\BibitemOpen
	\bibfield  {author} {\bibinfo {author} {\bibfnamefont {L.~S.}\ \bibnamefont
			{Brown}}\ and\ \bibinfo {author} {\bibfnamefont {G.}~\bibnamefont
			{Gabrielse}},\ }\bibfield  {title} {\bibinfo {title} {Precision spectroscopy
			of a charged particle in an imperfect {Penning} trap},\ }\href
	{https://doi.org/10.1103/PhysRevA.25.2423} {\bibfield  {journal} {\bibinfo
			{journal} {Physical Review A}\ }\textbf {\bibinfo {volume} {25}},\ \bibinfo
		{pages} {2423} (\bibinfo {year} {1982})}\BibitemShut {NoStop}%
	\bibitem [{\citenamefont {Cornell}\ \emph {et~al.}(1990)\citenamefont
		{Cornell}, \citenamefont {Weisskoff}, \citenamefont {Boyce},\ and\
		\citenamefont {Pritchard}}]{cornell_mode_1990}%
	\BibitemOpen
	\bibfield  {author} {\bibinfo {author} {\bibfnamefont {E.~A.}\ \bibnamefont
			{Cornell}}, \bibinfo {author} {\bibfnamefont {R.~M.}\ \bibnamefont
			{Weisskoff}}, \bibinfo {author} {\bibfnamefont {K.~R.}\ \bibnamefont
			{Boyce}},\ and\ \bibinfo {author} {\bibfnamefont {D.~E.}\ \bibnamefont
			{Pritchard}},\ }\bibfield  {title} {\bibinfo {title} {Mode coupling in a
			{Penning} trap: $\pi$ pulses and a classical avoided crossing},\ }\href
	{https://doi.org/10.1103/PhysRevA.41.312} {\bibfield  {journal} {\bibinfo
			{journal} {Physical Review A}\ }\textbf {\bibinfo {volume} {41}},\ \bibinfo
		{pages} {312} (\bibinfo {year} {1990})}\BibitemShut {NoStop}%
	\bibitem [{\citenamefont {Cornell}\ \emph {et~al.}(1989)\citenamefont
		{Cornell}, \citenamefont {Weisskoff}, \citenamefont {Boyce}, \citenamefont
		{Flanagan}, \citenamefont {Lafyatis},\ and\ \citenamefont
		{Pritchard}}]{cornell_single-ion_1989}%
	\BibitemOpen
	\bibfield  {author} {\bibinfo {author} {\bibfnamefont {E.~A.}\ \bibnamefont
			{Cornell}}, \bibinfo {author} {\bibfnamefont {R.~M.}\ \bibnamefont
			{Weisskoff}}, \bibinfo {author} {\bibfnamefont {K.~R.}\ \bibnamefont
			{Boyce}}, \bibinfo {author} {\bibfnamefont {R.~W.}\ \bibnamefont {Flanagan}},
		\bibinfo {author} {\bibfnamefont {G.~P.}\ \bibnamefont {Lafyatis}},\ and\
		\bibinfo {author} {\bibfnamefont {D.~E.}\ \bibnamefont {Pritchard}},\
	}\bibfield  {title} {\bibinfo {title} {Single-ion cyclotron resonance
			measurement of ${M}(\mathrm{CO}^+)/{M}(\mathrm{N}_{2}^+)$},\ }\href
	{https://doi.org/10.1103/PhysRevLett.63.1674} {\bibfield  {journal} {\bibinfo
			{journal} {Physical Review Letters}\ }\textbf {\bibinfo {volume} {63}},\
		\bibinfo {pages} {1674} (\bibinfo {year} {1989})}\BibitemShut {NoStop}%
	\bibitem [{\citenamefont {Froese~Fischer}\ \emph {et~al.}(2019)\citenamefont
		{Froese~Fischer}, \citenamefont {Gaigalas}, \citenamefont {Jönsson},\ and\
		\citenamefont {Bieroń}}]{froese_fischer_grasp2018fortran_2019}%
	\BibitemOpen
	\bibfield  {author} {\bibinfo {author} {\bibfnamefont {C.}~\bibnamefont
			{Froese~Fischer}}, \bibinfo {author} {\bibfnamefont {G.}~\bibnamefont
			{Gaigalas}}, \bibinfo {author} {\bibfnamefont {P.}~\bibnamefont {Jönsson}},\
		and\ \bibinfo {author} {\bibfnamefont {J.}~\bibnamefont {Bieroń}},\
	}\bibfield  {title} {\bibinfo {title} {{GRASP2018} -- {A} {Fortran} 95
			version of the {General} {Relativistic} {Atomic} {Structure} {Package}},\
	}\href {https://doi.org/10.1016/j.cpc.2018.10.032} {\bibfield  {journal}
		{\bibinfo  {journal} {Computer Physics Communications}\ }\textbf {\bibinfo
			{volume} {237}},\ \bibinfo {pages} {184} (\bibinfo {year}
		{2019})}\BibitemShut {NoStop}%
	\bibitem [{\citenamefont {Grant}(1970)}]{Grant1970}%
	\BibitemOpen
	\bibfield  {author} {\bibinfo {author} {\bibfnamefont {I.~P.}\ \bibnamefont
			{Grant}},\ }\bibfield  {title} {\bibinfo {title} {Relativistic calculation of
			atomic structures},\ }\href@noop {} {\bibfield  {journal} {\bibinfo
			{journal} {Adv.~Phys.}\ }\textbf {\bibinfo {volume} {19}},\ \bibinfo {pages}
		{747} (\bibinfo {year} {1970})}\BibitemShut {NoStop}%
	\bibitem [{\citenamefont {Desclaux}\ \emph {et~al.}(1971)\citenamefont
		{Desclaux}, \citenamefont {Mayers},\ and\ \citenamefont
		{O'Brien}}]{Desclaux1971}%
	\BibitemOpen
	\bibfield  {author} {\bibinfo {author} {\bibfnamefont {J.~P.}\ \bibnamefont
			{Desclaux}}, \bibinfo {author} {\bibfnamefont {D.~F.}\ \bibnamefont
			{Mayers}},\ and\ \bibinfo {author} {\bibfnamefont {F.}~\bibnamefont
			{O'Brien}},\ }\bibfield  {title} {\bibinfo {title} {Relativistic atomic wave
			functions},\ }\href@noop {} {\bibfield  {journal} {\bibinfo  {journal}
			{J.~Phys.~B}\ }\textbf {\bibinfo {volume} {4}},\ \bibinfo {pages} {631}
		(\bibinfo {year} {1971})}\BibitemShut {NoStop}%
	\bibitem [{\citenamefont {Grant}(2007)}]{grant2007relativistic}%
	\BibitemOpen
	\bibfield  {author} {\bibinfo {author} {\bibfnamefont {I.~P.}\ \bibnamefont
			{Grant}},\ }\href@noop {} {\emph {\bibinfo {title} {Relativistic quantum
				theory of atoms and molecules: theory and computation, Springer Series on
				Atomic, Optical, and Plasma Physics, Vol. 40}}}\ (\bibinfo  {publisher}
	{Springer},\ \bibinfo {address} {Berlin},\ \bibinfo {year}
	{2007})\BibitemShut {NoStop}%
	\bibitem [{\citenamefont {Froese-Fischer}\ \emph {et~al.}(2019)\citenamefont
		{Froese-Fischer}, \citenamefont {Gaigalas}, \citenamefont {J{\"o}nsson},\
		and\ \citenamefont {Biero{\'n}}}]{GRASP2018}%
	\BibitemOpen
	\bibfield  {author} {\bibinfo {author} {\bibfnamefont {C.}~\bibnamefont
			{Froese-Fischer}}, \bibinfo {author} {\bibfnamefont {G.}~\bibnamefont
			{Gaigalas}}, \bibinfo {author} {\bibfnamefont {P.}~\bibnamefont
			{J{\"o}nsson}},\ and\ \bibinfo {author} {\bibfnamefont {J.}~\bibnamefont
			{Biero{\'n}}},\ }\bibfield  {title} {\bibinfo {title} {Grasp2018 -- a fortran
			95 version of the general relativistic atomic structure package},\ }\href
	{https://doi.org/https://doi.org/10.1016/j.cpc.2018.10.032} {\bibfield
		{journal} {\bibinfo  {journal} {Comput. Phys. Commun.}\ }\textbf {\bibinfo
			{volume} {237}},\ \bibinfo {pages} {184 } (\bibinfo {year}
		{2019})}\BibitemShut {NoStop}%
	\bibitem [{\citenamefont {J\"{o}nsson}\ \emph
		{et~al.}(2023{\natexlab{a}})\citenamefont {J\"{o}nsson}, \citenamefont
		{Godefroid}, \citenamefont {Gaigalas}, \citenamefont {Ekman}, \citenamefont
		{Grumer}, \citenamefont {Li}, \citenamefont {Li}, \citenamefont {Brage},
		\citenamefont {Grant}, \citenamefont {Biero\'{n}},\ and\ \citenamefont
		{Froese~Fischer}}]{Jonsson2023-2}%
	\BibitemOpen
	\bibfield  {author} {\bibinfo {author} {\bibfnamefont {P.}~\bibnamefont
			{J\"{o}nsson}}, \bibinfo {author} {\bibfnamefont {M.}~\bibnamefont
			{Godefroid}}, \bibinfo {author} {\bibfnamefont {G.}~\bibnamefont {Gaigalas}},
		\bibinfo {author} {\bibfnamefont {J.}~\bibnamefont {Ekman}}, \bibinfo
		{author} {\bibfnamefont {J.}~\bibnamefont {Grumer}}, \bibinfo {author}
		{\bibfnamefont {W.}~\bibnamefont {Li}}, \bibinfo {author} {\bibfnamefont
			{J.}~\bibnamefont {Li}}, \bibinfo {author} {\bibfnamefont {T.}~\bibnamefont
			{Brage}}, \bibinfo {author} {\bibfnamefont {I.}~\bibnamefont {Grant}},
		\bibinfo {author} {\bibfnamefont {J.}~\bibnamefont {Biero\'{n}}},\ and\
		\bibinfo {author} {\bibfnamefont {C.}~\bibnamefont {Froese~Fischer}},\
	}\bibfield  {title} {\bibinfo {title} {An introduction to relativistic theory
			as implemented in grasp},\ }\href {https://doi.org/10.3390/atoms11010007}
	{\bibfield  {journal} {\bibinfo  {journal} {Atoms}\ }\textbf {\bibinfo
			{volume} {11}},\ \bibinfo {pages} {7} (\bibinfo {year}
		{2023}{\natexlab{a}})}\BibitemShut {NoStop}%
	\bibitem [{\citenamefont {Malyshev}\ \emph {et~al.}(2014)\citenamefont
		{Malyshev}, \citenamefont {Volotka}, \citenamefont {Glazov}, \citenamefont
		{Tupitsyn}, \citenamefont {Shabaev},\ and\ \citenamefont
		{Plunien}}]{PhysRevA.90.062517}%
	\BibitemOpen
	\bibfield  {author} {\bibinfo {author} {\bibfnamefont {A.~V.}\ \bibnamefont
			{Malyshev}}, \bibinfo {author} {\bibfnamefont {A.~V.}\ \bibnamefont
			{Volotka}}, \bibinfo {author} {\bibfnamefont {D.~A.}\ \bibnamefont {Glazov}},
		\bibinfo {author} {\bibfnamefont {I.~I.}\ \bibnamefont {Tupitsyn}}, \bibinfo
		{author} {\bibfnamefont {V.~M.}\ \bibnamefont {Shabaev}},\ and\ \bibinfo
		{author} {\bibfnamefont {G.}~\bibnamefont {Plunien}},\ }\bibfield  {title}
	{\bibinfo {title} {Qed calculation of the ground-state energy of
			berylliumlike ions},\ }\href {https://doi.org/10.1103/PhysRevA.90.062517}
	{\bibfield  {journal} {\bibinfo  {journal} {Phys. Rev. A}\ }\textbf {\bibinfo
			{volume} {90}},\ \bibinfo {pages} {062517} (\bibinfo {year}
		{2014})}\BibitemShut {NoStop}%
	\bibitem [{\citenamefont {Malyshev}\ \emph {et~al.}(2021)\citenamefont
		{Malyshev}, \citenamefont {Glazov}, \citenamefont {Kozhedub}, \citenamefont
		{Anisimova}, \citenamefont {Kaygorodov}, \citenamefont {Shabaev},\ and\
		\citenamefont {Tupitsyn}}]{PhysRevLett.126.183001}%
	\BibitemOpen
	\bibfield  {author} {\bibinfo {author} {\bibfnamefont {A.~V.}\ \bibnamefont
			{Malyshev}}, \bibinfo {author} {\bibfnamefont {D.~A.}\ \bibnamefont
			{Glazov}}, \bibinfo {author} {\bibfnamefont {Y.~S.}\ \bibnamefont
			{Kozhedub}}, \bibinfo {author} {\bibfnamefont {I.~S.}\ \bibnamefont
			{Anisimova}}, \bibinfo {author} {\bibfnamefont {M.~Y.}\ \bibnamefont
			{Kaygorodov}}, \bibinfo {author} {\bibfnamefont {V.~M.}\ \bibnamefont
			{Shabaev}},\ and\ \bibinfo {author} {\bibfnamefont {I.~I.}\ \bibnamefont
			{Tupitsyn}},\ }\bibfield  {title} {\bibinfo {title} {Ab initio calculations
			of energy levels in be-like xenon: Strong interference between
			electron-correlation and qed effects},\ }\href
	{https://doi.org/10.1103/PhysRevLett.126.183001} {\bibfield  {journal}
		{\bibinfo  {journal} {Phys. Rev. Lett.}\ }\textbf {\bibinfo {volume} {126}},\
		\bibinfo {pages} {183001} (\bibinfo {year} {2021})}\BibitemShut {NoStop}%
	\bibitem [{\citenamefont {J\"{o}nsson}\ \emph
		{et~al.}(2023{\natexlab{b}})\citenamefont {J\"{o}nsson}, \citenamefont
		{Gaigalas}, \citenamefont {Froese~Fischer}, \citenamefont {Biero\'{n}},
		\citenamefont {Grant}, \citenamefont {Brage}, \citenamefont {Ekman},
		\citenamefont {Godefroid}, \citenamefont {Grumer}, \citenamefont {Li},\ and\
		\citenamefont {Li}}]{Jonsson2023}%
	\BibitemOpen
	\bibfield  {author} {\bibinfo {author} {\bibfnamefont {P.}~\bibnamefont
			{J\"{o}nsson}}, \bibinfo {author} {\bibfnamefont {G.}~\bibnamefont
			{Gaigalas}}, \bibinfo {author} {\bibfnamefont {C.}~\bibnamefont
			{Froese~Fischer}}, \bibinfo {author} {\bibfnamefont {J.}~\bibnamefont
			{Biero\'{n}}}, \bibinfo {author} {\bibfnamefont {I.}~\bibnamefont {Grant}},
		\bibinfo {author} {\bibfnamefont {T.}~\bibnamefont {Brage}}, \bibinfo
		{author} {\bibfnamefont {J.}~\bibnamefont {Ekman}}, \bibinfo {author}
		{\bibfnamefont {M.}~\bibnamefont {Godefroid}}, \bibinfo {author}
		{\bibfnamefont {J.}~\bibnamefont {Grumer}}, \bibinfo {author} {\bibfnamefont
			{J.}~\bibnamefont {Li}},\ and\ \bibinfo {author} {\bibfnamefont
			{W.}~\bibnamefont {Li}},\ }\bibfield  {title} {\bibinfo {title} {Grasp manual
			for users},\ }\href {https://doi.org/10.3390/atoms11040068} {\bibfield
		{journal} {\bibinfo  {journal} {Atoms}\ }\textbf {\bibinfo {volume} {11}},\
		\bibinfo {pages} {68} (\bibinfo {year} {2023}{\natexlab{b}})}\BibitemShut
	{NoStop}%
	\bibitem [{\citenamefont {Lyu}\ \emph {et~al.}()\citenamefont {Lyu},
		\citenamefont {Sikora}, \citenamefont {Harman},\ and\ \citenamefont
		{Keitel}}]{lyu2023extreme}%
	\BibitemOpen
	\bibfield  {author} {\bibinfo {author} {\bibfnamefont {C.}~\bibnamefont
			{Lyu}}, \bibinfo {author} {\bibfnamefont {B.}~\bibnamefont {Sikora}},
		\bibinfo {author} {\bibfnamefont {Z.}~\bibnamefont {Harman}},\ and\ \bibinfo
		{author} {\bibfnamefont {C.~H.}\ \bibnamefont {Keitel}},\ }\bibfield  {title}
	{\bibinfo {title} {Extreme field calculations for {Penning} ion traps and
			corresponding strong laser field scenarios},\ }\href
	{https://doi.org/10.1080/00268976.2023.2252105} {\bibfield  {journal}
		{\bibinfo  {journal} {Molecular Physics}\ }\textbf {\bibinfo {volume} {0}},\
		\bibinfo {pages} {e2252105}}\BibitemShut {NoStop}%
	\bibitem [{\citenamefont {Wang}\ \emph {et~al.}(2021)\citenamefont {Wang},
		\citenamefont {Huang}, \citenamefont {Kondev}, \citenamefont {Audi},\ and\
		\citenamefont {Naimi}}]{wang_ame_2021}%
	\BibitemOpen
	\bibfield  {author} {\bibinfo {author} {\bibfnamefont {M.}~\bibnamefont
			{Wang}}, \bibinfo {author} {\bibfnamefont {W.~J.}\ \bibnamefont {Huang}},
		\bibinfo {author} {\bibfnamefont {F.~G.}\ \bibnamefont {Kondev}}, \bibinfo
		{author} {\bibfnamefont {G.}~\bibnamefont {Audi}},\ and\ \bibinfo {author}
		{\bibfnamefont {S.}~\bibnamefont {Naimi}},\ }\bibfield  {title} {\bibinfo
		{title} {The {AME} 2020 atomic mass evaluation ({II}). {Tables}, graphs and
			references},\ }\href {https://doi.org/10.1088/1674-1137/abddaf} {\bibfield
		{journal} {\bibinfo  {journal} {Chinese Physics C}\ }\textbf {\bibinfo
			{volume} {45}},\ \bibinfo {pages} {030003} (\bibinfo {year}
		{2021})}\BibitemShut {NoStop}%
	\bibitem [{\citenamefont {Tiesinga}\ \emph {et~al.}(2021)\citenamefont
		{Tiesinga}, \citenamefont {Mohr}, \citenamefont {Newell},\ and\ \citenamefont
		{Taylor}}]{tiesinga_codata_2021}%
	\BibitemOpen
	\bibfield  {author} {\bibinfo {author} {\bibfnamefont {E.}~\bibnamefont
			{Tiesinga}}, \bibinfo {author} {\bibfnamefont {P.~J.}\ \bibnamefont {Mohr}},
		\bibinfo {author} {\bibfnamefont {D.~B.}\ \bibnamefont {Newell}},\ and\
		\bibinfo {author} {\bibfnamefont {B.~N.}\ \bibnamefont {Taylor}},\ }\bibfield
	{title} {\bibinfo {title} {{CODATA} recommended values of the fundamental
			physical constants: 2018},\ }\href
	{https://doi.org/10.1103/RevModPhys.93.025010} {\bibfield  {journal}
		{\bibinfo  {journal} {Reviews of Modern Physics}\ }\textbf {\bibinfo {volume}
			{93}},\ \bibinfo {pages} {025010} (\bibinfo {year} {2021})}\BibitemShut
	{NoStop}%
	\bibitem [{\citenamefont {Berengut}\ \emph {et~al.}(2020)\citenamefont
		{Berengut}, \citenamefont {Delaunay}, \citenamefont {Geddes},\ and\
		\citenamefont {Soreq}}]{berengut_generalized_2020}%
	\BibitemOpen
	\bibfield  {author} {\bibinfo {author} {\bibfnamefont {J.~C.}\ \bibnamefont
			{Berengut}}, \bibinfo {author} {\bibfnamefont {C.}~\bibnamefont {Delaunay}},
		\bibinfo {author} {\bibfnamefont {A.}~\bibnamefont {Geddes}},\ and\ \bibinfo
		{author} {\bibfnamefont {Y.}~\bibnamefont {Soreq}},\ }\bibfield  {title}
	{\bibinfo {title} {Generalized {King} linearity and new physics searches with
			isotope shifts},\ }\href {https://doi.org/10.1103/PhysRevResearch.2.043444}
	{\bibfield  {journal} {\bibinfo  {journal} {Physical Review Research}\
		}\textbf {\bibinfo {volume} {2}},\ \bibinfo {pages} {043444} (\bibinfo {year}
		{2020})}\BibitemShut {NoStop}%
	\bibitem [{\citenamefont {Sun}\ \emph {et~al.}(2024)\citenamefont {Sun},
		\citenamefont {Valuev},\ and\ \citenamefont {Oreshkina}}]{sun_nuclear_2024}%
	\BibitemOpen
	\bibfield  {author} {\bibinfo {author} {\bibfnamefont {Z.}~\bibnamefont
			{Sun}}, \bibinfo {author} {\bibfnamefont {I.~A.}\ \bibnamefont {Valuev}},\
		and\ \bibinfo {author} {\bibfnamefont {N.~S.}\ \bibnamefont {Oreshkina}},\
	}\bibfield  {title} {\bibinfo {title} {Nuclear deformation effects in the
			spectra of highly charged ions},\ }\href
	{https://doi.org/10.1103/PhysRevResearch.6.023327} {\bibfield  {journal}
		{\bibinfo  {journal} {Physical Review Research}\ }\textbf {\bibinfo {volume}
			{6}},\ \bibinfo {pages} {023327} (\bibinfo {year} {2024})}\BibitemShut
	{NoStop}%
	\bibitem [{\citenamefont {Kahl}\ and\ \citenamefont
		{Berengut}(2019)}]{kahl_ambit_2019}%
	\BibitemOpen
	\bibfield  {author} {\bibinfo {author} {\bibfnamefont {E.}~\bibnamefont
			{Kahl}}\ and\ \bibinfo {author} {\bibfnamefont {J.}~\bibnamefont
			{Berengut}},\ }\bibfield  {title} {\bibinfo {title} {{AMBIT}: {A} programme
			for high-precision relativistic atomic structure calculations},\ }\href
	{https://doi.org/10.1016/j.cpc.2018.12.014} {\bibfield  {journal} {\bibinfo
			{journal} {Computer Physics Communications}\ }\textbf {\bibinfo {volume}
			{238}},\ \bibinfo {pages} {232} (\bibinfo {year} {2019})}\BibitemShut
	{NoStop}%
	\bibitem [{\citenamefont {Yerokhin}(2008)}]{yerokhin_hyperfine_2008}%
	\BibitemOpen
	\bibfield  {author} {\bibinfo {author} {\bibfnamefont {V.~A.}\ \bibnamefont
			{Yerokhin}},\ }\bibfield  {title} {\bibinfo {title} {Hyperfine structure of
			{Li} and $\mathrm{Be}^+$},\ }\href
	{https://doi.org/10.1103/PhysRevA.78.012513} {\bibfield  {journal} {\bibinfo
			{journal} {Physical Review A}\ }\textbf {\bibinfo {volume} {78}},\ \bibinfo
		{pages} {012513} (\bibinfo {year} {2008})}\BibitemShut {NoStop}%
	\bibitem [{\citenamefont {Naz\'{e}}\ \emph {et~al.}(2014)\citenamefont
		{Naz\'{e}}, \citenamefont {Verdebout}, \citenamefont {Rynkun}, \citenamefont
		{Gaigalas}, \citenamefont {Godefroid},\ and\ \citenamefont
		{Jönsson}}]{naze_isotope_2014}%
	\BibitemOpen
	\bibfield  {author} {\bibinfo {author} {\bibfnamefont {C.}~\bibnamefont
			{Naz\'{e}}}, \bibinfo {author} {\bibfnamefont {S.}~\bibnamefont {Verdebout}},
		\bibinfo {author} {\bibfnamefont {P.}~\bibnamefont {Rynkun}}, \bibinfo
		{author} {\bibfnamefont {G.}~\bibnamefont {Gaigalas}}, \bibinfo {author}
		{\bibfnamefont {M.}~\bibnamefont {Godefroid}},\ and\ \bibinfo {author}
		{\bibfnamefont {P.}~\bibnamefont {Jönsson}},\ }\bibfield  {title} {\bibinfo
		{title} {Isotope shifts in beryllium-, boron-, carbon-, and nitrogen-like
			ions from relativistic configuration interaction calculations},\ }\href
	{https://doi.org/10.1016/j.adt.2014.02.004} {\bibfield  {journal} {\bibinfo
			{journal} {Atomic Data and Nuclear Data Tables}\ }\textbf {\bibinfo {volume}
			{100}},\ \bibinfo {pages} {1197} (\bibinfo {year} {2014})}\BibitemShut
	{NoStop}%
	\bibitem [{\citenamefont {Plunien}\ \emph {et~al.}(1991)\citenamefont
		{Plunien}, \citenamefont {M\"{u}ller}, \citenamefont {Greiner},\ and\
		\citenamefont {Soff}}]{plunien_nuclear_1991}%
	\BibitemOpen
	\bibfield  {author} {\bibinfo {author} {\bibfnamefont {G.}~\bibnamefont
			{Plunien}}, \bibinfo {author} {\bibfnamefont {B.}~\bibnamefont {M\"{u}ller}},
		\bibinfo {author} {\bibfnamefont {W.}~\bibnamefont {Greiner}},\ and\ \bibinfo
		{author} {\bibfnamefont {G.}~\bibnamefont {Soff}},\ }\bibfield  {title}
	{\bibinfo {title} {Nuclear polarization in heavy atoms and superheavy
			quasiatoms},\ }\href {https://doi.org/10.1103/PhysRevA.43.5853} {\bibfield
		{journal} {\bibinfo  {journal} {Physical Review A}\ }\textbf {\bibinfo
			{volume} {43}},\ \bibinfo {pages} {5853} (\bibinfo {year}
		{1991})}\BibitemShut {NoStop}%
	\bibitem [{\citenamefont {Nefiodov}\ \emph {et~al.}(1996)\citenamefont
		{Nefiodov}, \citenamefont {Labzowsky}, \citenamefont {Plunien},\ and\
		\citenamefont {Soff}}]{nefiodov_nuclear_1996}%
	\BibitemOpen
	\bibfield  {author} {\bibinfo {author} {\bibfnamefont {A.~V.}\ \bibnamefont
			{Nefiodov}}, \bibinfo {author} {\bibfnamefont {L.~N.}\ \bibnamefont
			{Labzowsky}}, \bibinfo {author} {\bibfnamefont {G.}~\bibnamefont {Plunien}},\
		and\ \bibinfo {author} {\bibfnamefont {G.}~\bibnamefont {Soff}},\ }\bibfield
	{title} {\bibinfo {title} {Nuclear polarization effects in spectra of
			multicharged ions},\ }\href {https://doi.org/10.1016/0375-9601(96)00650-0}
	{\bibfield  {journal} {\bibinfo  {journal} {Physics Letters A}\ }\textbf
		{\bibinfo {volume} {222}},\ \bibinfo {pages} {227} (\bibinfo {year}
		{1996})}\BibitemShut {NoStop}%
	\bibitem [{\citenamefont {Valuev}\ and\ \citenamefont
		{Oreshkina}(2024)}]{valuev_full_2024}%
	\BibitemOpen
	\bibfield  {author} {\bibinfo {author} {\bibfnamefont {I.~A.}\ \bibnamefont
			{Valuev}}\ and\ \bibinfo {author} {\bibfnamefont {N.~S.}\ \bibnamefont
			{Oreshkina}},\ }\bibfield  {title} {\bibinfo {title} {Full leading-order
			nuclear polarization in highly charged ions},\ }\href
	{https://doi.org/10.1103/PhysRevA.109.042811} {\bibfield  {journal} {\bibinfo
			{journal} {Physical Review A}\ }\textbf {\bibinfo {volume} {109}},\ \bibinfo
		{pages} {042811} (\bibinfo {year} {2024})}\BibitemShut {NoStop}%
	\bibitem [{\citenamefont {Hanneke}\ \emph {et~al.}(2008)\citenamefont
		{Hanneke}, \citenamefont {Fogwell},\ and\ \citenamefont
		{Gabrielse}}]{hanneke_new_2008}%
	\BibitemOpen
	\bibfield  {author} {\bibinfo {author} {\bibfnamefont {D.}~\bibnamefont
			{Hanneke}}, \bibinfo {author} {\bibfnamefont {S.}~\bibnamefont {Fogwell}},\
		and\ \bibinfo {author} {\bibfnamefont {G.}~\bibnamefont {Gabrielse}},\
	}\bibfield  {title} {\bibinfo {title} {New {Measurement} of the {Electron}
			{Magnetic} {Moment} and the {Fine} {Structure} {Constant}},\ }\href
	{https://doi.org/10.1103/PhysRevLett.100.120801} {\bibfield  {journal}
		{\bibinfo  {journal} {Physical Review Letters}\ }\textbf {\bibinfo {volume}
			{100}},\ \bibinfo {pages} {120801} (\bibinfo {year} {2008})}\BibitemShut
	{NoStop}%
	\bibitem [{\citenamefont {Fan}\ \emph {et~al.}(2023)\citenamefont {Fan},
		\citenamefont {Myers}, \citenamefont {Sukra},\ and\ \citenamefont
		{Gabrielse}}]{fan_measurement_2023}%
	\BibitemOpen
	\bibfield  {author} {\bibinfo {author} {\bibfnamefont {X.}~\bibnamefont
			{Fan}}, \bibinfo {author} {\bibfnamefont {T.~G.}\ \bibnamefont {Myers}},
		\bibinfo {author} {\bibfnamefont {B.~A.~D.}\ \bibnamefont {Sukra}},\ and\
		\bibinfo {author} {\bibfnamefont {G.}~\bibnamefont {Gabrielse}},\ }\bibfield
	{title} {\bibinfo {title} {Measurement of the {Electron} {Magnetic}
			{Moment}},\ }\href {https://doi.org/10.1103/PhysRevLett.130.071801}
	{\bibfield  {journal} {\bibinfo  {journal} {Physical Review Letters}\
		}\textbf {\bibinfo {volume} {130}},\ \bibinfo {pages} {071801} (\bibinfo
		{year} {2023})}\BibitemShut {NoStop}%
	\bibitem [{\citenamefont {Pospelov}(2009)}]{Pospelov:2008zw}%
	\BibitemOpen
	\bibfield  {author} {\bibinfo {author} {\bibfnamefont {M.}~\bibnamefont
			{Pospelov}},\ }\bibfield  {title} {\bibinfo {title} {{Secluded U(1) below the
				weak scale}},\ }\href {https://doi.org/10.1103/PhysRevD.80.095002} {\bibfield
		{journal} {\bibinfo  {journal} {Phys. Rev. D}\ }\textbf {\bibinfo {volume}
			{80}},\ \bibinfo {pages} {095002} (\bibinfo {year} {2009})},\ \Eprint
	{https://arxiv.org/abs/0811.1030} {arXiv:0811.1030 [hep-ph]} \BibitemShut
	{NoStop}%
	\bibitem [{\citenamefont {Frugiuele}\ \emph {et~al.}(2017)\citenamefont
		{Frugiuele}, \citenamefont {Fuchs}, \citenamefont {Perez},\ and\
		\citenamefont {Schlaffer}}]{Frugiuele:2016rii}%
	\BibitemOpen
	\bibfield  {author} {\bibinfo {author} {\bibfnamefont {C.}~\bibnamefont
			{Frugiuele}}, \bibinfo {author} {\bibfnamefont {E.}~\bibnamefont {Fuchs}},
		\bibinfo {author} {\bibfnamefont {G.}~\bibnamefont {Perez}},\ and\ \bibinfo
		{author} {\bibfnamefont {M.}~\bibnamefont {Schlaffer}},\ }\bibfield  {title}
	{\bibinfo {title} {{Constraining New Physics Models with Isotope Shift
				Spectroscopy}},\ }\href {https://doi.org/10.1103/PhysRevD.96.015011}
	{\bibfield  {journal} {\bibinfo  {journal} {Phys. Rev. D}\ }\textbf {\bibinfo
			{volume} {96}},\ \bibinfo {pages} {015011} (\bibinfo {year}
		{2017})}\BibitemShut {NoStop}%
	\bibitem [{\citenamefont {Leeb}\ and\ \citenamefont
		{Schmiedmayer}(1992)}]{leeb_constraint_1992}%
	\BibitemOpen
	\bibfield  {author} {\bibinfo {author} {\bibfnamefont {H.}~\bibnamefont
			{Leeb}}\ and\ \bibinfo {author} {\bibfnamefont {J.}~\bibnamefont
			{Schmiedmayer}},\ }\bibfield  {title} {\bibinfo {title} {Constraint on
			hypothetical light interacting bosons from low-energy neutron experiments},\
	}\href {https://doi.org/10.1103/PhysRevLett.68.1472} {\bibfield  {journal}
		{\bibinfo  {journal} {Physical Review Letters}\ }\textbf {\bibinfo {volume}
			{68}},\ \bibinfo {pages} {1472} (\bibinfo {year} {1992})}\BibitemShut
	{NoStop}%
	\bibitem [{\citenamefont {Nesvizhevsky}\ \emph {et~al.}(2008)\citenamefont
		{Nesvizhevsky}, \citenamefont {Pignol},\ and\ \citenamefont
		{Protasov}}]{nesvizhevsky_neutron_2008}%
	\BibitemOpen
	\bibfield  {author} {\bibinfo {author} {\bibfnamefont {V.~V.}\ \bibnamefont
			{Nesvizhevsky}}, \bibinfo {author} {\bibfnamefont {G.}~\bibnamefont
			{Pignol}},\ and\ \bibinfo {author} {\bibfnamefont {K.~V.}\ \bibnamefont
			{Protasov}},\ }\bibfield  {title} {\bibinfo {title} {Neutron scattering and
			extra-short-range interactions},\ }\href
	{https://doi.org/10.1103/PhysRevD.77.034020} {\bibfield  {journal} {\bibinfo
			{journal} {Physical Review D}\ }\textbf {\bibinfo {volume} {77}},\ \bibinfo
		{pages} {034020} (\bibinfo {year} {2008})}\BibitemShut {NoStop}%
	\bibitem [{\citenamefont {Pokotilovski}(2006)}]{pokotilovski_constraints_2006}%
	\BibitemOpen
	\bibfield  {author} {\bibinfo {author} {\bibfnamefont {Y.~N.}\ \bibnamefont
			{Pokotilovski}},\ }\bibfield  {title} {\bibinfo {title} {Constraints on new
			interactions from neutron scattering experiments},\ }\href
	{https://doi.org/10.1134/S1063778806060020} {\bibfield  {journal} {\bibinfo
			{journal} {Physics of Atomic Nuclei}\ }\textbf {\bibinfo {volume} {69}},\
		\bibinfo {pages} {924} (\bibinfo {year} {2006})}\BibitemShut {NoStop}%
	\bibitem [{\citenamefont {Barbieri}\ and\ \citenamefont
		{Ericson}(1975)}]{barbieri_evidence_1975}%
	\BibitemOpen
	\bibfield  {author} {\bibinfo {author} {\bibfnamefont {R.}~\bibnamefont
			{Barbieri}}\ and\ \bibinfo {author} {\bibfnamefont {T.~E.~O.}\ \bibnamefont
			{Ericson}},\ }\bibfield  {title} {\bibinfo {title} {Evidence against the
			existence of a low mass scalar boson from neutron-nucleus scattering},\
	}\href {https://doi.org/10.1016/0370-2693(75)90073-8} {\bibfield  {journal}
		{\bibinfo  {journal} {Physics Letters B}\ }\textbf {\bibinfo {volume} {57}},\
		\bibinfo {pages} {270} (\bibinfo {year} {1975})}\BibitemShut {NoStop}%
	\bibitem [{\citenamefont {Delaunay}\ \emph
		{et~al.}(2017{\natexlab{b}})\citenamefont {Delaunay}, \citenamefont
		{Frugiuele}, \citenamefont {Fuchs},\ and\ \citenamefont
		{Soreq}}]{Delaunay:2017dku}%
	\BibitemOpen
	\bibfield  {author} {\bibinfo {author} {\bibfnamefont {C.}~\bibnamefont
			{Delaunay}}, \bibinfo {author} {\bibfnamefont {C.}~\bibnamefont {Frugiuele}},
		\bibinfo {author} {\bibfnamefont {E.}~\bibnamefont {Fuchs}},\ and\ \bibinfo
		{author} {\bibfnamefont {Y.}~\bibnamefont {Soreq}},\ }\bibfield  {title}
	{\bibinfo {title} {{Probing new spin-independent interactions through
				precision spectroscopy in atoms with few electrons}},\ }\href
	{https://doi.org/10.1103/PhysRevD.96.115002} {\bibfield  {journal} {\bibinfo
			{journal} {Phys. Rev. D}\ }\textbf {\bibinfo {volume} {96}},\ \bibinfo
		{pages} {115002} (\bibinfo {year} {2017}{\natexlab{b}})}\BibitemShut
	{NoStop}%
	\bibitem [{\citenamefont {Gebert}\ \emph {et~al.}(2015)\citenamefont {Gebert},
		\citenamefont {Wan}, \citenamefont {Wolf}, \citenamefont {Angstmann},
		\citenamefont {Berengut},\ and\ \citenamefont
		{Schmidt}}]{gebert_precision_2015}%
	\BibitemOpen
	\bibfield  {author} {\bibinfo {author} {\bibfnamefont {F.}~\bibnamefont
			{Gebert}}, \bibinfo {author} {\bibfnamefont {Y.}~\bibnamefont {Wan}},
		\bibinfo {author} {\bibfnamefont {F.}~\bibnamefont {Wolf}}, \bibinfo {author}
		{\bibfnamefont {C.~N.}\ \bibnamefont {Angstmann}}, \bibinfo {author}
		{\bibfnamefont {J.~C.}\ \bibnamefont {Berengut}},\ and\ \bibinfo {author}
		{\bibfnamefont {P.~O.}\ \bibnamefont {Schmidt}},\ }\bibfield  {title}
	{\bibinfo {title} {Precision {Isotope} {Shift} {Measurements} in {Calcium}
			{Ions} {Using} {Quantum} {Logic} {Detection} {Schemes}},\ }\href
	{https://doi.org/10.1103/PhysRevLett.115.053003} {\bibfield  {journal}
		{\bibinfo  {journal} {Physical Review Letters}\ }\textbf {\bibinfo {volume}
			{115}},\ \bibinfo {pages} {053003} (\bibinfo {year} {2015})}\BibitemShut
	{NoStop}%
	\bibitem [{\citenamefont {Raffelt}(2012)}]{raffelt_limits_2012}%
	\BibitemOpen
	\bibfield  {author} {\bibinfo {author} {\bibfnamefont {G.}~\bibnamefont
			{Raffelt}},\ }\bibfield  {title} {\bibinfo {title} {Limits on a
			{CP}-violating scalar axion-nucleon interaction},\ }\href
	{https://doi.org/10.1103/PhysRevD.86.015001} {\bibfield  {journal} {\bibinfo
			{journal} {Physical Review D}\ }\textbf {\bibinfo {volume} {86}},\ \bibinfo
		{pages} {015001} (\bibinfo {year} {2012})}\BibitemShut {NoStop}%
	\bibitem [{\citenamefont {Grifols}\ and\ \citenamefont
		{Mass\'{o}}(1986)}]{grifols_constraints_1986}%
	\BibitemOpen
	\bibfield  {author} {\bibinfo {author} {\bibfnamefont {J.~A.}\ \bibnamefont
			{Grifols}}\ and\ \bibinfo {author} {\bibfnamefont {E.}~\bibnamefont
			{Mass\'{o}}},\ }\bibfield  {title} {\bibinfo {title} {Constraints on
			finite-range baryonic and leptonic forces from stellar evolution},\ }\href
	{https://doi.org/10.1016/0370-2693(86)90509-5} {\bibfield  {journal}
		{\bibinfo  {journal} {Physics Letters B}\ }\textbf {\bibinfo {volume}
			{173}},\ \bibinfo {pages} {237} (\bibinfo {year} {1986})}\BibitemShut
	{NoStop}%
	\bibitem [{\citenamefont {Grifols}\ \emph {et~al.}(1989)\citenamefont
		{Grifols}, \citenamefont {Mass\'{o}},\ and\ \citenamefont
		{Peris}}]{grifols_constraint_1989}%
	\BibitemOpen
	\bibfield  {author} {\bibinfo {author} {\bibfnamefont {J.~A.}\ \bibnamefont
			{Grifols}}, \bibinfo {author} {\bibfnamefont {E.}~\bibnamefont {Mass\'{o}}},\
		and\ \bibinfo {author} {\bibfnamefont {S.}~\bibnamefont {Peris}},\ }\bibfield
	{title} {\bibinfo {title} {Constraint on the {Higgs}-boson mass from nuclear
			scattering data},\ }\href {https://doi.org/10.1103/PhysRevLett.63.1346}
	{\bibfield  {journal} {\bibinfo  {journal} {Physical Review Letters}\
		}\textbf {\bibinfo {volume} {63}},\ \bibinfo {pages} {1346} (\bibinfo {year}
		{1989})}\BibitemShut {NoStop}%
	\bibitem [{\citenamefont {{Michael Bordag}}\ \emph {et~al.}(2009)\citenamefont
		{{Michael Bordag}}, \citenamefont {{Galina Leonidovna Klimchitskaya}},
		\citenamefont {{Umar Mohideen}},\ and\ \citenamefont {{Vladimir Mikhaylovich
				Mostepanenko}}}]{michael_bordag_advances_2009}%
	\BibitemOpen
	\bibfield  {author} {\bibinfo {author} {\bibnamefont {{Michael Bordag}}},
		\bibinfo {author} {\bibnamefont {{Galina Leonidovna Klimchitskaya}}},
		\bibinfo {author} {\bibnamefont {{Umar Mohideen}}},\ and\ \bibinfo {author}
		{\bibnamefont {{Vladimir Mikhaylovich Mostepanenko}}},\ }\href
	{https://academic.oup.com/book/359} {\emph {\bibinfo {title} {Advances in the
				{Casimir} {Effect}}}}\ (\bibinfo  {publisher} {Oxford University Press},\
	\bibinfo {year} {2009})\BibitemShut {NoStop}%
\end{thebibliography}
\end{document}


\title{Supplemental Material: \\Nonlinear calcium King plot constrains new bosons and nuclear properties}

 \author{Alexander Wilzewski}
 \thanks{These authors contributed equally.}
 \author{Lukas J. Spie{\ss}}
 \author{Malte Wehrheim}
 \author{Shuying Chen}
 \author{Steven A. King}
 \author{Peter Micke}
 \author{Melina Filzinger}
 \author{Martin R. Steinel}
 \author{Nils Huntemann}
 \author{Erik Benkler}
 \author{Piet O. Schmidt$^\star$}
 \affiliation{Physikalisch-Technische Bundesanstalt, Bundesallee 100, 38116 Braunschweig, Germany\\
 $^\star$Also at Institut f\"{u}r Quantenoptik, Leibniz Universit\"{a}t Hannover, Welfengarten 1, 30167 Hannover, Germany}%
 
\author{Luca I. Huber}
\thanks{These authors contributed equally.}
\author{Jeremy Flannery}
\author{Roland Matt}
\thanks{Current address Oxford Ionics, Kidlington, OX5 1GN , UK}
\author{Martin Stadler}
\author{Robin Oswald}
\author{Fabian Schmid}
\author{Daniel Kienzler}
\author{Jonathan Home}
\author{Diana P. L. \surname{Aude Craik}}
\affiliation{Institute for Quantum Electronics, Department of Physics, Eidgenössische Technische Hochschule Z\"{u}rich, Otto-Stern-Weg 1, 8093 Zurich, Switzerland}%

\author{Menno Door}
\thanks{These authors contributed equally.}
\author{Sergey Eliseev}
\author{Pavel Filianin}
\author{Jost Herkenhoff}
\author{Kathrin Kromer}
\author{Klaus Blaum}
\affiliation{Max-Planck-Institut f\"{u}r Kernphysik, Saupfercheckweg 1, 69117 Heidelberg, Germany}

\author{Vladimir A. Yerokhin}
\author{Igor A. Valuev}
\author{Natalia S. Oreshkina}
\affiliation{Max-Planck-Institut f\"{u}r Kernphysik, Saupfercheckweg 1, 69117 Heidelberg, Germany}

\author{Chunhai Lyu}
\author{Sreya Banerjee}
\author{Christoph H. Keitel}
\author{Zolt\'{a}n Harman}
\affiliation{Max-Planck-Institut f\"{u}r Kernphysik, Saupfercheckweg 1, 69117 Heidelberg, Germany}

\author{Julian C. Berengut}%
\affiliation{School of Physics, University of New South Wales, Sydney, NSW 2052, Australia\\
and UNSW Nuclear Innovation Centre, UNSW Sydney, Kensington, NSW 2052, Australia}

\author{Anna Viatkina}%
\author{Jan Gilles}
\author{Andrey Surzhykov}
\affiliation{Physikalisch-Technische Bundesanstalt, Bundesallee 100, 38116 Braunschweig, Germany\\
and Institut f\"ur Mathematische Physik, Technische Universit\"at Braunschweig, D-38116 Braunschweig, Germany}

\author{Michael K. Rosner}
\author{Jos\'{e} R. {Crespo L\'{o}pez-Urrutia}}
\affiliation{Max-Planck-Institut f\"{u}r Kernphysik, Saupfercheckweg 1, 69117 Heidelberg, Germany}

\author{Jan Richter$^{\mathsection}$}
\thanks{These authors contributed equally.}
\author{Agnese Mariotti}
\thanks{These authors contributed equally.}
\author{Elina Fuchs$^\mathsection$}

\affiliation{Institut f\"{u}r Theoretische Physik, Leibniz Universit\"{a}t Hannover, Appelstra{\ss}e 2, 30167 Hannover, Germany\\
$^\mathsection$Also at Physikalisch-Technische Bundesanstalt, Bundesallee 100, 38116 Braunschweig, Germany}

\date{\today}%

\maketitle

\section{\label{sec:intro}Frequency measurements of highly charged calcium}
In the following, we provide details on the isotope shift measurements of highly charged \Ca{}{14+}.
First, we briefly introduce the experimental apparatus where we put an emphasis on the modifications compared to our previous work \cite{king_optical_2022} required for the calcium measurements.
We then summarize the absolute frequency measurement campaigns that provided the data for the isotope shifts.
Finally, we discuss systematic frequency shifts and evaluate their uncertainties.
\subsection{Overview of experimental apparatus}
    Highly charged calcium is produced in a compact electron-beam ion-trap (EBIT) \cite{micke_heidelberg_2018}. 
    In the electron beam of the EBIT, which has an energy of about \SI{1}{\kilo\electronvolt}, a megakelvin hot plasma of ions in different charge states forms.
    These temperatures do not allow for precision spectroscopy on the level of optical atomic clocks. 
    To cool the ions, we transfer them through an ion-optical beamline to a cryogenic linear Paul trap \cite{leopold_cryogenic_2019}. 
    The charge states separate in time-of-flight in the beamline and we select the one of interest before further slowing down and cooling in a pulsed drift tube and subsequent injection into the Paul trap \cite{schmoger_deceleration_2015}.
    There, a laser-cooled \Be{}{+}-ion crystal sympathetically cools down the HCIs to millikelvin temperatures \cite{schmoger_coulomb_2015}. 
    Thereafter, we eject all but one \Be{}{+} ion, so that a \Be{}{+}-\Ca{}{14+} two-ion crystal suitable for quantum-logic spectroscopy (QLS) is prepared \cite{micke_coherent_2020}. 
    We then use the coherent control over \Ca{}{14+} to stabilize the clock laser frequency to the optical transition in the HCI of interest and measure its absolute frequency through an optical clock comparison with the \Yb octupole clock at PTB \cite{king_optical_2022}.

\subsection{Experimental hardware changes and upgrades as compared to \Ar{}{13+} measurements}
    In the following we describe the changes and upgrades of the experimental apparatus for the \Ca{}{14+} frequency measurements as compared to our previous measurements using \Ar{}{13+} \cite{king_optical_2022}.
    A complete and detailed overview can be found in Ref.~\cite{wilzewski_isotope_2024}.
    \subsubsection{Laser ablation source for EBIT \label{sec:ebit_laser_ablation}}
        For injection of atoms into the EBIT from solid targets, we constructed a laser ablation source, based on the design described in Ref.~\cite{schweiger_production_2019}. 
        It can hold up to five different targets and switch between the different calcium isotopes by rotating a shaft with a rotation vacuum-feed-through. 
        Isotopes \Ca{40,42,44,48}{} were procured in the form of metallic flakes with an isotopic enrichment of over 90\% each. 
        A vacuum compatible glue (\textit{Vacseal II}, Space Environment Labs - Vacseal Inc.) fixes them on the tips of M1.6 grub screws, which are then mounted in five compatible M1.6 threads on the tip of the rotatable shaft.
        The rare isotope \Ca{46}{} was in carbonate form with maximal enrichment of 25\%.
        For fixing the powder-like carbonate, we first applied a thin film of glue on the screw tip and then pressed the carbonate powder into it.\\
        The targets are electrically insulated from the vacuum chamber.
        This allows to apply a bias voltage to them, which is necessary to not distort the electron beam of the EBIT when the target holder is brought close to it.\\ 
        A pulsed laser enters the vacuum chamber through a vacuum-viewport opposite to the targets and is focused to a beam diameter of $\approx \SI{30}{\micro\meter}$ at the target position by a $f=$\SI{250}{\milli\meter} lens that is positioned out-of-vacuum.
        
    \subsubsection{Clock laser system}
        A commercially available external-cavity diode laser (ECDL) system at a wavelength of \SI{1140}{\nano\meter} with an integrated frequency-doubling cavity produces \SI{500}{\milli\watt} of optical power at the desired \SI{570}{\nano\meter} wavelength. 
        For pre-stabilization of the laser frequency, we employed a cubic cavity \cite{webster_force-insensitive_2011}, with a geometry and mounting scheme optimized for minimal sensitivity to vibrations. 
        Our cavity design is similar to that published in \cite{dawel_coherent_2024} with the only difference that the cavity mirrors have a single-wavelength coating with highest reflectivity at \SI{1140}{\nano\meter} instead of a dual-wavelength coating. 
        The cavity spacer is made of ultra-low expansion (ULE) glass, with a measured zero-crossing of the coefficient of thermal expansion (CTE) at \SI{31.5}{\celsius}. 
        The \SI{50}{\milli\meter} cavity length results in a free spectral range (FSR) of \SI{3}{\giga\hertz}. 
        The cavity is placed in a vacuum chamber and two heat shields thermally isolate it resulting in a thermal time constant of about one day for the cavity. 
        Additionally, an active temperature stabilization with Peltier elements keeps the cavity temperature close to the zero-crossing point of the CTE.
        With this, we observe a frequency drift rate of about \SI{300}{\hertz\per\hour}.
        To control the frequency offset of the laser from the cavity, a broadband fiber electro-optical modulator (ixBlue, NIR-MPX-LN-02-00-P-P-FA-FA-LIL) generates sidebands onto the laser frequency and one of the sidebands is locked to the cavity with the Pound-Drever-Hall technique~\cite{leopold_tunable_2016}.
        For further stabilization, part of the fundamental light is sent to a frequency comb, which is locked to the ultra-stable Si2 cavity at PTB \cite{matei_15text_2017}.
        Phase-locking our clock laser to the frequency comb transfers the stability of Si2 to our clock laser. 
        Stabilized in this way, the laser has a \SI{}{\hertz}-level linewidth suitable for quantum logic spectroscopy (QLS).

     \subsubsection{Experimental sequence}
        The pre-stabilized clock laser is further stabilized to the \CaPtoP transition of a single \Ca{}{14+} ion~\cite{peik_laser_2005}.
        For this, the laser interrogates the ion several times to determine its frequency offset to the atomic transition.
        We call such an interrogation cycle a \textit{servo cycle}.
        A servo cycle comprises probing all three Zeeman components of the \CaPtoP transition with two counter-propagating lasers.
        This results in six independent \textit{Zeeman servos} that follow their respective transitions.
        Each Zeeman servo interrogates the two half-maxima of the atomic transition four times and from the difference of excitation probabilities on the half-maxima an error signal is derived.
        This error signal is then used by a digital feedback loop to steer the laser back on resonance with the ion~\cite{peik_laser_2005}.
        The sequence to interrogate the ion is structured as follows:
        First, the \Be{}{+}-\Ca{}{14+} ion crystal is Doppler cooled on the $^2\mathrm{S}_{1/2} \rightarrow ^2\mathrm{P}_{3/2}$ transition of \Be{}{+} at \SI{313}{\nano\meter}.
        Then the crystal is cooled to the ground state of both axial modes of motion employing Raman sideband cooling on the \Be{}{+} ion~\cite{monroe_resolved-sideband_1995}.
        This is followed by a sequence of algorithmic ground-state cooling~\cite{king_algorithmic_2021} to cool the two weakly-coupled radial modes, in which the \Ca{}{14+} (\Be{}{+}) ion has a large (small) amplitude of motion, close to their respective ground states.
        The \Ca{}{14+} ion is then interrogated by a laser pulse of \SI{15}{\milli\second} length for optimal stability~\cite{peik_laser_2005} and finally its internal state is read out using QLS~\cite{micke_coherent_2020}.
        One such interrogation takes \SI{56}{\milli\second}.
        In total, the \Ca{}{14+} is probed 48 times during a servo cycle; the six Zeeman servos probe the atomic transition each eight times.
        Therefore, one servo cycle takes \SI{2.7}{\second}.
        The average frequency of the six Zeeman servos is free of a linear Doppler shift, linear Zeeman shift and quadrupole shift, as discussed below. 
        The center frequency of the clock laser is steered to this average frequency using a second digital feedback loop~\cite{king_optical_2022} and part of the such stabilized laser is sent to an optical frequency comb for an optical clock comparison.
        More details on the clock laser stabilization scheme can be found in Ref.~\cite{wilzewski_isotope_2024}.
        
\subsection{Measurement campaigns}
    \begin{figure*}[t]
        \includegraphics[width=0.99\textwidth]{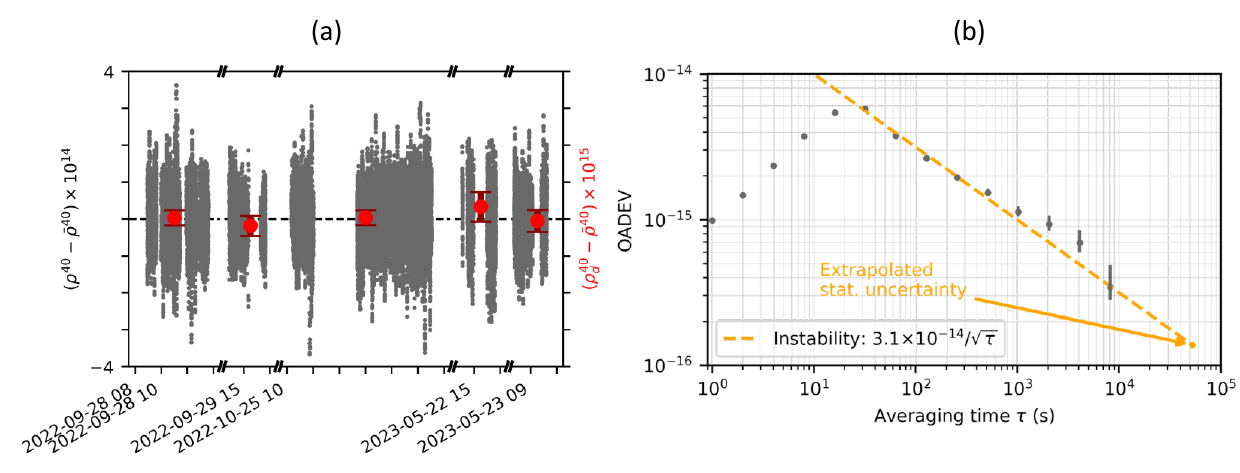}
        \caption{(a) The time trace shows $\approx \SI{50000}{\second}$ frequency ratio data points $\rho_{40}$ as their deviation from the mean ratio $\bar{\rho}_{40}$. Daily mean values $\rho_d$ are overlaid in red (note different axis scale). (b) The overlapping Allan deviation (OADEV) of the ratio data shown in (a) decreases with the characteristic $\tau^{-1/2}$ dependence for white frequency noise originating from quantum projection noise. The dashed line is a fit to the data (reduced $\chi^2=0.7$) for averaging times longer than \SI{30}{\second}, which is the time constant of the employed digital feedback loop. An extrapolation of the fit to the total measurement time yields the statistical uncertainty of the measurement. Figure adapted from Ref.~\cite{wilzewski_isotope_2024}.}
        \label{fig:40Ca_data}
    \end{figure*}
    We measured the frequency ratios by simultaneous operation of the HCI and \Yb\ optical clocks at PTB. 
    The ratio
    \begin{equation}
        \rho^A = \nu^A_{570}/\nu_{\text{Yb}}
    \end{equation}
    is directly measured on the same frequency comb, where $\nu_\text{Yb}$ is the frequency of the \Yb\ clock laser and $\nu^A_{570}$ is the frequency of the clock laser stabilized to the \CaPtoP transition in \Ca{A}{14+} with $A$ labeling one of the five stable and even calcium isotopes.
    Note that we correct for the systematic frequency shifts (discussed in Sec.~\ref{sec:Ca14+_systematics}) from the frequency ratio measured on the frequency comb.
    The absolute frequency of the \Yb\ octupole clock transition $\bar{\nu}_{\text{Yb}} = \SI{642121496772645.10(8)}{\hertz}$ was previously measured by comparison with the caesium primary standard at PTB with a fractional uncertainty of \SI{1.3e-16}{} \cite{lange_improved_2021}.
    Thus, the frequency of the transition in \Ca{A}{14+} follows from the ratio measurement:
    \begin{equation}
        \bar{\nu}^A =  \bar{\rho}^A \cdot \bar{\nu}_{\text{Yb}},
    \end{equation}
    where $\bar{\rho}^A$ is the mean of the acquired frequency ratio data.
    Table~\ref{tab:Ca_frequencies} shows the results for these measurements.
    We took $t_\text{meas} > \SI{30000}{\second}$ of data for each isotope over the course of eight months starting with the \Ca{40}{14+} measurement.
    After the last measurement (\Ca{46}{}), another short (\SI{10000}{\second}) \Ca{40}{} run agreed with the first \Ca{40}{} data taken eight months prior.
    As an example, Fig.~\ref{fig:40Ca_data}(a) shows the time trace of the measured frequency ratio $\rho^{40}$.
    The daily mean values $\rho_d$, shown in red in Fig.~\ref{fig:40Ca_data}(a), all agree within their uncertainties.
    Figure~\ref{fig:40Ca_data}(b) shows the corresponding overlapping Allan deviation (OADEV) calculated for the time trace.
    The instability decreases for averaging times $\tau > 30s$ which is the servo time constant of the used digital feedback loop \cite{king_optical_2022, peik_laser_2005}. 
    The orange dashed line is a fit to the OADEV data for $\tau > \SI{30}{\second}$ with a $\tau^{-1/2}$ scaling characteristic for white frequency noise as expected for quantum projection noise.
    We obtain the overall statistical uncertainties of the ratio measurements (shown in parentheses in Table~\ref{tab:Ca_frequencies}) by ex\-tra\-po\-lat\-ing the fitted instability to the total measurement length (see Fig.~\ref{fig:40Ca_data}(b)).   
    Note that the uncertainties of the absolute frequencies for \Ca{}{14+} include the uncertainty of the \Yb\ frequency and they are of comparable magnitude.
    However, for the determination of the isotope shifts, the uncertainty in the \Yb\ reference frequency is common-mode suppressed with a reproducibility $\ll 10^{-17}$ \cite{sanner_optical_2019} much smaller than the relative uncertainty of the ratio difference $\rho^{A} - \rho^{A^\prime}$.
    The \Ca{}{14+} systematic uncertainty discussed in the following section does not affect the overall uncertainty, which is dominated by the statistical uncertainty.
    {\renewcommand{\arraystretch}{1.4}
    \begin{table*}
    \caption{\label{tab:Ca_frequencies} Mean frequency ratios $\bar{\rho}^A$ and calculated absolute frequencies $\bar{\nu}^A$ with 1-$\sigma$ uncertainties in parentheses (systematic and statistical uncertainty added in quadrature). The fourth column lists the length of the frequency data time series in seconds $t_\text{meas}$ and the dates when the measurements were performed.}
    \begin{ruledtabular}
    \begin{tabular}{crrr}     
     $A$ & $\bar{\rho}^A$ & $\bar{\nu}^A$ in \SI{}{\hertz}   &  $t_\text{meas}$ and date   \\ \hline
     
     40 & 0.819 646 249 827 850 37(11) & 526 312 476 763 544.68(11)& \SI{50000}{\second} in Sep22, Oct22, May23 \\
     
     42 & 0.819 647 089 370 559 88(12) & 526 313 015 851 965.92(11) & \SI{40000}{\second} in Jan23 \\
     
     44 & 0.819 647 854 582 926 82(10) & 526 313 507 211 276.32(09) &  \SI{70000}{\second} in Jan23, Feb23 \\
     
     46 & 0.819 648 556 456 664 67(17)& 526 313 957 899 491.42(13) &\SI{30000}{\second} in May23\\
     
     48 & 0.819 649 199 888 398 19(17) & 526 314 371 060 839.22(113)  & \SI{30000}{\second} in Feb23 \\
    \end{tabular}
    \end{ruledtabular}
    \end{table*}
    } 
    
\subsection{Systematic shifts and their uncertainties \label{sec:Ca14+_systematics}}
    The statistical uncertainty of the fractional frequency measurements was larger than \SI{1e-16}{}.
    Therefore, any shift and its uncertainty that is at the \SI{1e-18}{} level or below is negligible, which is the case for most of the shifts discussed in the following.\\ 
    Because of the similarity of the \CaPtoP transition in \Ca{}{14+} to the \ArPtoP transition in \Ar{}{13+} as well as a similar charge state and atomic masses, many of the experimental parameters (Paul trap voltages and RF drive power, laser powers, laser pulse lengths, quantization magnetic field, etc.) are similar to those published in \cite{king_optical_2022}.
    The methods for evaluation and mitigation of the systematic effects did not change conceptually but have a slight mass dependence. 
    Tab.~\ref{tab:sys_uncertainty} summarizes the leading systematic shifts found for the \Ca{42}{} isotope. 
    We chose to show \Ca{42}{} because it has the largest systematic uncertainty originating from excess micromotion. 
    The uncertainty budgets of all other isotopes are similar, but yield smaller overall systematic uncertainties.\\
    The second-order Doppler shift caused by excess micromotion is the largest systematic shift and uncertainty. 
    It is dominated by uncompensatable axial micromotion due to a defect of the employed trap.
    Therefore, we automatically and periodically halt the experiment during clock operation and perform a measurement of the excess micromotion level in three orthogonal directions using the sideband ratio method on the \Ca{}{14+} clock transition \cite{keller_precise_2015,berkeland_minimization_1998}.
    If the level of observed micromotion started to increase, we manually compensated micromotion before resuming the frequency measurement.
    The largest uncertainty in this shift is estimated for \Ca{42}{}, see Tab.~\ref{tab:sys_uncertainty}, which stems from less frequent micromotion compensation.\\
    The overall second-order Doppler shift caused by the secular motion  is $< \SI{2e-18}{}$, as shown in Tab.~\ref{tab:sys_uncertainty}. 
    This is dominated by the phonon number in two weakly-coupled radial modes, which are cooled using algorithmic cooling \cite{king_algorithmic_2021}.
    Their temperature is continuously measured during clock operation and the average phonon number was always below two phonons.\\
    To detect a potential first-order Doppler shift the ion is probed with two counter-propagating beams. We note that in the combined data of all isotope measurements (\SI{270000}{\second}), we find a frequency difference of the two beams of \SI{-55(30)}{\milli\hertz}. 
    By averaging the frequency measured along the two counter-propagating beams, this shift is suppressed below the measured value and negligible.\\
    By averaging over all three Zeeman components the linear Zeeman and the electric quadrupole shift of the excited state is suppressed below \SI{1e-18}{}.\\
    Averaging over the Zeeman components also suppresses shifts originating from the tensor polarizability of the $^3\text{P}_1$-state.\\
    The scalar polarizability of the clock transition is expected to be small due to the absence of low-lying electric-dipole allowed transitions connecting to the clock states.
    However, in contrast to \Ar{}{13+} \cite{yu_investigating_2019}, which has no transition connecting to the clock state with wavelength shorter than \SI{50}{\nano\meter}, \Ca{}{14+} features transitions at longer wavelengths.
    This potentially leads to a larger polarizability than that of \Ar{}{13+}, for which the static scalar polarizability was estimated to be \SI{-3e-45}{\joule\meter^2\per\volt^2} \cite{yu_investigating_2019}. 
    Most prominently, the transitions within the ground-state fine-structure manifold, $^3P_1 \rightarrow ^3P_2$ and $^3P_0 \rightarrow ^3P_2$ are at wavelength of \SI{544}{\nano\meter} and \SI{278}{\nano\meter} \cite{kramida_nist_nodate}, respectively.
    These transitions are not E1-allowed and therefore feature small oscillator strengths (converting the decay rates of Ref.~\cite{kramida_nist_nodate} into an oscillator strength yields $f\approx 10^{-6}$).
    All other transitions connecting to the clock states are below \SI{110}{\nano\meter} with the lowest-lying E1-allowed transition at \SI{20}{\nano\meter} \cite{jonsson_energies_2011}.
    Only considering the two optical transitions, $^3P_1 \rightarrow ^3P_2$ and $^3P_0 \rightarrow ^3P_2$, the oscillator strength converts to a differential static polarizability of $\approx \SI{-5e-42}{\joule\meter^2\per\volt^2}$.
    From the excess micromotion measurements, we find that the residual electric field caused by the trap drive is $E \approx \SI{100}{\volt\per\meter}$. 
    This would lead to a negligible shift of $\delta\nu/\nu \approx \SI{1e-18}{}$. %
    Furthermore, electric fields originating from black-body radiation are strongly suppressed in the cryogenic environment, rendering this shift negligible.\\
    The second-order Zeeman shift was recently estimated from ab-initio atomic structure calculations.
    Calculations are confirmed by measurements of the second-order Zeeman shifts of the $m_J = 0 \rightarrow m_{J} = 0$ transition \cite{gilles_quadratic_2024} leading to a shift of $\approx \SI{200}{\micro\hertz}$ in our DC magnetic field of \SI{23}{\micro\tesla}, which is negligible. 
    In Ref.~\cite{spies_excited-state_2024} the second-order Zeeman shift induced by an AC magnetic field from currents on the electrodes of the Paul trap was bounded to less than \SI{1}{\micro\tesla} corresponding to less than \SI{100}{\nano\hertz} for our typical RF powers.\\
    The probe laser induces a shift by off-resonantly coupling different Zeeman components through an AC Zeeman effect, depending on the clock laser polarization and power. 
    We estimate the shift the same way as for \Ar{}{13+} by measuring the transition frequency for various laser powers.
    The frequency shift is then extrapolated to the laser power used during clock operation for one or both beams before each measurement.
    We found a relative shift of $ \lesssim \SI{2(4)e-18}{}$ in the worst case.
    As the shift is polarization-dependent, we adjusted the polarization of the lasers for equal coupling strengths before the first measurement run (\Ca{40}{14+}), which suppresses this shift to a negligible level when averaging over all Zeeman components.
    Note that with this method we also account for any shifts caused by the clock laser coupling the upper clock state $^3P_1$ to the $^3P_2$ state whose transition wavelength at \SI{544}{\nano\meter} is close to the \SI{570}{\nano\meter} clock transition wavelength.\\
    {\renewcommand{\arraystretch}{1.4}
    \begin{table}[h]
    \caption{\label{tab:sys_uncertainty}%
    Summary of relative systematic shift $\delta\nu/\nu$ with uncertainties $\sigma_{\delta \nu}/\nu$ of \Ca{42}{14+} frequency measurement in units \SI{1e-18}{}.}\begin{ruledtabular}
    \begin{tabular}{cccc}
    \multicolumn{1}{c}{Type of shift} & \multicolumn{1}{c}{$\delta \nu/\nu$} & \multicolumn{1}{c}{$\sigma_{\delta \nu}/\nu$} \\
    \hline
    \hline
    Excess micromotion & -560  & 53 \\
    
    Secular motion & -1.5 & 0.9 \\
    
    First-order Doppler & 0 & $<$ 1 \\
   
    Electric quadrupole & 0 & $<$ 1 \\
    
    First-order Zeeman & 0 & $<$1 \\
    
    Second-order Zeeman & $<$1 & $<$1 \\
    
    Black-body radiation & $\ll 1$ & $\ll 1$ \\

    Trap-drive-induced AC Stark & $< 1$ & $< 1$ \\

    Laser-induced AC Zeeman & 2 & 3.5 \\

    Gravitational red-shift & -163.3 & 0.8 \\
    \hline
    Total & -723 & 53 \\
    
    \end{tabular}
    \end{ruledtabular}
    \end{table}
    Since the \Ca{}{14+} ion is at the same height in the gravitational field as the \Ar{}{13+} ion, the height difference to the \Yb\ clock did not change. 
    After the \Ar{}{13+} measurement campaign, the height of the HCI clock was determined more precisely.
    The updated value of \SI{11.136(5)}{\meter} \cite{denker_ergebnisse_2022} leads to a fractional gravitational redshift of \SI{1.633(8)e-16}{}, consistent with the previous measurement.\\
    All other frequency shifts (fibre noise, accousto-optical phase chirp, line-pulling effects, servo error, background gas collisions) discussed in Ref.~\cite{king_optical_2022} were checked for and are negligible for the same reasons as discussed therein.\\

\section{Improved isotope shift measurements of \CaStoD{5/2} transition in \Ca{}{+}}\label{sec:Ca+ systematics}
To measure the frequency shift of all stable and even isotopes of \Ca{}{+}, we co-trap a single pair of two different isotopes and perform correlation spectroscopy \cite{roos_precision_2005, manovitz_precision_2019, chwalla_precision_2007}. The frequency interrogated is the differential isotope shift of the narrow-linewidth $S_{1/2}\rightarrow D_{5/2}$ electric-quadrupole transition at \SI{729}{\nano\meter}. We first summarize the experimental setup and then describe the procedure, which is a Ramsey interferometry experiment with two ions. Finally, we discuss the systematic shifts and uncertainties. A more in-depth discussion of the various systematic effects is in preparation and will be part of a future publication.

\subsection{Experimental setup}
All measurements are performed in a micro-fabricated segmented linear Paul trap composed of stacked silica glass wafers in a cryogenic system at a temperature of 6~K \cite{matt_roland_parallel_2023}. We load ion pairs consisting of one of each isotope, which are aligned along the axis of the trap, in which the primary confinement is provided by the static potential. Perpendicular to this direction, the confinement is provided by a radio-frequency pseudopotential. The ions are addressed by a beam propagating at \SI{90}{\degree} to the trap axis at the transition wavelength of \SI{729}{\nano\meter}. Transitions in both isotopes are driven simultaneously by the same \SI{729}{\nano\meter} laser field, which has frequency tones imposed at the resonant frequencies of both ions simultaneously, produced by phase modulation using a broadband electro-optical modulator (EOM). The EOM is driven with two microwave frequencies, one at $f_\textrm{offset}$ to address the lower frequency species, the other one at $f_\textrm{offset} + f_\textrm{synth}$. $f_\textrm{offset}$ was chosen to be 12~MHz. $f_\textrm{synth}$ is produced by a microwave (MW) synthesizer with low-phase noise (Anritsu MG3692C), which is locked to a 10 MHz Rubidium clock (SRS FS725) disciplined by a GPS receiver (HP 58503B).
Since each tone follows the same path, any fluctuations of the optical path are common-mode and do not introduce systematic shifts in the measured isotope shift.
All lasers are pulsed by switching acousto-optical modulators (AOM) on and off. 
Loading of different isotopes is achieved by photo-ionizing neutral calcium from an evaporative oven with natural abundance of calcium isotopes \cite{lucas_isotope-selective_2004}. The frequencies of the ionization, cooling and repump lasers are set according to the chosen isotope, leading to isotope-selective loading and cooling. In this way we are able to load all required calcium isotopes including \Ca{46}{+}, which has a natural abundance of only 0.004\%. Occasionally, sympathetic cooling leads to the co-trapping of undesired isotopes, which we remove by parametric excitation\cite{schmidt_mass-selective_2020}.

\subsection{Experimental sequence and calibrations}
The basic pulse sequence of a single experimental cycle includes Doppler cooling, state preparation into the desired sublevel of the $S_{1/2}$ manifold, the Ramsey interferometry sequence described in Sec.~\ref{sec:Ca+_correlation_spec} and quantum state detection by state-dependent scattering. Scattered photons from each ion are spatially resolved on an EMCCD camera.
The experiment is interleaved with calibrations of the frequency of the detection and cooling lasers at \SI{397}{\nano\meter}, calibrations of the microwave power sent to the EOM, Rabi-frequency calibrations and micromotion compensations by parametric excitation. We take care to only calibrate micromotion after completing a measurement of the phase for all 5 Ramsey waiting times in both configurations to correctly cancel spatially-dependent systematic effects (see Sec.~\ref{sec:Ca+ systematics}). The calibration of the \SI{397}{\nano\meter} lasers and the micromotion calibrations are performed once every hour, with micromotion calibration being delayed as long as the full measurement in the two configurations is not completed. Therefore, the time between micromotion calibration procedures typically varies between 60 minutes and 90 minutes, in exceptional cases the time between calibrations has been 4 hours. Calibrations of the microwave power to the EOM and the Rabi frequencies are triggered whenever anomalies in the fitted two-ion Ramsey signal are detected, such as the contrast of the signal dropping below 0.2 or the signal offset leaving the range between 0.4 and 0.6.
Throughout all Ramsey experiments and calibrations we check for lasers unlocking and unintentional swapping of the ion crystal  by adding a photon detection during the cooling procedure and only turning on the cooling laser which cools the heavier ion during this detection window. In case any laser unlocks, the experiment is paused until the laser is locked again at the correct frequency and any corrupted data is retaken. Similarly, if the ion crystal is detected to be in the wrong configuration, the experiment is paused and the ions are swapped by applying a suitable sequence of electrode voltages~\cite{splatt_deterministic_2009}. All these procedures are fully automated.

\subsection{Method}
\subsubsection{Correlation Spectroscopy \label{sec:Ca+_correlation_spec}}
In the following, we denote the selected Zeeman state in the $S_{1/2}$-manifold with $g$ and the selected state in the $D_{5/2}$-manifold with $e$. Indices 1 and 2 refer to the two different isotopes, one of them being \Ca{40}{+} and the other one being \Ca{42}{+}, \Ca{44}{+}, \Ca{46}{+} or \Ca{48}{+}. We denote the four possible Bell states as:  $|\Phi^\pm\rangle = \frac{1}{\sqrt{2}} (|g_1 g_2 \rangle \pm |e_1 e_2 \rangle)$ and $|\Psi^\pm\rangle = \frac{1}{\sqrt{2}} (|g_1 e_2 \rangle \pm |e_1 g_2 \rangle)$.\\
We first initialize both ions in the ground state and after applying $R_Y(\frac{\pi}{2})$ to both ions we obtain the state:

\begin{align}
    |\psi(t=0) \rangle &= \frac{1}{\sqrt{2}} ( |g_1\rangle + |e_1 \rangle) \otimes \frac{1}{\sqrt{2}} ( |g_2\rangle + |e_2 \rangle) 
    \\&= \frac{1}{2}  (|g_1 g_2 \rangle + |g_1e_2 \rangle + |e_1 g_2\rangle + |e_1 e_2\rangle ).
    \\&= \frac{1}{\sqrt{2}}  (|\Phi^+\rangle + |\Psi^+\rangle ).
\end{align}
After a waiting time $\tau$, which is assumed to be much longer than the single ion coherence time $\tau_c$, $|\Phi^+ \rangle$ has decohered to a mixed state of equal parts of $|\Phi^+ \rangle$ and $|\Phi^- \rangle$. In contrast, $|\Psi^+ \rangle$ is in a decoherence-free subspace (DFS) and evolves as:
\begin{align}
    |\psi (\tau) \rangle &=  \frac{1}{\sqrt{2}}(|g_1 e_2 \rangle + e^{i \omega_\textrm{IS} \tau}|e_1 g_2 \rangle)\\
    &= \cos\left(\frac{1}{2}\omega_\textrm{IS} \tau \right) |\Psi^+\rangle + i \sin\left(\frac{1}{2} \omega_\textrm{IS} \tau \right) |\Psi^- \rangle
\end{align}
i.e., the relative populations of these two orthogonal Bell states oscillate at the isotope shift frequency $\omega_\textrm{IS}$.\\
To measure in this basis after a Ramsey waiting time $\tau$, we apply another $R_Y(\frac{\pi}{2})$ rotation to both ions. This maps (up to a global phase) the symmetric odd parity Bell state to the even parity Bell state, while the anti-symmetric Bell state is invariant under global rotations:
\begin{align}
R_Y\left(\frac{\pi}{2}\right)|\Psi^+\rangle&= -|\Phi^-\rangle\\
R_Y\left(\frac{\pi}{2}\right)|\Psi^-\rangle&=  |\Psi^-\rangle
\end{align}
Finally, the even parity population $P_\textrm{even}$ is measured. From the even parity signal $P_\textrm{even}= \frac{1}{2} (1 - \cos(\omega_\textrm{IS}\tau))$ the isotope shift frequency can be extracted.\\
The rotation $R_Y(\frac{\pi}{2})$ transforms $|\Phi^+ \rangle$ and $|\Phi^- \rangle$ into $|\Phi^+\rangle$ and $|\Psi^+\rangle$, respectively. Hence, the parity of the non-DFS part of the signal averages to 0.5.

\subsubsection{Two-tone addressing}
The expression $P_\textrm{even}= \frac{1}{2} (1 - \cos(\omega_\textrm{IS}\tau))$ is strictly correct only if both ions have the same reference laser frequency. As described earlier, we address the two ions in practice with different tones: $\omega_L$ and $\omega_L + \omega_\textrm{synth}$, where $\omega_L$ is the laser frequency of the first tone and $\omega_\textrm{synth}$ is the frequency of the MW synthesizer. Importantly, the MW synthesizer has very low phase noise, such that $\omega_\textrm{synth}$ is highly stable and introduces minimal statistical noise. In this case the parity signal becomes:
\begin{equation}
    P_\textrm{even}= \frac{1}{2} (1 - \cos((\omega_\textrm{IS}-\omega_\textrm{synth})\tau))
\end{equation}
In order to extract $\omega_\textrm{IS}$, we can, in principle, only scan the Ramsey waiting time $\tau$. The disadvantage of this approach is that the correct fitting of $\omega_\textrm{IS}$ then depends on having a correct physical model which includes pulse fidelity and dark state decay. A more robust approach is to scan the MW synthesizer frequency between $\omega_\textrm{synth}$ and $\omega_\textrm{synth} + 2\pi/\tau$, effectively scanning the relative phase of the second Ramsey pulse between 0 and $2 \pi$. While experimental imperfections will lead to reduced contrast and some offset, the fitted phase $\phi_0 = (\omega_\textrm{IS} - \omega_\textrm{synth})\tau$ remains unaffected. Extraction of $\omega_\textrm{IS}$ then requires at least two such phase scans at different Ramsey waiting times $\tau$. For higher robustness, we perform phase scans at Ramsey waiting times of 30~ms, 50~ms, 100~ms, 300~ms and 325~ms. Fig. \ref{fig:Ca_plus_phase_fit} shows an example of such a Ramsey phase scan and Fig. \ref{fig:Ca_plus_frequency_fit} shows an example of the phase evolution as a function of the waiting time and the extracted frequency deviation from the synthesizer frequency.\\
\begin{figure}
    \includegraphics[width=0.49\textwidth]{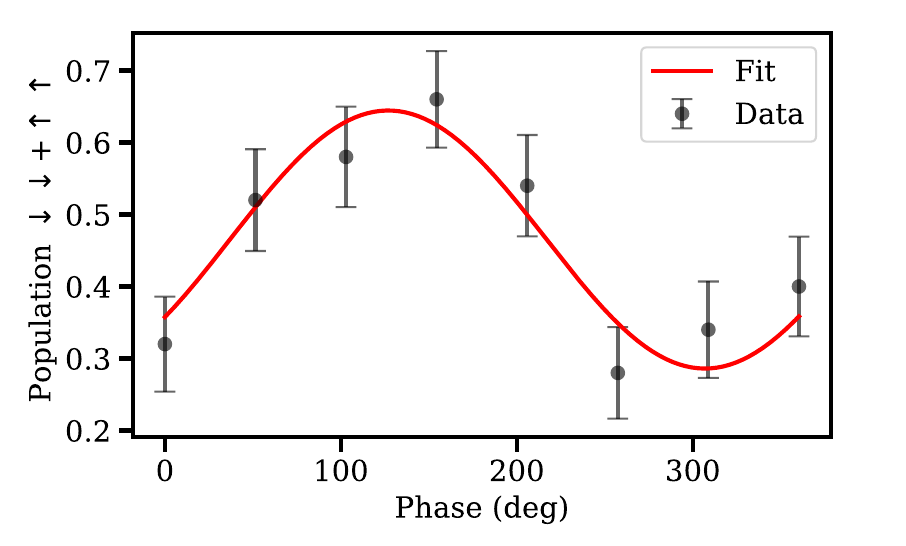}
    \caption{Example of a phase scan of a \Ca{40}{+}-\Ca{48}{+} two-ion crystal at a Ramsey waiting time of 325 ms measured on the $|S_{1/2}, m_J=+1/2\rangle \rightarrow |D_{5/2}, m_J=+1/2\rangle$ transition.}
    \label{fig:Ca_plus_phase_fit}
\end{figure}
\begin{figure}
    \includegraphics[width=0.49\textwidth]{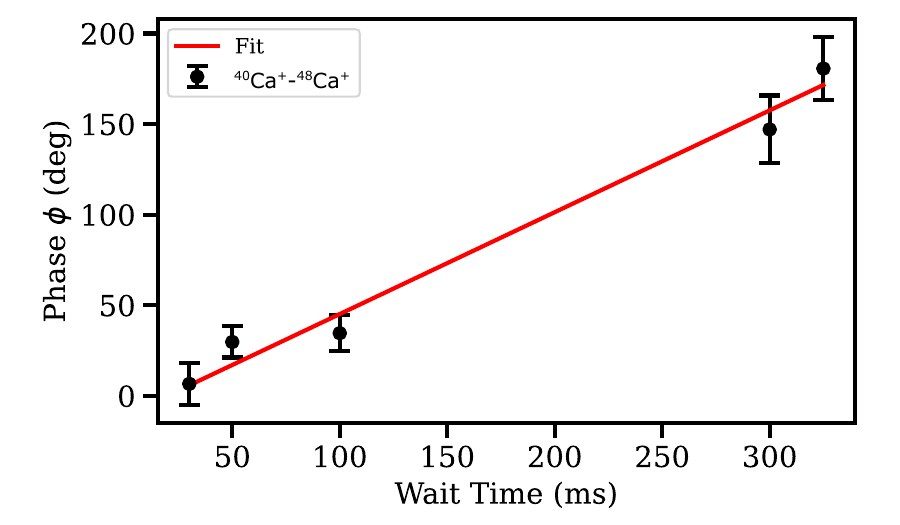}
    \caption{Example of a phase fit of a \Ca{40}{+}-\Ca{48}{+} two-ion crystal for the $|S_{1/2}, m_J=-1/2\rangle \rightarrow |D_{5/2}, m_J=-1/2\rangle$ transition. From the slope of the fit the frequency difference of the two isotopes can be extracted.}
    \label{fig:Ca_plus_frequency_fit}
\end{figure}
In principle, remaining single ion coherence could lead to an additional signal which could alter the fitted $\phi_0$. In our system, the coherence time $\tau_c \approx 70$~ms of the chosen transitions is indeed longer than the shortest Ramsey waiting time. Such remaining coherence can be taken into account in the fitting process. However, we saw minimal change in the final extracted $\omega_\textrm{IS}$; the shift is about an order of magnitude smaller than the statistical error of our experiment. This can be explained by the contribution to the signal from the $|\Phi^+\rangle$ component drifting over time. 
By averaging over many scans, which are spaced over hours, this additional effect is averaged away, same as the shot-to-shot averaging observed for waiting times $\gg \tau_c$.

\subsubsection{Averaging over transitions and ion crystal configurations}
To eliminate the influence of the differential first-order Zeeman shift on the isotope shift measurement, we average over two transitions with opposite magnetic sensitivities: $|S_{1/2}, m_J=-1/2\rangle \rightarrow |D_{5/2}, m_J=-1/2\rangle$~ and $|S_{1/2}, m_J=+1/2\rangle \rightarrow |D_{5/2}, m_J=+1/2\rangle$~. Additionally, we perform all measurements in both ion crystal configurations, by swapping the positions of the two ions. While systematic shifts are expected to be the same for the two ion locations, averaging over both configurations reduces the susceptibility to spatially dependent systematic shifts from the inter-ion distance (around $4.5$~$\mu$m) to tens of nanometers. The remaining position difference originates from imperfect alignment of the radio-frequency pseudopotential axis and the static potential axis. Since the radio-frequency pseudopotential is mass dependent, also the trap potential minimum exhibits mass-dependence. We do not find the isotope shift measurement to depend in a statistically significant way on the configuration, and therefore we expect remaining position-dependence for the configuration-averaged result to be negligible.

\subsection{Systematic shifts and uncertainties}
Tab.~\ref{tab:sys_uncertainty_ETH} shows the estimated systematic shifts and uncertainties. We show the isotope pairs with the smallest (\Ca{40}{+}-\Ca{42}{+}) and the largest mass-difference (\Ca{40}{+}-\Ca{48}{+}). The systematic errors of the other isotope pairs are of similar size.
{\renewcommand{\arraystretch}{1.4}
\begin{table*}
    \caption{\label{tab:sys_uncertainty_ETH}%
    Summary of relative systematic shift $\delta\nu$ with uncertainties $\sigma_{\delta \nu}$ for the isotope shift measurement of \Ca{40}{+}-\Ca{42}{+} and of \Ca{40}{+}-\Ca{48}{+} in units of \SI{}{mHz}.}\begin{ruledtabular}
    \begin{tabular}{cccccc}
    \multicolumn{1}{c}{Isotope pair} & \multicolumn{2}{c}{ \Ca{40}{+}-\Ca{42}{+}} & \multicolumn{2}{c}{ \Ca{40}{+}-\Ca{48}{+}} \\
    \multicolumn{1}{c}{Type of shift} & \multicolumn{1}{c}{$\delta \nu_{40-42}$} & \multicolumn{1}{c}{$\sigma_{\delta \nu_{40-42}}$} & \multicolumn{1}{c}{$\delta \nu_{40-48}$} & \multicolumn{1}{c}{$\sigma_{\delta \nu_{40-48}}$} \\
    \hline
    \hline
    Clock uncertainty & 0  & 11.7  & 0  & 39.7\\
    
    Magnetic field gradient fluctuations & 0 & 5.0 & 0 & 5.5 \\
    
    AC Stark shift during Ramsey pulses & 0 & 4.5  & 0 & 4.5\\
   
    Excess micromotion & 7.3 & 6.1 & 24.1 & 29.9  \\
    
    Intrinsic micromotion & -1.6 & 1.4 & 0.3 & 3.3 \\
    
    AC Stark shift due to light-leakage & $<1$ & $<1$  & $<1$ & $<1$\\

    Magnetic field drift & $\ll 1$  & $\ll 1$  & $\ll 1$  & $\ll 1$ \\

    Electric quadrupole shift & $\ll 1$  & $\ll 1$  & $\ll 1$  & $\ll 1$  \\
    
    Second-order Zeeman & $\ll 1$  & $\ll 1$ & $\ll 1$  & $\ll 1$ \\
    
    Black-body radiation & $\ll 1$ & $\ll 1$ & $\ll 1$ & $\ll 1$ \\

    \hline
    Total & 5.7 & 14.9 & 24.4 & 50.3 \\
    
    \end{tabular}
    \end{ruledtabular}
    \end{table*}
}

The systematic uncertainty is dominated by the reference clock uncertainty and excess micromotion\cite{berkeland_minimization_1998}. To a smaller extent intrinsic micromotion, a possible drift of the AC Stark shift during the Ramsey pulses and a possible drift of the magnetic field gradient add to the overall systematic uncertainty.\\
In order to estimate the error due to uncertainty of the reference clock, we fit a frequency noise spectrum which reproduces the Allan deviation provided by the manufacturer of the GPS receiver\cite{noauthor_58503b_2003} and create a time simulation of the frequency deviation. Monte Carlo trajectory simulations were seen to produce a standard deviation between \SI{11.7}{\milli\hertz} (\Ca{40}{+}-\Ca{42}{+} isotope shift measurement) and \SI{39.7}{\milli\hertz} (\Ca{40}{+}-\Ca{48}{+} isotope shift measurement). We do not take into account the frequency error of the rubidium clock since this would require accurate modelling of the locking dynamics of the rubidium clock. This does not create an unaccounted systematic error: on the time scales relevant to our experiment (from a few minutes up to a day) the rubidium clock promises to provide a lower Allan deviation than the GPS receiver such that we consider the error of the GPS receiver to be an upper bound to the clock error.\\
The systematic uncertainty due to excess micromotion primarily originates from trap charging by one of the cooling and detection beams at a wavelength of \SI{397}{\nano\meter}. The primary difference is between times when sequences are running which have a high percentage of these lasers being on (calibrations) and those which have a low percentage (Ramsey experiments). We refer to these as high and low duty cycle operation modes. By switching the duty cycle between high and low and operating at the same duty cycle for one hour at a time multiple times, the characteristic time of this charging and discharging was measured to be $15$~min. Since various calibrations and Ramsey experiments have different \SI{397}{\nano\meter} laser duty cycles, we bound the error by measuring the difference in stray fields for a maximum \SI{397}{\nano\meter} duty cycle and a minimum \SI{397}{\nano\meter} duty cycle. This charging creates an additional static electric field of about \SI{14}{\volt\per\meter}.  This is an upper bound since calibrations typically do not last for longer than 15 minutes.
In addition to trap charging, micromotion compensation is also limited by the precision with which static electric stray fields can be determined, typically around \SI{0.7}{\volt\per\meter}. Furthermore, we notice drifts of the micromotion compensation values from one calibration to the next. These drifts are typically on the order of a few \SI{}{\volt\per\meter} and reach \SI{11}{\volt\per\meter} in the worst case. To bound the error due to such a drift, we assume that the whole measurement was performed at a miscalibration of the size of the drift and calculate the corresponding systematic shift. Finally, we infer the tilt of the ion crystal compared to the radio-frequency pseudopotential axis by parametrically exciting the two-isotope crystal in both configurations. We parametrically excite the in-phase radial motional modes which primarily couple to the lighter ion. In the presence of a tilt, the micromotion compensation field therefore differs based on the location of the location of the lighter ion. Conversely, the difference in the micromotion compensation values allow us to determine the amount of crystal tilt. We don't compensate such a tilt but instead determine the maximum systematic shift it could produce.\\
While averaging over transitions with opposite magnetic field sensitivities cancels contributions from the first-order Zeeman shift, the measurement is still susceptible to drifts in the magnetic field or its gradient if such drifts happen on a time scale comparable to the switching between different transitions. By measuring the transition shift for a pair of \Ca{40}{+} ions we characterize the drift of the magnetic field gradient by continuously performing correlation spectroscopy using a magnetic field sensitive transition over more than 10 hours. We do not observe any evidence of magnetic field gradient fluctuations beyond the level of statistical fluctuations. Given the observed fluctuations, which include statistical fluctuations, we obtain a mean power spectral density for the magnetic field difference between the two ion locations of \SI{1.7e-20}{\tesla\squared\per\hertz} in the frequency band between \SI{24}{\micro\hertz} and \SI{1.3}{\milli\hertz}, the maximum is \SI{7.8e-20}{\tesla\squared\per\hertz}. We estimate a possible systematic error by creating Monte Carlo time simulations with the same power spectral density and for each time simulation we determine the systematic frequency offset due to these magnetic field gradient fluctuations. The final systematic error is the root-mean square of these simulations.\\
Similar to averaging over magnetic transitions, averaging over both possible ion crystal configurations leads to a high level of suppression of the differential second-order Zeeman shift \cite{chwalla_absolute_2009}, position-dependent AC-Stark shifts from light leakage, and differential electric quadrupole shifts \cite{roos_designer_2006} such that these effects have a negligible contribution to the overall error budget. Calculations using Multi-Configuration Dirac-Fock method further show that the quadrupole moments of different isotopes differ by less than $10^{-5}$ and can therefore be assumed to be the same across all considered isotopes.\\ %
Finally, the correlated Ramsey scheme used here is insensitive to both the common-mode AC Stark shift during the entirety of the Ramsey waiting time as well as isotope-dependent Stark shifts during the Ramsey pulses. The first point is clear since we only measure the relative frequency of the two isotopes, the second point can be explained as follows: a light-induced AC Stark shift during the Ramsey pulses will indeed lead to an additional phase shift in a single Ramsey phase scan. However, by taking phase scans at various waiting times and only extracting the change in phase, this additional phase drops out as a common-mode shift. Our experiment is however susceptible to AC Stark shifts which are both isotope dependent and Ramsey time dependent. For instance, it is conceivable that the change in laser duty cycle could lead to a change in power or polarization of the laser beam. To bound such an error, we measure the differential AC Stark shift by performing Rabi spectroscopy both, for short (\SI{30}{\milli\second}) and long (\SI{300}{\milli\second}), waiting times before performing a coherent laser pulse. The main contribution to the AC Stark shift is expected to originate from a spectator EOM sideband at $-f_\textrm{offset}$, which primarily shifts \Ca{40}{+}. Hence, we expect the systematic uncertainty to be similar for all isotope pairs. For the $|S_{1/2}, m_J=+1/2\rangle \rightarrow |D_{5/2}, m_J=+1/2\rangle$ transition, this differential AC Stark shift change was measured to be \SI{33(28)}{\hertz}, while for the $|S_{1/2}, m_J=-1/2\rangle \rightarrow |D_{5/2}, m_J=-1/2\rangle$ it was \SI{23(37)}{\hertz}. The ions are only susceptible to such a frequency shift during the pulse times, which are much shorter than the Ramsey waiting time, such that we estimate the systematic error of such a differential AC Stark shift to be \SI{4.5}{\milli\hertz}.\\
We suppress all but one of the laser beams with two sequential AOMs to achieve an optical suppression when in the off state of at least 100~dB. The $D_{3/2}$-to-$P_{1/2}$ repump laser, which is only suppressed by a single AOM, is far detuned from any transition involving $S_{1/2}$ or $D_{5/2}$ levels, such that to a high degree any leakage light creates a common-mode shift for different isotopes and is suppressed in the isotope shift measurement. 

\subsubsection{Consistency checks}
While isotope shift measurements between \Ca{40}{+} and \Ca{44,46,48}{+} show good agreement with a previous measurement in the literature \cite{knollmann_part-per-billion_2019}, we find significant deviation for the pair \Ca{40}{+}-\Ca{42}{+}. In order to test the consistency of our results, we additionally measured the isotope shift of the pair \Ca{42}{+}-\Ca{48}{+}. This allows for an additional independent measurement of the \Ca{40}{+}-\Ca{42}{+} isotope shift by subtracting the \Ca{42}{+}-\Ca{48}{+} isotope shift from the \Ca{40}{+}-\Ca{48}{+} isotope shift. Adding statistical and systematic error in quadrature, the difference between the direct measurement of \Ca{40}{+}-\Ca{42}{+} isotope shift and the indirect measurement of \Ca{40}{+}-\Ca{42}{+} isotope shift deviates by less than $1.9\sigma$ from zero. 

\section{Calculations}
This section describes how the calculations of electronic binding energies required for a determination of the nuclear mass ratios were carried out. Further, it describes the calculations of the electronic isotope-shift constants. Additionally, it provides information on the calculation and model assumptions with which nuclear polarization is treated. For all the calculated values an uncertainty is estimated, that we include in any of the constraints on beyond-Standard-Model physics. The sections concludes by estimating that other higher-order Standard-Model effects other than the second-order mass shift and nuclear polarization are expected to contribute much less than these two leading effects.
\subsection{Calculations of the electronic binding energies in calcium}
From PENTATRAP experiments, the mass ratios between various Ca isotope ions have been measured precisely at the $10^{-11}$-level. This corresponds to a mass difference below 1~eV/c$^2$. To derive the nuclear mass ratios to a similar precision, accurate electron binding energies of a neutral Ca atom and highly charged Ca ions are needed. 
In principle, the total electron binding energies of Ca ions can be obtained by summing the ionization potentials (IPs) of all the charge states listed in the NIST atomic database~\cite {kramida_nist_nodate}. However, the corresponding accuracies do not reach the level required in this work. Taking neutral Ca and Ne-like Ca$^{10+}$, for example, this method would result in a value of $18,510(5)$ and $17,506(4)$\,eV, respectively, for the binding energy of the 20 and 10 electrons. Both of them have a ten-times-larger uncertainty than achieved in the measurement of mass ratios. Therefore, advanced atomic structure calculations are performed to reduce the uncertainties in the electron binding energies.\\ 
The calculations are performed via the \textit{ab initio} fully relativistic multiconfiguration Dirac--Hartree--Fock (MCDHF) and relativistic configuration interaction (RCI) methods~\cite{Grant1970,Desclaux1971,grant2007relativistic} implemented in the GRASP2018 code~\cite{GRASP2018,Jonsson2023-2}. In this method, the many-electron atomic state function (ASF)
is constructed as a linear combination of configuration state functions (CSFs) with common total angular momentum ($J$), magnetic ($M$), and parity ($P$) quantum \mbox{numbers:} $|\Gamma P J M\rangle = \sum_{k} c_k |\gamma_k P J M\rangle$.
Each CSF $|\gamma_k P J M\rangle$ is built from products of one-electron orbitals (Slater
determinants), $jj$-coupled to the appropriate angular symmetry and parity, and $\gamma_k$ represents orbital occupations, together with orbital and intermediate
quantum numbers necessary to uniquely define the CSF. $\Gamma$ collectively denotes \mbox{all} the $\gamma_k$ involved in the representation of the ASF. $c_k$ is the corresponding mixing coefficient. We first solve the MCDHF equations self-consistently~\cite{Grant1970,Desclaux1971,grant2007relativistic} to obtain optimized $c_k$ and radial orbital wavefunctions under the Dirac--Coulomb Hamiltonian.\\ 
Our work starts with the single-configuration Dirac--Hartree--Fock calculation of $^{40}$Ca ions, based on which one can access the contributions from field shift (or finite nuclear size effect), mass shift (or nuclear recoil effect), Breit interaction, frequency-dependent transverse photon interaction, and QED effect.\\ 
As the field shifts are dominated by electrons in the inner-shell $1s$ and $2s$ orbitals, ions with more than 4 electrons share a similar field shift of around $-0.030(1)$~eV, with the uncertainty arising from the uncertainty in the nuclear radius. The mass shift is $-0.183(3)$~eV for Ca$^{16+}$, and slightly grows to $-0.199(3)$~eV for Ca$^{10+}$, bearing an accuracy on the order of $(m_e/M)(\alpha Z)m_ec^2$. Here, $m_e$ is the electron mass, $M$ the nuclear mass, $\alpha$ the fine structure constant, $Z$ the nuclear charge, and $c$ the speed of light. For Ca$^{10+}$, the Breit interaction and frequency-dependent transverse interaction further contribute around $-4.95$ and $0.03$~eV, respectively, to the total binding energies. We have also performed QED calculations. With a value of around $-3.35(3)$~eV for Ca$^{10+}$, the QED effect only varies slightly between the ions considered in this work. The accuracy of these QED terms is obtained by comparing the difference between our result and the more accurate result of Be-like Ca$^{16+}$ calculated via \textit{ab initio} methods~\cite{PhysRevA.90.062517,PhysRevLett.126.183001}. Nevertheless, while all these contributions can be accounted for with sufficient precision, the main challenge of the calculations (thus the uncertainties of the final results) is devoted to the electron correlation effect.\\ 
\begin{table*}[ht]
    \footnotesize
    \centering
    \caption{Ionization potentials of $^{40}$Ca ions. In the case of Ca$^{16+}$, Ca$^{7+}$, and Ca$^{5+}$, the calculated values in this work are different from those in the NIST database. Considering that the accuracy of our calculations has always been benchmarked by accurate experimental IPs of nearby ions, we are confident that our results are more reliable. All values are in units of eV.}
\begin{ruledtabular}
\begin{tabular}{l|l|l|l|l|l|l|l|l|l|l}
ions & Ca$^{19+}$ &  Ca$^{18+}$ & Ca$^{17+}$ & Ca$^{16+}$ & Ca$^{15+}$ & Ca$^{14+}$ & Ca$^{13+}$ & Ca$^{12+}$ & Ca$^{11+}$ & Ca$^{10+}$  \\
 valence & $1s^1$ &  $1s^2$ & $2s^1$ & $2s^2$  & $2p^1$ & $2p^2$ & $2p^3$ & $2p^4$ & $2p^5$ & $2p^6$  \\
 \hline

 NIST & 5469.86358(6) &  5128.8576(3) & 1157.726(7) & 1086.8(4) & 973.7(3) & 894.0(4) & 817.2(6) & 728.6(11) & 658.2(9) & 591.60(12)  \\

this work & $-$ & $-$ & $-$ & 1087.28(3) & 973.78(2) & 894.72(3) & 816.82(3)  & 728.04(2)  & 658.02(2)  &  591.40(2)  \\
  \hline
  \hline
ions & Ca$^{9+}$ &  Ca$^{8+}$ & Ca$^{7+}$ & Ca$^{6+}$ & Ca$^{5+}$ & Ca$^{4+}$ & Ca$^{3+}$ & Ca$^{2+}$ & Ca$^{+}$ & Ca  \\
 valence & $3s^1$ & $3s^2$  & $3p^1$ & $3p^2$ & $3p^3$ & $3p^4$ & $3p^5$ & $3p^6$ & $4s^1$ &  $4s^2$  \\
 \hline

 NIST & 211.275(4) &  188.54(6) & 147.24(12) & 127.21(25) & 108.78(25) & 84.34(8) & 67.2732(21) & 50.9132(3) & 11.871719(4) & 6.1131555(3)  \\

this work & 211.270(4) &  188.55(2) & 147.83(3) & 127.35(3) & 107.65(3) & 84.39(3) &   $-$ & $-$ & $-$ &  $-$ 
    \label{tab.ip}
    \end{tabular}
    \end{ruledtabular}
\end{table*}
\begin{table*}[ht]
    \footnotesize
    \centering
    \caption{Total electron binding energies for neutral and highly charged calcium isotope ions. All values are in units of eV.}
\begin{ruledtabular}
\begin{tabular}{l|l|l|l|l|l|l|l|l}
isotope ions & $^{40}$Ca & $^{40}$Ca$^{10+}$ &  $^{40}$Ca$^{11+}$ & $^{40}$Ca$^{12+}$ &  $^{40}$Ca$^{13+}$ & $^{40}$Ca$^{14+}$ & $^{40}$Ca$^{15+}$ \\
 \hline

 NIST & 18,510(5) & 17,507(4) &  16,915(4) & 16,257(3) & 15,528(2) & 14,711(1) & 13,817(1)   \\

this work & 18,509.73(21) & 17,506.52(16) &  16,915.12(14) & 16,257.10(12) & 15,529.06(10) & 14,712.24(8) & 13,817.51(6)   \\
\hline
\hline
isotope ions &   &   &  $^{44}$Ca$^{11+}$ & $^{48}$Ca$^{12+}$ &   &  & $^{42}$Ca$^{15+}$  & $^{46}$Ca$^{15+}$ \\
\hline
this work &   &   &  16,915.14(14) & 16,257.14(12) &   &  & 13,817.52(6)  & 13,817.53(6) 

    \label{tab.EHCIs}
    \end{tabular}
    \end{ruledtabular}
\end{table*}
To calculate the electron correlation energies accurately, we have systematically expanded the size of the CSF list by allowing single and double excitations of the electrons from the $1s,2s,2p,3s,3p$ occupied orbitals to highly-lying correlation orbitals. The final sizes of the lists range from tens of thousands of CSFs in ions such as Ca$^{10+}$, Ca$^{11+}$, Ca$^{14+}$ and Ca$^{15+}$, to hundreds of thousands of CSFs in ions such as Ca$^{12+}$ and Ca$^{13+}$, and to millions of CSFs in ions such as Ca$^{3+}$, Ca$^{4+}$, and Ca$^{5+}$. These correlation orbitals are added and optimized layer-by-layer~\cite{Jonsson2023} up to $n=10$ ($n$ is the principle quantum number), with the highest orbital angular momentum in each layer being of $n-1$. 
As the increment of the correlation energy at each layer decreases exponentially as a function of $n$~\cite{lyu2023extreme}, it allows us to extrapolate the electron correlation energy to $n=\infty$.\\ 
We note that, though one can directly calculate the total electron binding energies of each Ca ion, the above procedure in calculating correlation energies has only considered the single and double electron exchanges. The unaccounted high-order correlation effects render it difficult to evaluate the error bars at the sub-eV level. Therefore, our calculations are aimed at improving the accuracies of the single-electron IPs of Ca at different ionization stages. In this way, one can make use of the experimental IPs of nearby elements to benchmark the systematic uncertainties of the calculations. The summation of the refined IPs will result in total binding energies for each ion with much higher accuracy.\\ 
In the case of O-like Ca$^{12+}$, the IP for this ion listed in the NIST database is 728.6(11)~eV, which has the largest error bar in the IPs of all Ca ions. To refine this value, we calculated the total binding energy of the 8 and 7 electrons in Ca$^{12+}$ and Ca$^{13+}$ ions, respectively. Their difference would thus give rise to the IP of Ca$^{12+}$. For calculations based on a point-like nucleus with infinite mass, one gets a value of 726.593~eV. When the finite nuclear size and mass are considered, these two effects add up 0.14(1)~meV and $-3.22(5)$~meV, respectively to the calculated IP, with another 18.2(2)-meV corrections from the QED effects. While these three corrections are small, further consideration of the Breit interaction and the frequency-dependent transverse interaction would contribute $-0.289$~eV and $-0.012$~eV, respectively. The uncertainty of these two terms will be accounted for together with the calculation of the residual Coulomb--Breit interaction, i.e., the so-called correlation energy. When only single and double electron exchanges are considered, we obtain a contribution of 1.717(1)~eV from the correlation energy, and then a calculated IP of 728.024~eV for Ca$^{12+}$. In order to assert the accuracy of this result, we performed similar IP calculations for both O-like F$^+$ and Ne$^{2+}$, which have a similar electron configuration as Ca$^{12+}$. These two ions have accurate experimental IPs of 34.97081(12) and 63.4233(3)~eV, respectively, which are found to be 0.060 and 0.034~eV larger than our calculated values. As the differences are mainly a result of the unaccounted high-order electron correlations whose effects become smaller in heavier O-like ions, it indicates that the value of the high-order correlation correction to the IP of Ca$^{12+}$ must be smaller than 0.034~eV. This allows us to conservatively assert a 0.017(17)-eV correction (i.e., half of 0.034 eV) to the theoretical IP of Ca$^{12+}$, and obtain a refined value of 728.04(2)~eV, where the 0.017-eV error bar is rounded up to 0.02 eV.\\ 
With a similar procedure, we have improved the accuracies of the IPs for ions from Ca$^{4+}$ up to Ca$^{16+}$ accordingly. The results are presented in Table~\ref{tab.ip}, where the values from the NIST database are also shown for comparison. While most of the calculated IPs agree with their NIST values, three ions, namely Ca$^{16+}$, Ca$^{7+}$, and Ca$^{5+}$ have IPs differ from their NIST values by up to 5 standard deviations. Considering that each of our calculations has been benchmarked by a few other nearby ions, we are confident that our values are more reliable than those in the NIST database. Take P-like Ca$^{5+}$ for instance, the calculated 107.65(3)-eV IP is 1.2~eV smaller than the NIST value. However, when applying the same theoretical procedure to P-like S$^+$ and neutral P atom whose IPs have been accurately measured, the calculated IPs are only 0.11 and 0.12~eV, respectively, smaller than their experimental values. Therefore, one can conclude that there must be an error in the IP of P-like Ca$^{5+}$ in the NIST database~\cite{kramida_nist_nodate}.\\ 
Based on the IPs listed in Table~\ref{tab.ip}, we derive a total electron binding energy of  $18\,509.73(21)$~eV for neutral $^{40}$Ca atom. For the highly charged calcium ions employed in the mass-ratio measurement, their values are listed in Table~\ref{tab.EHCIs}. With an uncertainty of 0.2~eV or less, our calculated values are up to 25-fold more accurate than those obtained from the NIST database. Here, the field shift and mass shift for different isotopes have already been accounted for. As a result, the total binding energies of $^{42}$Ca$^{15+}$ and  $^{46}$Ca$^{15+}$ are 0.0082(1) and 0.0235(1)~eV, respectively, larger than that of $^{40}$Ca$^{15+}$. In the case of $^{44}$Ca$^{11+}$ and $^{48}$Ca$^{12+}$, the corresponding isotope shifts of their total binding energies are 0.0179(1) and 0.0352(1)~eV, respectively, with respect to their $^{40}$Ca counterparts. 

\subsection{Isotope-shift constants for Ca$^{14+}$ \label{sec:calculated_isotope_shift_constants}}
We performed calculations of the mass-shift (MS) and field-shift (FS) isotope-shift constants for the $(1s)^2(2s)^2(2p)^2$
\CaPtoP transition in the C-like ion of \Ca{}{14+}.
The MS constant $K$ is separated into the first-order $K^{(1)}$ and second-order $K^{(2)}$ terms,
\begin{align}
    K = K^{(1)} + \frac{m_e}{m}\,K^{(2)}\,,
\end{align}
with $m_e$ and $m$ being the electron and nuclear masses, respectively. 
For the FS, we restrict our calculations to the first-order constant, $F^{(1)}$.
Calculations were carried out by the relativistic configuration-interaction (CI) method.
Our implementation of the method was described in Ref.~\cite{yerokhin_hyperfine_2008}. 
The generalization of the CI method for calculations of the second-order MS was developed in our previous investigation~\cite{yerokhin_nonlinear_2020}. 
The CI expansion in our calculation included single, double, and most important triple, quadruple, and quintuple excitations from the multiple reference state of the type $1s^22s^22p^2 + 1s^22p^4$. 
Our large-scale computations included up to two million configuration-state functions (CSFs). 
Uncertainty estimations are based on a careful analysis of results obtained with many (about 40) different sets of CSFs, which allowed us to evaluate small corrections due to extensions of the configuration space in different directions.
The obtained numerical results for \Ca{}{14+} are presented in Table~\ref{tab:IS}. 
Our results for the FS constants $K^{(1)}$ and $F^{(1)}$ are in good agreement with previous calculations by the multiconfigurational Dirac-Fock (MCDF) method \cite{naze_isotope_2014}, $K^{(1)} = 464$~\SI{}{\giga\hertz}$\times$amu and $F^{(1)} = 8.1$~MHz$\times$fm$^{-2}$.
The relatively large uncertainty of our FS constant $F^{(1)}$ is explained by large cancellation between the two transition states $^3P_0$ and $^3P_1$, which are separated only by the fine structure. The five-digit cancellation leads to a very small FS for the transition and, consequently, a large relative uncertainty.
\begin{table*}
\caption{
Isotope-shift constants for \Ca{}{14+} and \Ca{}{+}. Results for \Ca{}{+} are from Ref.~\cite{viatkina_calculation_2023}.
\label{tab:IS}
}
\begin{ruledtabular}
\begin{tabular}{llcccc}
  & \multicolumn{1}{c}{Units}
  & \multicolumn{1}{c}{\Ca{}{14+}, \CaPtoP}
  & \multicolumn{1}{c}{Uncertainty}
  & \multicolumn{1}{c}{\Ca{}{+}, \CaStoD{5/2}}
  & \multicolumn{1}{c}{Uncertainty}
\\
  \hline\\[-2pt]
$K^{(1)}$  & GHz$\times$amu       &    462      &  $\pm$ 5    &  2448    &  $\pm$ 10\% \\
$K^{(2)}$  & GHz$\times$amu$^2$   &    $-$1.0   &  $\pm$ 0.1  &   0.59   &  $\pm$ 65\%  \\
$F^{(1)}$  & MHz$\times$fm$^{-2}$ &     8.4     &  $\pm$ 3.1  &  378     &   $\pm$ 5\% \\
$F^{(2)}$  & MHz$\times$fm$^{-4}$ &             &             & \!\!\! $-$127  &   $\pm$ 5\% \\
\end{tabular}
\end{ruledtabular}
\end{table*}

\subsection{Nuclear polarization}
The nuclear polarization correction to energy levels was calculated following the method developed in Refs.~\cite{plunien_nuclear_1991,nefiodov_nuclear_1996} and its generalization to the case of many-electron atoms in Ref.~\cite{viatkina_calculation_2023}.
Our calculations include the contribution of the lowest-lying electric-quadrupole nuclear excited state and the nuclear giant-resonances contribution. For the details of the method we refer to Ref.~\cite{viatkina_calculation_2023}.
As was demonstrated in Ref.~\cite{viatkina_calculation_2023}, it is advantageous to calculate not only the nuclear polarization correction but its ratio to the field shift, as this ratio is remarkably stable with respect to details of the treatment of the electron correlation.
Because of the large cancellations (to five digits) of the field shift between the transition states in \Ca{}{14+}, we do not perform calculations for separate states as in Ref.~\cite{viatkina_calculation_2023}, but directly for the transition energy, thus avoiding precision losses due to this cancellation. 
For consistency, we recalculate the nuclear polarization correction for \Ca{}{+}, using exactly the same numerical method as for \Ca{}{14+}.
Following Ref.~\cite{viatkina_calculation_2023}, we perform our calculations by the many-body perturbation theory (MBPT) approach. 
Two methods were used for treatment of the electron correction. 
The first one, referred to as the Dirac-Fock with core relaxation (DF-CR) method, includes all MBPT diagrams with one-photon exchange between the electrons. The second method, referred to as DF-CR+RPA method, includes in addition to DF-CR the sequence of ladder diagrams delivered by the so-called random phase approximation (RPA), see Ref.~\cite{viatkina_calculation_2023} for details. 
Our numerical results for the nuclear polarization correction are presented in Table~\ref{tab:np}, expressed in terms of the ratio function $g_{ab}^A$, defined as
\begin{align} \label{ea:gab}
E_{\rm np}^A(a)-E_{\rm np}^A(b)
 = - \langle r^2\rangle^A \, [F^{(1)}(a)-F^{(1)}(b)]\, g_{ab}^{A}
 \, 10^{-3}\,.
\end{align}
Here, $A$ labels the isotope, $a$ and $b$ are the upper and the lower transition states, respectively; $E_{\rm np}^A(a)$ is the nuclear polarization corrections for the state $a$, $F^{(1)}(a)$ is the first-order FS constant for the state $a$, and $\langle r^2_{A}\rangle$ is the mean-square nuclear charge radius.
The uncertainty of our results in Tab.~\ref{tab:np} comes in two ways: (i) from the incomplete treatment of the electron correlation and (ii) from the approximate nuclear model. 
The electron-correlation uncertainty was estimated by taking the difference of the results obtained with the DF-CR and DF-CR+RPA methods.
As already noted in Ref.~\cite{viatkina_calculation_2023}, for \Ca{}{+} the variation of the ratios $g_{ab}$ obtained with the two methods is very small, $<1\%$.
For \Ca{}{14+}, it is larger but still within about 2\%. 
By contrast, a much larger source of uncertainty is the nuclear model. 
The major issues are that the nuclear excitation spectrum is not available and that only the Coulomb part of the electron-nucleus interaction is taken into account by the model \cite{plunien_nuclear_1991,nefiodov_nuclear_1996}. 
We estimate the nuclear-model uncertainties to be 20\% for energy corrections for single isotopes and 50\% for the isotope-shift differences.
These uncertainties are supposed to be highly correlated between \Ca{}{14+} and \Ca{}{+}.
\begin{table}[b]
\caption{
Nuclear polarization for \Ca{}{14+} and \Ca{}{+}, in terms of $g_{ab}$ defined by Eq.~(\ref{ea:gab}).
\label{tab:np}
}
\begin{ruledtabular}
\begin{tabular}{ldd}
  $A$
  & \multicolumn{1}{c}{\Ca{}{14+}, \CaPtoP}
  & \multicolumn{1}{c}{\Ca{}{+}, \CaStoD{5/2}}
\\
  \hline
40  &   0.18599   &  0.36478 \\
42  &   0.20301   &  0.39663 \\
44  &   0.20979   &  0.41277 \\
46  &   0.20573   &  0.41257 \\
48  &   0.20989   &  0.42492 \\
\end{tabular}
\end{ruledtabular}
\end{table}
\subsection{Electronic coefficients $X_i$ of beyond-Standard-Model physics contribution}
To generate bounds on the beyond-Standard-Model (BSM) physics coupling, we calculated the electronic BSM coefficients $X_i$, for the transition labeled $i$, as differences between the $X(a)$ coefficients of the involved states $a$. To compute the latter, we used a combination of configuration interaction (CI) and many-body perturbation theory (MBPT), implemented in the AMBiT code~\cite{kahl_ambit_2019}.\\
First, we performed a Dirac-Fock calculation for a given number of electrons, resulting in the wavefunctions of the one-electron orbitals in the Dirac-Fock core. For the $^3P_0 \to$ $^3P_1$ transition in \Ca{}{14+} all six electrons in the $1s^2 2s^2 2p^2$ configuration were included.
All remaining valence and excited orbitals were constructed as B-spline solutions of the Dirac-Fock operator. The CI-space was constructed starting from the leading configuration $1s^2 2s^2 2p^2$, by including single- and double-electron excitations, involving one-electron orbitals with maximal principal quantum number $n=12$ and orbital angular momentum $l=3$. 
Additionally, single and double hole excitations from the $2s$ shell were included, while the $1s$-electrons were treated as the ``frozen core'', from which no hole excitations enter the CI-space. Core-valence correlations were taken into account via second-order MBPT corrections to the CI matrix elements including one-electron orbitals up to $n=35$ and $l=4$.
For \Ca{}{+} the Dirac-Fock calculation was performed for the 18 core electrons in the $1s^2 2s^2 2p^6 3s^2 3p^6$ configuration. The CI-space is built up starting from the leading configurations $3p^6 4s$, $3p^6 4p$ and $3p^6 3d$, by including single electron excitations of the valence electron to orbitals up to $n=10$ and $l=4$.
 The MBPT corrections for \Ca{}{+} include orbitals up to $n=30$ and $l=4$.\\
Finally, after adding the Yukawa potential
    \begin{equation}
        V_{\phi}= \lambda \frac{\hbar c}{r}e^{- \frac{c}{\hbar} m_\phi r}
    \end{equation}
to the Dirac-Fock potential, where $\lambda$ is a dimensionless scaling parameter and $m_\phi$ is the mass of the new boson, we extracted the $X(a)$ coefficients of the states by taking the numerical derivative of the eigenenergies $E(a,\lambda)$ with respect to the parameter $\lambda$:
    \begin{equation}
        X(a) = \frac{\partial E(a,\lambda)}{\partial \lambda}\,.
    \end{equation}
For this method, we estimate a 10\% uncertainty on the calculated electronic coefficients $X_i$ of the \CaPtoP transition in \Ca{}{14+}.
\subsection{Higher-order effects}
Here we address the higher-order effects contributing to the nonlinearity of the King plot,
beyond the second-order MS and the nuclear polarization discussed in the previous
sections. These effects were studied in detail
for \Ca{}{+} in
Ref.~\cite{viatkina_calculation_2023}. In Tab.~\ref{tab:ho} we compile results 
for individual effects contributing to the King-plot nonlinearities for the
isotope shift between \Ca{40}{+} and \Ca{42}{+}
and the transition of interest, \CaStoD{5/2}. 
The second-order MS and the nuclear polarization are
by far the two dominant effects; all other higher-order effects are much smaller. 
Taking into account that for the \CaPtoP transition of \Ca{}{14+} all
FS effects are highly suppressed due to the large cancellation between the
two states, we conclude that the higher-order effects can safely
be neglected for \Ca{}{14+} as well as for \Ca{}{+} at our current measurement precision. 
Nuclear deformation effects are also expected to not contribute significantly, as inferred from recent estimations of this effect in hydrogen-like highly charged ions~\cite{sun_nuclear_2024}.
\begin{table}
\caption{
Comparison of different higher-order effects that contribute to the isotope shift of the \CaStoD{5/2} transition between \Ca{40}{+} and \Ca{42}{+}. Values are units of \SI{}{\kilo\hertz} and taken from Ref.~\cite{viatkina_calculation_2023}.
\label{tab:ho}
}
\begin{ruledtabular}
\begin{tabular}{lc}
second-order MS & $-$34 \\
Nuclear polarization       & $-$178 \\
FS, higher-order corr.  & $-$3.3 \\
FS, second order & $-$5.7 \\
Cross    & 0.16 \\
\end{tabular}
\end{ruledtabular}
\end{table}
\section{Analysis}
This section contains further information on the generalized King plot analysis, that we use to constrain beyond-Standard-Model physics. We show that all three measurements, namely the isotope shifts in \Ca{}{14+} and \Ca{}{+}, as well as the mass ratio measurements, were simultaneously necessary to reveal the NL of the Ca King plot. Furthermore, we show how additional precision isotope shift measurements and improved calculations would further tighten these constraints and provide a comparison of our constraints with those derived with other methods. We also give insight in what causes the losses of sensitivity to beyond-Standard-Model physics of the King plot method in the exclusion plot.

\subsection{Constraining beyond-Standard-Model physics with the generalized King plot}
\label{Sec: Constraining beyond-Standard-Model physics with GKP}
Including explicitly the BSM contribution, the general IS expansion introduced in the main text reads
    \begin{align}
        \delta\nu^{AA^\prime}_i = & K_i\mu^{AA^\prime} + F_i \delta \langle r^2 \rangle^{A A^\prime} \notag
        \\& + \sum_{\ell=1}^\infty G_{i}^{(\ell)} \eta_{\,(\ell)}^{A A^\prime} + \alpha_\mathrm{BSM} X_i \gamma^{AA^\prime}\,.
    \end{align}
Assuming a factorization into nuclear ($\eta_{\,(\ell)}^{A A^\prime}$) and electronic $(G_{i}^{(\ell)})$ parts, an additional transition allows the elimination of one higher-order term (analogous to the elimination of the nuclear charge radii in the standard KP), without requiring detailed knowledge of the nuclear dependence of this term.
Given $m$ transitions, $m-1$ nuclear parameters (including $\delta\langle r^2 \rangle^{AA^\prime}$) can be traded for IS measurements, obtaining a higher-dimensional equivalent of the King relation, the so-called generalized King plot (GKP)~\cite{berengut_generalized_2020}:
    \begin{align}
        \vec{\delta\bar\nu}_j = \mathcal{K}_j\vec{1} + \sum_{\substack{k=1\\ k\neq j}}^{m-1} \mathcal{F}^k_j\vec{\delta\bar\nu}_k\, + \alpha_\mathrm{BSM}\mathcal{X}_j\vec{\bar\gamma},
    \label{eq:GKP_plus_NP}
    \end{align}
where the bar stands for division by $\mu^{AA^\prime}$, e.g. $\delta \bar{\nu}_{i,j}^{A A^\prime} = \delta \nu_{i,j}^{A A^\prime} / \mu^{A A^\prime}$, the $(m+1)$-vectors are intended in isotope-pair space and $\mathcal{K}_j,\,\mathcal{F}^k_j$ and $\mathcal{X}_j$ represent combinations of electronic coefficients (explicit form in Ref.~\cite{berengut_generalized_2020}). Eq.~\eqref{eq:GKP_plus_NP} represents a system of $(m+1)$ equations, whose solution yields an analytical expression for the BSM coupling~\cite{berengut_generalized_2020}:
    \begin{align}\label{eq:yeyn_volumes}
        \alpha_\mathrm{BSM} &= \frac{V_\mathrm{exp}}{V_\mathrm{th}} \\
        &= \frac{\det(\vec{\delta\bar{\nu}}_1,\dots,\vec{\delta\bar{\nu}}_m,\vec{1})}{\frac{1}{(m-1)!}\varepsilon_{i_1\ldots i_{m}}\varepsilon_{a_1\ldots a_{m+1}}X_{i_1}\bar\gamma^{a_1}\delta\bar{\nu}^{a_2}_{i_2}\ldots \delta\bar{\nu}^{a_{m}}_{i_{m}}}\, \nonumber,
    \end{align}
where $V_\mathrm{exp}$ and $V_\mathrm{th}$ are the experimentally observed NL volume and theoretically predicted NL volume for $\alpha_\mathrm{BSM}=1$, respectively. Further, $\varepsilon$ denotes the Levi-Civita symbol and the dependence on the new boson mass $m_\phi$ is contained in the electronic coefficients $X_i$.
We apply this method for $m=3$ transitions, using the frequency shifts $\nu_{570}$, $\nu_{729}$ and $\nu_\mathrm{DD}$, measured in four isotope pairs, to eliminate the leading higher-order contribution. Under the assumption that the remaining NL can be attributed to BSM, we obtain $2\sigma$-upper bounds, presented as the blue line in Fig.~\ref{fig:exclusion_plot}.
Here, the uncertainty estimation includes the experimental uncertainties of $\delta\bar{\nu}_i^{AA^\prime}$ as well as a $10\%$ uncertainty of the calculated $X_i$.\\
However, as described in the main text, we expect contributions from two higher-order SM sources on NL, namely the second-order MS and nuclear polarization. 
We account for the second-order MS by subtracting the theoretical prediction from the experimental IS data:
    \begin{equation}
        \widetilde{\delta\nu}_i^{AA^\prime}=\delta\nu_i^{AA^\prime}-K_i^{(2)}\,\mu_{(2)}^{A A^\prime},
    \end{equation}
with the electronic coefficients of the second-order MS $K_i^{(2)}$ (see Sec.~\ref{sec:calculated_isotope_shift_constants}) and $\mu_{(2)}^{A A^\prime}~=~\frac{m_A^2 - m_{A^\prime}^2}{m_A^2 m_{A^\prime}^2}$. 
The remaining NL is treated then within the GKP approach.
The uncertainty of $K_i^{(2)}$ increases that of $\widetilde{\delta\nu}_i^{AA^\prime}$, compared to $\delta\nu_i^{AA^\prime}$, affected purely by experimental uncertainties. However, taking into account that the uncertainty on $K_i^{(2)}$ is correlated across all isotope pairs, its impact on the total uncertainty $\sigma\left[\alpha_\mathrm{BSM}\right]$ is strongly suppressed. Therefore, using $\mathrm{MS}^{(2)}$-corrected IS data in the GKP method yields an improvement of the bounds shown as a red solid line in Fig.~\ref{fig:exclusion_plot}, identical to the final bound shown in Fig.~3 of the main manuscript.

\begin{figure}
    \centering
    \includegraphics[width=\linewidth]{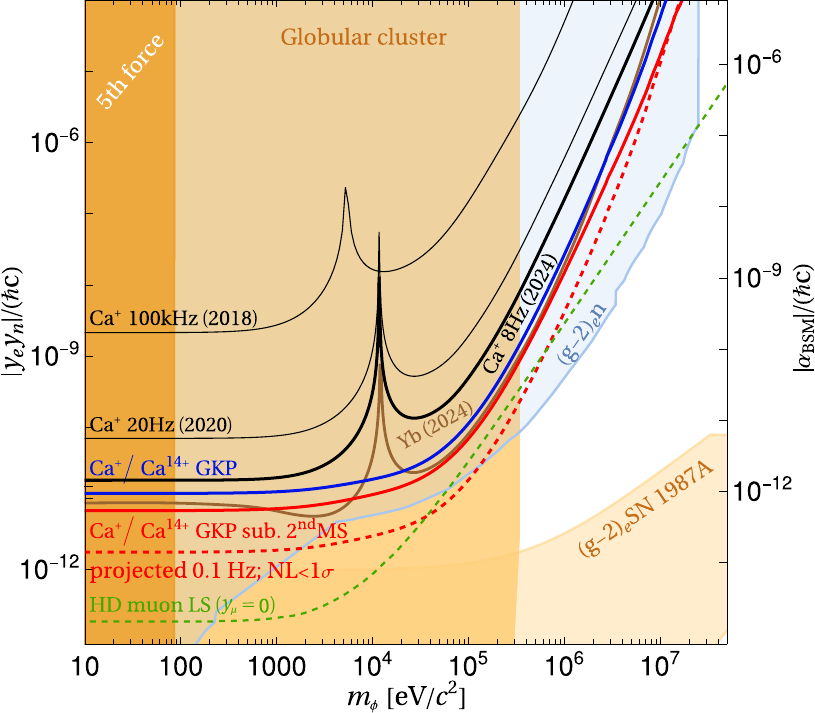}
    \caption{Shown are various bounds on the beyond-Standard-Model (BSM) physics coupling $y_ey_n = 4 \pi \, \alpha_\mathrm{BSM}$: previous bounds from King plots in \Ca{}{+} as solid black lines~\cite{gebert_precision_2015,solaro_improved_2020,chang_systematic-free_2024} and a recent bound from Yb~\cite{door_search_2024} (brown). 
    The solid blue curve results from our GKP analysis~\cite{berengut_generalized_2020} with the \CaStoD{5/2}, \CaDtoD~\cite{solaro_improved_2020, chang_systematic-free_2024} and \CaPtoP transitions, under the assumptions that the remaining nonlinearity is attributed to BSM physics.
    The solid red curve presents the results of our GKP analysis, after subtracting the theoretical prediction for the second-order MS from the experimental IS data, as described in the main text and in Sec.~\ref{Sec: Constraining beyond-Standard-Model physics with GKP}.
    Projections on improved future bounds from reduced theoretical and experimental uncertainties are depicted by the red dashed line, with more details given in Sec.~\ref{Sec: projection for improved limits}.
    The excluded region from bounds on $y_e$ from $(g-2)_e$~\cite{fan_measurement_2023} combined with constraints on $y_n$ from neutron scattering experiments~\cite{leeb_constraint_1992, nesvizhevsky_neutron_2008, pokotilovski_constraints_2006, barbieri_evidence_1975}  is shown in light blue. Bounds from other experiments and astrophysical observations are shown as shaded areas: constraints from supernova SN1987A~\cite{raffelt_limits_2012} (light orange); fifth force searches~\cite{michael_bordag_advances_2009} (dark orange); globular clusters~\cite{grifols_constraints_1986, grifols_constraint_1989} (orange) and IS between H and D using the nuclear charge radius extracted via Lamb shift (LS) in muonic atoms~\cite{Delaunay:2017dku}(purple).}
    \label{fig:exclusion_plot}
\end{figure}

\subsection{Comparison of nonlinearity significances}
To analyze the main contributions that led to the first observation of NL in a Ca King plot, we compare, in Tab.~\ref{tab:nonlinearities} the significance of the NL, $V_\mathrm{exp}/\sigma\left[V_\mathrm{exp}\right]$, for different data sets, with $V_\mathrm{exp}$ defined as in the numerator of Eq.~\eqref{eq:yeyn_volumes}. First we combine, in a minimal KP analysis (two transitions and three isotope pairs), old isotope shift data from Refs.~\cite{chang_systematic-free_2024,knollmann_part-per-billion_2019} with the nuclear masses derived from either the atomic masses of Ref.~\cite{wang_ame_2021} or our newly measured nuclear mass ratios. Both combinations show no significant NL. Then, we combine our new IS measurements of the $\nu_{570}$ and $\nu_{729}$ transitions with the nuclear masses derived from Ref.~\cite{wang_ame_2021}, which leads to a NL slightly larger than $2\sigma$. Finally, the combination of the new IS data with the new mass ratios leads to a NL significance of $V_\mathrm{exp}/\sigma\left[V_\mathrm{exp}\right]\approx 900$, revealing the importance of all three new measurements. \\
We include the $\nu_\mathrm{DD}$ transition in a GKP analysis with three transitions and four isotope pairs, in order to account for one higher-order SM term. As shown in Table~\ref{tab:nonlinearities}, the significance of the NL is reduced to about $4\sigma$, in this case. 
This reduction of NL significance in the GKP analysis is due to the suppression of an additional higher-order term and the increased experimental uncertainty in the $\nu_{\mathrm{DD}}$ transition. With this GKP analysis we would exclude all parameter space above the blue line in Fig.~\ref{fig:exclusion_plot}.
Finally, we subtract the theory prediction of the second-order MS from the IS data before performing the GKP analysis. The last row of Table~\ref{tab:nonlinearities} shows the significance of the NL for the scenario of a $10\%$ uncertainty on $K_i^{(2)}$ in \Ca{}{14+} and $100\%$ in \Ca{}{+}. In this case, the significance falls below $1\sigma$, corresponding to a linear GKP with which we can exclude the parameter space above the red solid line in Fig.~\ref{fig:exclusion_plot}. 
{\renewcommand{\arraystretch}{1.4}
    \begin{table}[hb]
    \caption{\label{tab:nonlinearities}Comparison of the significance of nonlinearity $V_\mathrm{exp}/\sigma\left[V_\mathrm{exp}\right]$ for different data sets. Here, $V_\mathrm{exp}$ is defined as in Eq.~\eqref{eq:yeyn_volumes}. In the first two rows, old isotope shift data from Refs.~\cite{chang_systematic-free_2024,knollmann_part-per-billion_2019} is combined either with the nuclear masses, derived from the atomic masses of Ref.~\cite{wang_ame_2021} or with our newly measured nuclear mass ratios. In the third and fourth row, said mass measurements are combined with our isotope shift data of the $\nu_{570}$ and $\nu_{729}$ transitions. The fifth column shows the reduction of nonlinearity by including the $\nu_{\mathrm{DD}}$ transition for a GKP analysis, as described in the main text.
    This reduction of NL significance is due to the inclusion of an additional higher-order term and the increased experimental uncertainty in the $\nu_{\mathrm{DD}}$ transition.
    The last column presents the NL of the GKP after subtraction of the second order MS from the same IS data, assuming $K_i^{(2)}$ to have an uncertainty of $10\%$, in \Ca{}{14+} and $100\%$, in \Ca{}{+}.}
        \begin{ruledtabular}
        \begin{tabular}{l l r}
            Transitions & Masses & $V_\mathrm{exp}/\sigma\left[V_\mathrm{exp}\right]$ \\
            \hline
            $\nu_{732}$, $\nu_{729}$ \cite{chang_systematic-free_2024,knollmann_part-per-billion_2019} & \cite{wang_ame_2021} &  $0.3$\\
            $\nu_{732}$, $\nu_{729}$ \cite{chang_systematic-free_2024,knollmann_part-per-billion_2019} & this work & $0.3$ \\
            $\nu_{570}$, $\nu_{729}$  & \cite{wang_ame_2021} & $2.4$ \\
            $\nu_{570}$, $\nu_{729}$  & this work & $892$\\
            $\nu_{570}$, $\nu_{729}$, $\nu_{\mathrm{DD}}$ & this work & $4.2$\\
            $\nu_{570}$, $\nu_{729}$, $\nu_{\mathrm{DD}}$ ($\mathrm{MS}^{(2)}$ subtracted) & this work & $0.6$
        \end{tabular}
        \end{ruledtabular}
    \label{tab:my_label}
    \end{table}
}
\subsection{Projections for improved limits on beyond-Standard-Model physics}
\label{Sec: projection for improved limits}
In future works, our constraints on the BSM coupling $\alpha_\mathrm{BSM}$ might be further enhanced by improving both experimental and theoretical uncertainties, under the condition that the significance of the observed nonlinearity does not increase. We analyzed the potential effects of improvements in the experimental uncertainty on the $\nu_\mathrm{DD}$ transition to 0.1 Hz, and in the theoretical uncertainties on the electronic factors $K^{(2)}_i$ in \Ca{}{+} to 10\% (as in \Ca{}{14+}). The bounds for these optimized conditions are estimated by assuming a 1$\sigma$ NL and are shown in Fig.~\ref{fig:exclusion_plot} as a red dashed line.
The projected constraints show an enhancement factor of approximately 3.5 with respect to our current limits (red, solid), breaching the $(g-2)_en$ constraints in the mass range $10^3 - 10^5$ eV. Notice, however, that if the improved precision results in an increase of the significance of NL, no improvement is expected. 
\subsection{Comparison with other bounds}
Our data and analysis improves previous constraints derived from King plots of Ca~\cite{gebert_precision_2015, berengut_probing_2018,solaro_improved_2020, chang_systematic-free_2024} and the recent Yb-derived~\cite{door_search_2024} constraint for most of the parameter space. Prominently, we can also exclude the parameter space at $m_\phi=10^4 \SI{}{\electronvolt}/c^2$, where the other analyses lose sensitivity (see Sec.~\ref{sec:exclusion_peaks} for more insights on the occurrences of the peaks in exclusion plots).\\   
The King-plot analysis only relies on the assumption that BSM physics manifests as a new boson coupling to electrons and neutrons.  The only other bound on the new boson, free from additional assumptions, in the mass range we explore is derived from the magnetic moment of the electron~\cite{fan_measurement_2023} and neutron scattering experiments~\cite{leeb_constraint_1992, nesvizhevsky_neutron_2008, pokotilovski_constraints_2006, barbieri_evidence_1975, Frugiuele:2016rii} (blue shaded region in Fig.~\ref{fig:exclusion_plot}).   
In comparison, other laboratory bounds require additional assumptions about BSM physics. For example, bounds derived from the measured IS between hydrogen (H) and deuterium (D), (purple shaded region in Fig.~\ref{fig:exclusion_plot}), require input of the nuclear charge radii, whose extraction from the Lamb shift (LS) in muonic atoms may be affected by a BSM contribution to the field shift coefficient and a possible coupling of muons to $\phi$ and, therefore, $y_\mu=0$ is assumed~\cite{Delaunay:2017dku}. 
Our constraints also serve as independent laboratory probes of bounds based on astrophysical observations~\cite{raffelt_limits_2012,grifols_constraints_1986,grifols_constraint_1989} (orange shaded regions in Fig.~\ref{fig:exclusion_plot}).\\
\subsection{Occurrence of peaks in the exclusion plot and high-mass behavior \label{sec:exclusion_peaks}}
A notable difference in the exclusion plot lines from our combination of transitions in \Ca{}{14+} and \Ca{}{+} compared to the older lines, that only involved transitions in \Ca{}{+}, is the absence of ``peaks'' for specific values of the mediator mass $m_\phi$.
To gain a general understanding of this behavior, we start by analyzing the algebra of the minimal ($m=2$ in Eq.~\eqref{eq:GKP_plus_NP}) KP, which does not take higher-order terms into account.
In this case the King relation takes the form
    \begin{equation}
        \delta \bar{\nu}_j^{AA^\prime}=\frac{F_j}{F_i}\delta \bar{\nu}_i^{AA^\prime}+K_{ji}+\alpha_\mathrm{BSM} X_{ji} \bar{\gamma}^{AA^\prime}, 
        \label{eq.: King relation nonlinear}
    \end{equation}
with $K_{ji} = K_j - \frac{F_j}{F_i} K_i$ and $X_{ji} = X_j - \frac{F_j}{F_i} X_i$. 
From this equation it follows that, when the ratio of the BSM coefficients $X_j/X_i$ is equal to the ratio of field shift coefficients $F_j/F_i$, the last term, representing the BSM modification, becomes zero. In this case, all sensitivity to the BSM coupling is lost and a peak is formed in the exclusion plot line. 
In Fig.~\ref{fig: X21 plot}, the mass dependence of $X_{ji}$ is visualized for the combinations of the \CaStoD{5/2} transition in \Ca{}{+} ($\nu_{729}$) with either the \CaPtoP transition in \Ca{}{14+} ($\nu_{570}$) or the \CaStoD{3/2} transition in \Ca{}{+} ($\nu_{732}$). 
The pure \Ca{}{+} line crosses zero at $m_\phi \approx \SI[print-unity-mantissa = false]{e4}{\eV}$, thus explaining the corresponding peak in the exclusion plot.\\ 
Moreover, this analysis provides insight into the behavior of the bounds in the high-mass limit.
In this mass regime, when the range of the Yukawa potential becomes smaller than the nucleus, the ratio of the BSM coefficients $X_j/X_i$ becomes equal to the ratio of field shift coefficients $F_j/F_i$ such that both lines in Fig.~\ref{fig: X21 plot} approach zero for large masses. 
This behavior is the origin of the strong loss of sensitivity in the high-mass limit. Since a slight deviation between the experimental and theoretical results of the field shift ratio $F_j/F_i$ can substantially alter the high-mass behavior of the bounds on BSM physics, the BSM electronic coefficients are rescaled such that the ratio $X_j/X_i$ approaches the experimental field shift ratio $F_j^{\mathrm{exp}}/F_i^{\mathrm{exp}}$ in the high-mass limit.\\
In this work, the bounds were obtained by using the GKP relation, Eq.~\eqref{eq:GKP_plus_NP}, where non-trivial combinations of electronic coefficients appear. The occurrence of peaks in an $m$-dimensional GKP is still connected to the zeroes of $\mathcal{X}_m$, but the prediction of peak locations in the exclusion plot requires knowledge of the electronic coefficients for the FS and BSM terms, as well as those for $(m-2)$ higher-order corrections.

\begin{figure}
    \centering
    \includegraphics[width=\linewidth]{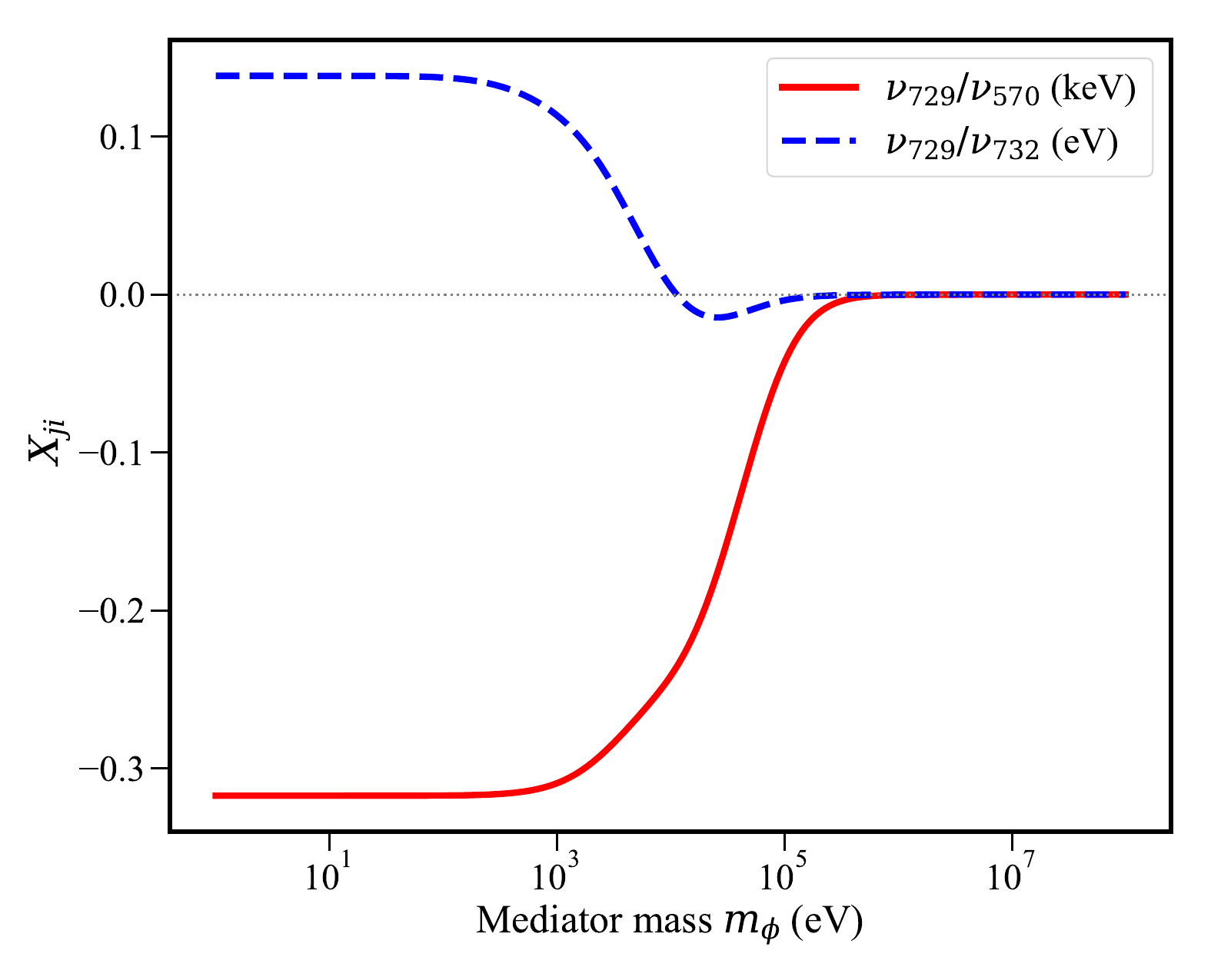}
    \caption{Beyond-Standard-Model physics coefficient $X_{ji}$ in the King relation from Eq. \eqref{eq.: King relation nonlinear} for the combination of the \CaStoD{5/2} transition in \Ca{}{+} ($\nu_{729}$) with either the \CaPtoP transition in \Ca{}{14+} ($\nu_{570}$; red, solid) or with the \CaStoD{3/2} transition in \Ca{}{+} ($\nu_{732}$; blue, dashed).}
    \label{fig: X21 plot}
\end{figure}

\FloatBarrier